\documentclass[preprint,showpacs,preprintnumbers,amsmath,amssymb,floatfix,aps,nofootinbib,aip]{revtex4-1}

\usepackage{graphics}% Include figure files
\usepackage{dcolumn}% Align table columns on decimal point
\usepackage{bm} 
\usepackage{gensymb}
\usepackage{textcomp}

\begin{document}

\preprint{}

\title{Potential barrier heights at metal on oxygen-terminated diamond interfaces}

\affiliation{Univ. Grenoble Alpes, Inst. NEEL, F-38042 Grenoble, France
\\
CNRS, Inst. NEEL, F-38042 Grenoble, France}    

\author{P. Muret}
\email{pierre.muret@neel.cnrs.fr}
\affiliation{Univ. Grenoble Alpes, Inst. NEEL, F-38042 Grenoble, France
\\
CNRS, Inst. NEEL, F-38042 Grenoble, France}    

\author{A. Traor\'{e}}
\affiliation{Univ. Grenoble Alpes, Inst. NEEL, F-38042 Grenoble, France
\\
CNRS, Inst. NEEL, F-38042 Grenoble, France}    
\author{A. Mar\'{e}chal}
\affiliation{Univ. Grenoble Alpes, Inst. NEEL, F-38042 Grenoble, France
\\
CNRS, Inst. NEEL, F-38042 Grenoble, France}    
\author{D. Eon}
\affiliation{Univ. Grenoble Alpes, Inst. NEEL, F-38042 Grenoble, France
\\
CNRS, Inst. NEEL, F-38042 Grenoble, France}
    
\author{J. Pernot}
%\altaffiliation{Institut Universitaire de France, 103 Boulevard Saint-Michel, F-75005 Paris, France}

\affiliation{Univ. Grenoble Alpes, Inst. NEEL, F-38042 Grenoble, France
\\
CNRS, Inst. NEEL, F-38042 Grenoble, France
\\
Institut Universitaire de France, 103 Boulevard Saint-Michel, F-75005 Paris, France}   

\author{J. C. Pin\~{e}ro}
\author{M. P. Villar}
\author{D. Araujo}
\email{daniel.araujo@@uca.es}
\affiliation{Dpto. Ciencias de los Materiales, Universidad de C\'{a}diz, 11510 Puerto Real (C\'{a}diz), Spain} 

\date{\today}

\begin{abstract}
Electrical properties of metal-semiconductor (M/SC) and metal/oxide/SC structures built with Zr or ZrO$_2$ deposited on oxygen-terminated surfaces of (001)-oriented diamond films, comprised of a stack of lightly p-doped diamond on a heavily doped layer itself homoepitaxially grown on a Ib substrate, are investigated experimentally and compared to different models. In Schottky barrier diodes, the interfacial oxide layer evidenced by high resolution transmission electron microscopy and electron energy losses spectroscopy before and after annealing, and barrier height inhomogeneities accounts for the measured electrical characteristics until flat bands are reached, in accordance with a model which generalizes that of R.T. Tung [Phys. Rev. \textbf{B 45}, 13509 (1992)] and permits to extract physically meaningful parameters of the three kinds of interface: (a) unannealed ones, (b) annealed at 350\textcelsius, (c) annealed at 450\textcelsius~with characteristic barrier heights of 2.2-2.5~V in case (a) while as low as 0.96 V in case (c). Possible models of potential barriers for several metals deposited on well defined oxygen-terminated diamond surfaces are discussed and compared to experimental data. It is concluded that interface dipoles of several kinds present at these compound interfaces and their chemical evolution due to annealing are the suitable ingredients able to account for the Mott-Schottky behavior when the effect of the metal work function is ignored, and to justify the reverted slope observed regarding metal work function, in contrast to the trend always reported for all other metal-semiconductor interfaces.
\end{abstract}

\pacs{73.20.–r, 73.30.+y, 73.40.Ns, 73.40.Qv, 85.30.De} % PACS, the Physics and Astronomy Classification Scheme.
\keywords{metal-semiconductor interfaces, Schottky diode, diamond}%Use showkeys class option if keyword
\maketitle

\section{Introduction}
Metal-semiconductor interfaces are necessary to implement sensors and electronic devices, and diamond does not escape to this need. Building well defined interfaces at nanometer scale is specially necessary for electrical rectifiers, which can gain high benefits from the bulk properties of boron doped diamond like its very high thermal conductivity, hole mobility and electrical breakdown field \cite{Volpe1, Volpe2} if specific properties of the interface are also fulfilled. These are mainly good adhesion, thermal and chemical stability at elevated temperatures, compatible with the superior possibilities of diamond, and potential barriers ensuring both low electrical losses under forward voltage and minimal reverse currents even at high temperatures. Several attempts have been proposed, relying either on materials which react with the bare surface of diamond to form carbides, like silicon \cite{Vescan} or refractory metals and compounds \cite{Muret1, Liao, Craciun}, even hetero-epitaxially grown on diamond \cite{Saby1}, or metals on either the oxygenated or hydrogenated  surface of diamond \cite{Evans, Teraji1, Baumann, Kawarada, Aoki}. On bare diamond surfaces, potential barriers close to 2~V, stable at temperatures higher than 500\textcelsius~have been obtained but reverse current densities and ideality factors were much greater than expected from the thermionic mechanism alone at least at room and moderate temperatures, and some of these contacts were ohmic. These poor rectifying current-voltage characteristics have been assigned to inhomogeneous and defective interfaces, specially with carbide forming metals \cite{Evans, Alvarez}. Lower potential barriers are implemented on the hydrogen-terminated diamond surfaces due to the dipole H$^{+\vartheta}-$C$^{-\vartheta}$ responsible for the negative electron affinity of these surfaces, where $\vartheta$ is the averaged fraction of the elementary charge per atom. But the chemical stability is not guaranteed at temperatures as high as in other interfaces and reverse current densities turn out not to be lower than $10^{-7}$~A/cm$^2$ at room temperature, still higher than the thermionic limit. Improvement has been achieved with the help of an oxide layer deposited on the H-terminated surface to build field effect transistors \cite{Daicho,KawaradaFET} but this solution is not relevant for Schottky barrier diodes firstly because it is very difficult to control accurately and in a reproducible way an oxide interlayer made of foreign species with thickness in the subnanometer range. Secondly, the Schottky barrier heights are known to be much smaller than 1~V and even zero for noble metals \cite{Tsugawa} on the hydrogenated diamond surface, thus inducing far too high reverse currents in rectifiers. Because of an inverted electric dipole at the oxygenated surface of diamond, the largest potential barriers are generally obtained with either noble (Au) or transition metals deposited on this type of surface \cite{Teraji1}, but some authors have shown that thermal treatments up to 500 or 600\textcelsius~are able to decrease the barrier height down to 1.2 V \cite{Ikeda, Teraji2}, probably because of the cancellation of the electric dipole O$^{-\vartheta'}-$C$^{+\vartheta'}$, still preserving a good adhesion of the metallic layer. More recently, we have shown that a less electronegative and more easily oxidizable metal like Zr was able to reach even more beneficial properties for the Schottky contact, namely an even lower potential barrier height while maintaining the reverse current density after an anneal at 450~\textcelsius ~close to $3\times 10^{-10}$~A/cm$^2$ at room temperature \cite{Traore} and an ideal current-voltage behavior for most diodes. However, a comprehensive picture of metal oxygenated-diamond interfaces did not emerge clearly in the last fifteen years, neither for the transport properties of carriers through interfaces, nor for the question of how the potential barrier height is determined, contrary to the case of hydrogen-terminated diamond surfaces \cite{Tsugawa}. In the present paper, mechanisms of carrier flow and significance of barrier heights are investigated on the one hand with more details and analysed with the help of several models. 
\\       
On the other hand, the set of works quoted above demonstrates that both chemistry and physics must be both considered when dealing with Schottky barriers. This is specially true in the case of diamond where a large dispersion of barrier heights occurred for any metal \cite{Evans} in literature. Much work was devoted in the past to general models able to explain and predict band misalignments at interfaces compounded of various materials, and a recent review of the various approaches, including a mixed physical-chemical point of view, has been published by R. T. Tung \cite{Tung2014}. But, as often the case for more than one decade in the literature excepted in ref.\cite{Tsugawa}, no experimental data regarding diamond interfaces has been brought and discussed. The present study aims at filling this gap, giving evidence of the presence of an oxide interfacial layer and discussing the relevance of various models of barrier height regarding data presented in this article and others available in literature for oxygenated diamond interfaces. In the second section, a detailed study by high resolution transmission electron microscopy (HRTEM) and electron energy loss spectroscopy (EELS) is developed to show the characteristics of the interfacial layer at a sub-nanometer scale, the more noticeable being the systematic presence of oxygen at interface before and after annealing. In the third section, a first level analysis of the current-voltage characteristics relevant to the depleted semiconductor interface is presented and representative parameters of the junction are derived, confirming the presence of an interfacial layer in-between Zr and diamond and the large decrease of the barrier height after annealing. In the fourth section, potential barrier inhomogeneities at different scales are analyzed and evidence of positively charged centers in the interfacial oxide is given. The consequences upon current-voltage characteristics are discussed and it is demonstrated that the junctions experience a change from depletion to accumulation regime at sufficient forward bias. Two models able to take into account the influence of the barrier height inhomogeneities are developed and matched to experimental data, allowing to derive meaningful parameters of each interface which indicate that large barrier inhomogeneities prevail before annealing while they almost completely vanish after sufficient annealing. The previous results are used in the fifth section to address the problem of whether such an accurate control of the interface and metal choice can allow to tailor and predict the potential barrier heights of Schottky junctions on diamond. Gathering all the data available for well characterized interfaces of various metals on oxygen-terminated diamond permits to unveil a completely unusual relationship between potential barrier heights and metal work functions. Modeling this new result needs to consider the contribution of all electrical dipoles which may be present at these compound interfaces. Conclusions are drawn in the sixth section.
               
\section{Nanostructure of Zirconium on oxygen-terminated diamond interfaces} \label{experimental}
Schottky junctions are implemented by electron-gun deposition in ultra high vacuum of Zr on a stack comprising a lightly boron doped homoepitaxial diamond layer grown on a heavily doped one, itself grown on a Ib substrate. The surface of the lightly doped layer is submitted to a photo-chemical UV-ozone treatment for two hours at room temperature, so that oxygen terminations prevail on the diamond surface. Other details of the diamond growth and interface preparation are given in reference \cite{Traore}. Three types of sample are elaborated, the first one being annealed at temperatures not exceeding 300\textcelsius, while the second and third ones are annealed respectively at 350\textcelsius~and 450\textcelsius; and the respective interfaces are labeled (a), (b) and (c). Identification of the oxide layer and evaluation of its thickness are carried out by a combination of high resolution transmission electronic microscopy (HRTEM), electronic energy loss spectroscopy (EELS), and high angle angular dark field-TEM (HAADF-TEM) techniques. HRTEM, HAADF and EELS measurements were performed using the beam of a JEOL 2010F STEM microscope directed on electron-transparent specimens of the interfaces labeled (a) and (c), fabricated by a lift-off technique in a focused ion beam (FIB) FEI Quanta 200-3D dual beam microscope.  
  
\begin{figure}
\includegraphics{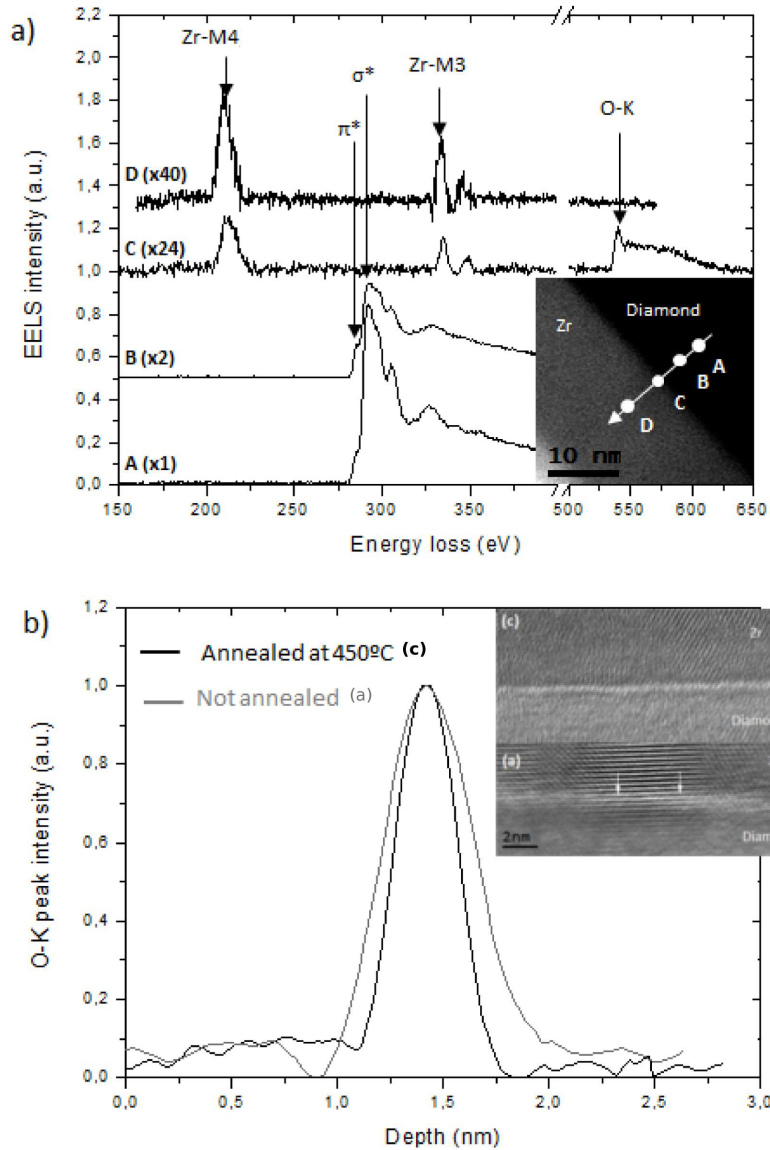}
\caption{\label{fig0} a) EELS spectra recorded in an interface of kind (c) along the linescan drawn in the HRTEM image shown in the inset. The peak Zr-M2 is also visible near 345 eV but Zr-M5 normally detected in a 10 nm thick ZrO$_{2}$ layer near 190 eV is missing in the interfacial layer; b) Normalized intensity profile of the oxygen related peak across the Zr/diamond interface before (grey curve) and after annealing at 450\textcelsius~for 30 minutes (black curve), corresponding to interfaces of kind (a) and (c) as labeled respectively in the text. In the inset, HRTEM filtered images of the Zr/oxygen-terminated diamond interfaces of kind (a, unannealed) and (c, annealed at 450\textcelsius) are shown. Superimposed contrasts of diamond and Zr can be observed in the Zr/diamond interface of kind (a), with white arrows to guide the eyes through the diamond/Zr interface.}
\end{figure}
The characterization of the oxide layer have been carried out in three steps: chemical contrast TEM (CTEM) identification of the metal/oxide/diamond interface, scanning TEM-EELS study of the previously identified oxide layer and HRTEM imaging of the interface.
In a first approach, bright-field (BF) images in CTEM mode are used to identify the interface. BF imaging variations in the electron absorption and dependence on the diffraction angles allow to distinguish between diamond and Zr material. In a second step, EELS spectra are acquired to chemically identify the oxide layer. During EELS spectra acquisition, HAADF-TEM operation mode is used to identify the diamond/Zr interface, insofar contrasts in HAADF imaging are due to different Z-number of the elements in the sample. Secondly, EELS profiles are carried out, revealing the oxide layer through the observation of the EELS oxygen-related peak (see Fig.~\ref{fig0}b). Thirdly, once CTEM and EELS techniques have been applied, HRTEM images of the previously identified oxide layer are acquired. Fig.~\ref{fig0}a shows EELS spectra recorded across the diamond/Zr interface after the thermal treatment, at the positions indicated in the inset. Presence of oxygen is revealed at the diamond/Zr interface (position C). Moreover, the simultaneous observation of the Zr-M2 to 4 and the O-K peaks at position C discloses the formation of an oxide layer at interface, however slightly different from bulk ZrO$_{2}$ since the peak Zr-M5 is missing. In one side of the interface, a diamond-related shape of the carbon core-loss is evidenced, while typical Zr-related peaks are observed in the other side. Fig.~\ref{fig0}b shows normalized intensity profiles of the oxygen-related peak (see spectra at position C) acquired across the diamond/oxide/metal interfaces before and after thermal treatment. To improve signal/noise ratio, ten spectra were averaged. The latter profile exposes how oxygen distribution changes after thermal treatment, showing a real oxide thickness reduction (FWHM varies from 0.6nm to 0.3nm) while the amplitude is increased by a factor 1.5, suggesting an augmented oxygen density. Note that the FWHM of the EELS linescan results from the convolution with the probe size of 0.2nm. Simultaneously, the EELS analysis of the oxygen content in the Zr layer indicates an averaged concentration of 5~\% in the as-deposited metal while it decreases to 1~\% after annealing. Finally, analysis of HRTEM images like shown in the inset of Fig.~\ref{fig0}b indicates some in-depth roughness in the interfaces of kind (a) whereas a smoother interface takes place after annealing in contacts of kind (c), leaving however an abrupt transition across the interface between carbon atoms and those of the interlayer in both cases.
\\
Titanium and Zirconium are well known to incorporate oxygen and possibly nitrogen when they condense onto a substrate after evaporation in vacuum, an effect which is clearly evidenced by a pressure decrease by about one order of magnitude inside the deposition chamber. Consequently, it is not surprising to find 5~\% of oxygen in the as-deposited Zr layer. Nitrogen is not detected. Increasing the temperature leads to augmented internal pressure and elastic energy of the Zr lattice, so that it becomes less expensive in term of total internal energy to form an oxide at the metal interfaces, both because the internal pressure is decreased and the enthalpy of oxide formation is negative. Such effects explain why the oxygen content of the Zr layer is decreased after annealing. More paradoxical is the sharpening of the oxide interlayer shown by EELS spectra because an increase of the oxygen content at interface should be expected from the evolutions described previously. At least three reasons can reconcile this apparent contradiction: (i) the beam spot (with an approximate size of 0.2nm) averages in-depth roughness during the EELS capture and, as a consequence, the recorded O-K profile of Fig.~\ref{fig0}b shows a widened profile in the interface of kind (a) before thermal treatment; (ii) we know that oxygen termination coverage has a lower density than the last carbon atom layer of the diamond lattice even after the most efficient treatment like UV-ozone oxidation used in this study \cite{Arnault}, a fact which correlates very well with the previous statement since a natural non-uniformity results before annealing; (iii) as demonstrated in the following sections, a strong change in electronic properties of the interface follows the 450\textcelsius~anneal, correlated with the electron affinity variation of the diamond surface induced by the cancellation of the oxygen terminations. However, because oxygen is still present at interface after annealing, one is forced to conclude that bonding has been deeply modified, with a probable transformation from a physisorbed metal layer on the oxygen terminated diamond surface likely through the hydroxyl terminations (see subsection \ref{variousterminations}), towards stronger bonds between carbon and oxygen atoms, characterized both by shorter bond lengths and a higher and more uniform density. Taking into account the convolution with the probe size, the oxygen thickness detected by EELS after 450\textcelsius~anneal in interfaces of kind (c) suggests that the oxide interlayer comprises two planes of oxygen atoms separated by a Zr atom plane, since the distance between two oxygen planes is 0.256~nm in cubic zirconia, thus defining a full unit cell of ZrO$_2$ as the whole interlayer thickness. Next sections are devoted to determine and understand how oxygen species at the interface influence the electrical properties.

\section{Current-voltage characteristics of the Zirconium-Oxygen-terminated diamond interfaces} \label{secmodel1}
\subsection{Theoretical background of the relationship between current and voltage}
Current-voltage characteristics and mechanisms of carrier transport through the potential barrier present in metal-semiconductor junctions have been discussed extensively in  review articles and textbooks \cite{Tung2014, Rhoderick1, Rhoderick2, Monchbook}. The \textit{z} direction being perpendicular to the $x-y$ plane of the interface, the current can be calculated in the semi-classical framework of the Maxwell-Boltzmann distribution of majority carrier velocities, by : 
\begin{equation}
I_{z}=\iiint{e_0\, v_z(E) \, D(E) \, dE \, dx \, dy}
\label{Igeneral}
\end{equation} 
$e_0$ being the elementary charge, $v_z(E)$ the carrier velocity in the \textit{z} direction and $D(E)$ the majority carrier distribution just at energies $E$ above the point where the potential barrier culminates. If the barrier is homogeneous, at a constant potential $ \Phi_B $ throughout the \textit{x-y} plane, the current density can be written from the previous expression, in the case of a voltage $V_j$ applied to the junction \cite{Rhoderick1, Crowell}: 
\begin{equation}
J_{z,\ref{Jhomogeneous}}=\frac{e_0 N_M v_{coll}}{1+v_{coll}/v_{diff}} \exp \left(-e_0 \Phi_{B}/k_B T \right) \left[ \exp (e_0 V_j/k_B T) -1 \right]
\label{Jhomogeneous}
\end{equation}  
with $N_M$ being the effective density of states at the majority carrier band edge, $ k_B $ the Boltzmann constant, $ T $ the absolute temperature, $v_{coll}$ the collection velocity at interface $ \sqrt{\frac{k_B T}{2 \pi m^*}} $, due to the thermionic effect and equal to one quarter of the thermal velocity, $m^*$ the conduction effective mass of majority carriers, and $ v_{diff} $ their diffusion velocity. The previous expression takes into account the two transport mechanisms of majority carriers which work in series, due respectively to the thermoionic velocity at interface and diffusion velocity inside the depletion zone, whereas it neglects bulk recombination. The diffusion velocity simplifies into the product of the majority carrier mobility by the electric field at interface if the band bending exceeds some $ k_B T/e_0 $ units and is generally not the limiting factor for the direct current, as confirmed in the samples under study from the comparison of the two velocities. The numerator of the first factor in Eq. (\ref{Jhomogeneous}) can be rewritten $ A^* T^2 $ where $ A^* = \left( 4 \pi e_0 k_B^2 m_0 / h^3  \right) \times (m^*/m_0) $, the first quantity inside the parenthesis being the Richardson constant for emission of electrons into vacuum with a mass $m_0$ at rest, $ A_R = $ 120~A~cm$^{-2}$~K$^{-2}$.    
Because $ v_{diff} $ decreases with the increase of the forward bias voltage $ V_j $, the pre-exponential factor of Eq. (\ref{Jhomogeneous}) becomes voltage dependent. Other reasons for such an effect involving also the factor $ \exp \left(-e_0 \Phi_{B}/k_B T \right) $, are (i)~the image force lowering of the barrier, (ii)~tunneling through the top of the barrier possibly assisted by the electric field, (iii)~voltage drop inside an interfacial layer between the metal and semiconductor, (iv)~charging interface states in equilibrium with the semiconductor and (v)~a spatially inhomogeneous barrier as discussed further. However, the second effect is noticeable only in heavily doped semiconductors, a case which is not achieved in this study since the boron concentration of the lightly doped layer is close to $ N_A = 5 \times 10^{15}$cm$^{-3}$. As one almost always wants to characterize the Schottky contact with a single potential barrier, the more common approximation consists in keeping only the zero bias voltage barrier height $ \Phi_{B}^0 $ in the first exponential factor of Eq. (\ref{Jhomogeneous}) and gathering all the voltage dependent effects into the last factor. Since all these effects contribute to a slope of the logarithm of the current which is smaller than $ e_0 / k_B T $, an ideality factor $n$ larger than one is introduced in the denominator of the exponent of $ \exp (e_0 V_j/k_B T) $ and often considered as a constant at least for a given temperature. Such an hypothesis is strictly valid in the case of an interfacial layer where the voltage drop is a constant fraction $1 - 1/n $ of the band bending $ U = \Phi_B^0 - V_j - V_{dop} $ (with $ e_0 V_{dop} $ being the energy difference between the Fermi level and majority carrier band edge in the bulk semiconductor) mainly determined by the oxide capacitance and density of interface states \cite{Card} when mechanisms (ii) and (iii) are combined. But it is not adequate if the interfacial layer either allows tunneling of carriers as the main transport mechanism or does not display any additional barrier for majority carriers because of band alignments similar to those occurring in type II heterostructures, or in the case of barrier inhomogeneities and even with image force lowering. Other effects, which are due to quantum reflexion, transmission and phonon scattering of carriers at the interface are voltage independent and generally very weakly temperature dependent so that it is convenient to lump them into an effective Richardson constant $ A^{**} $ which is weaker than $ A^{*} $ by some tens of percent in the case of intimate Schottky contacts \cite{Andrews}, yielding the standard value of 90~A~cm$^{-2}$~K$^{-2}$ usually considered for diamond. This effective Richardson constant can be further severely decreased because of the attenuation by a factor $ \alpha $ of the collection velocity $v_{coll}$ due to additional scattering of carriers by charged centers and roughness, and transport mechanisms through the interfacial layer when they exist. Finally, the non ideal current density is rewritten as : 
\begin{equation}
J_{z,\ref{Jnonideal}}= \alpha A^{**} T^2 \exp \left( -\frac{e_0 \Phi_{B}^0}{k_B T} \right) \left[ \exp \left(\frac{e_0 V_j}{ n k_B T} \right) -1 \right]
\label{Jnonideal}
\end{equation}  

\begin{figure}
\includegraphics{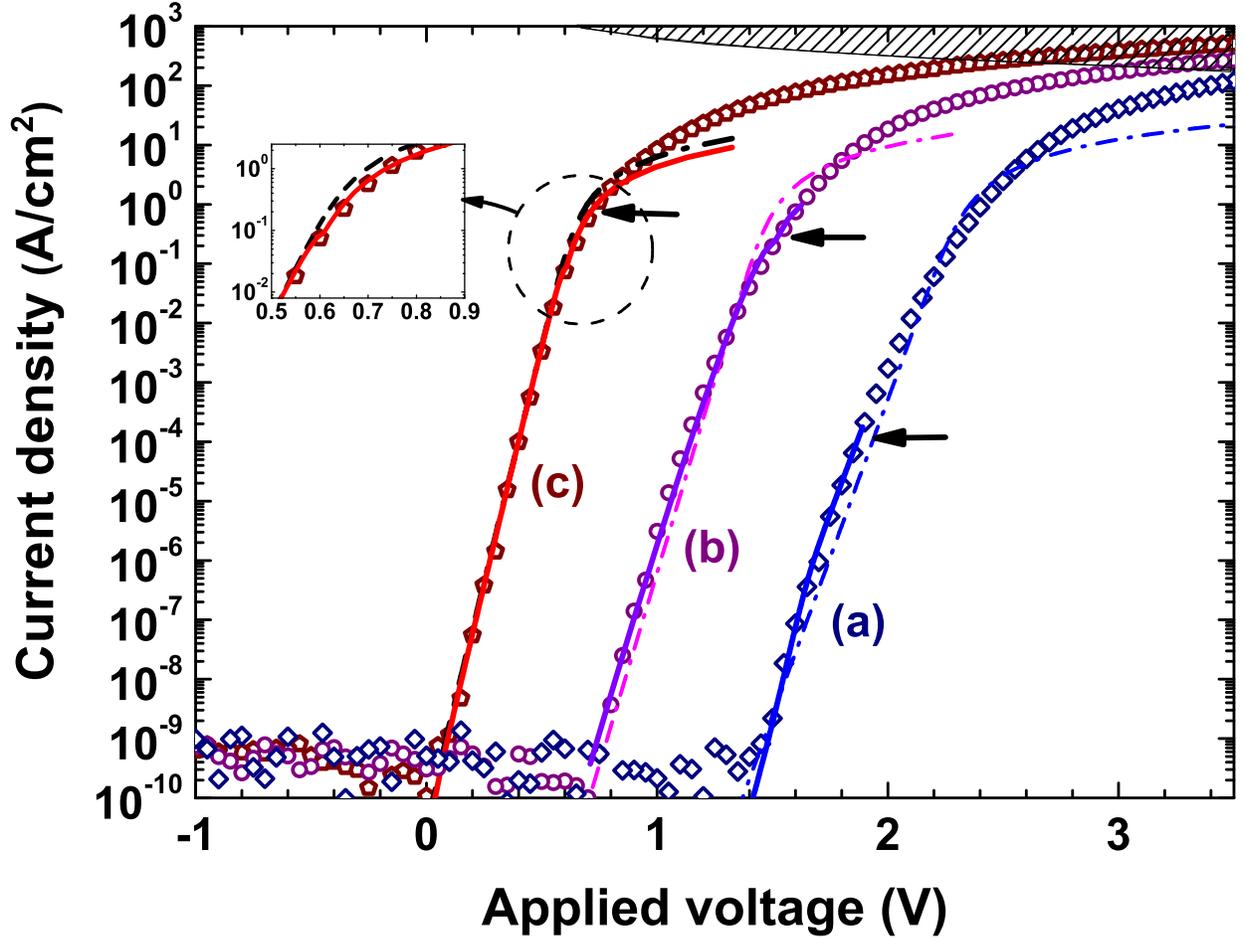}
\caption{\label{fig1} Current density as a function of applied voltage at 300 K in the three interfaces made of Zr deposited on oxygen terminated diamond (a) with annealing temperature not exceeding 300~\textcelsius; (b) annealed at 350~\textcelsius; (c) annealed at 450~\textcelsius. Dash-dot lines are calculated with the help of Eq. (\ref{Jnonideal}),  $ \alpha A^{**}$ values included in table \ref{Table1} which are obtained from fits achieved in the two following figures, and  adjusted constant values of the apparent barrier height and ideality factor in each case, that are respectively: (a) 1.83~V and 1.57 ; (b) 1.47~V and 1.29 ; (c) 0.97~V and 1.05. Voltage takes into account the voltage drop in a constant series resistance of 600~$\Omega$ in case (a) and 300 ~$\Omega$ in cases (b) and (c), in good agreement with the theoretical ones, but current densities above a few A/cm$^{2}$ turn out to be higher because of the conductivity increase due to hole injection from the heavily doped layer. Full lines rely on models which incorporate barrier height lowering and are described in section \ref{inhom-models}, with Eq.~(\ref{Jpatchy}) for the cases (b) and (c), while Eq.~(\ref{Jinhom}) is used for the case (a). Horizontal arrows mark the flat band situation in each junction. The effect of image force lowering is displayed in the inset for the case (a). In the hatched area, self-heating becomes important.}
\end{figure} 

\begin{figure}
\includegraphics{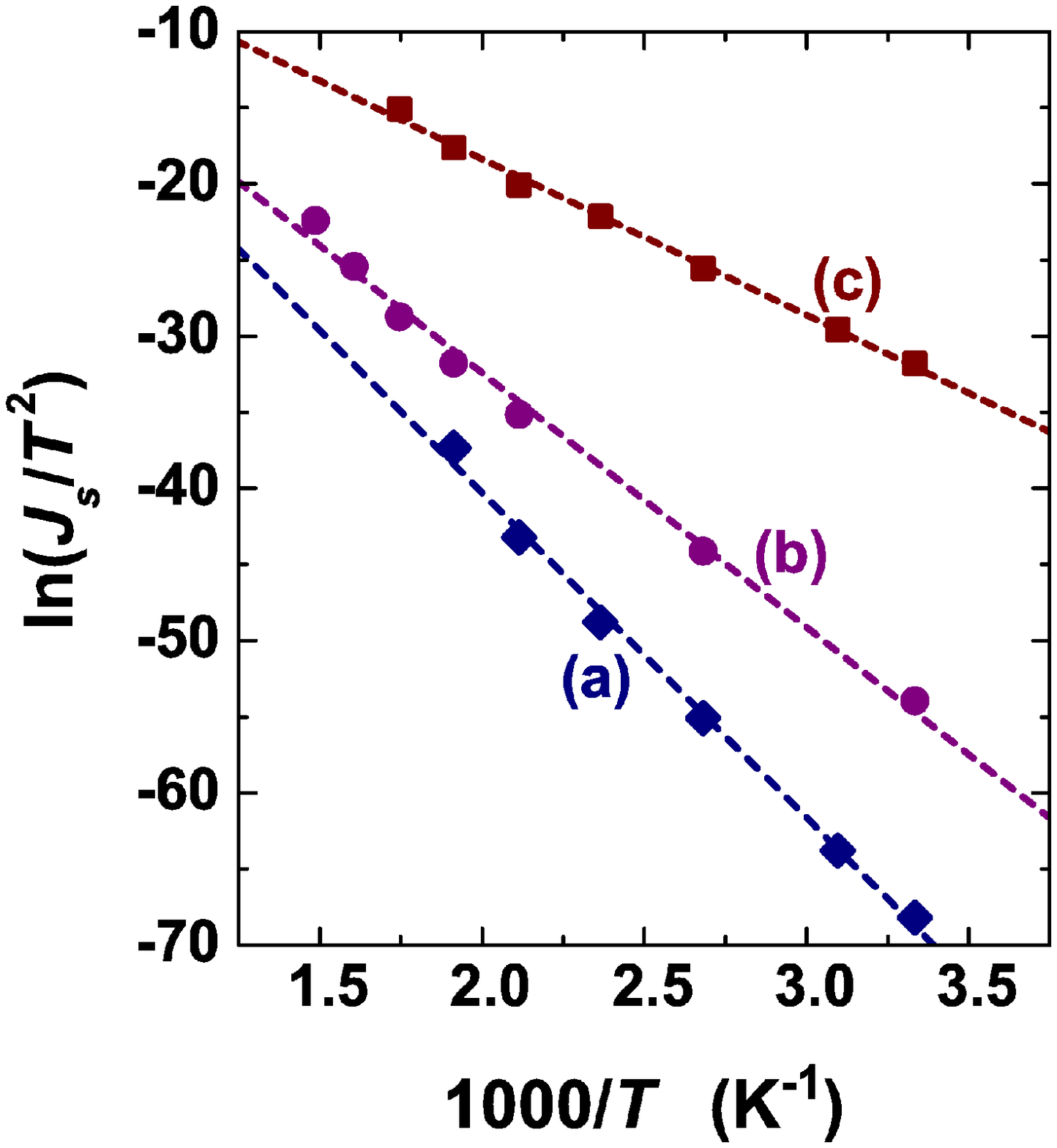}
\caption{\label{fig2} Richardson plot for the three same interfaces as in Fig. \ref{fig1}. Averaged barrier heights deduced from the slope of each straight line are respectively: (a) 1.84~V; (b) 1.47~V; (c) 0.95~V; very close to that used in Fig. \ref{fig1}.}
\end{figure}     

Experimental current-voltage characteristics can then be accounted for by such an expression with the ideality factor $n$ as an adjustable parameter, provided $\alpha A^{**} $ is known and the series resistance $ R_s $ is adjusted in order to fit the additional voltage drop  $ R_s I_z $ appearing in the total measured voltage  $ V = V_j + R_s I_z $. This is done in Fig.~\ref{fig1} for the three interfaces near room temperature. Two kinds of deviation between experimental data and the theoretical laws may appear as it can be seen in Fig. \ref{fig1} ; the first one is due to a non constant ideality factor as long as the voltage drop in the series resistance is negligible (typically below 0.5~A/cm$^2$ in these diodes at room temperature) and the second one is due to less common transport mechanisms which are discussed in the following. It is worthwhile to notice that the rectification ratios are more than twelve decades and that no generation-recombination current is detected on these current-voltage characteristics, which would have involved a symmetrical behavior of the absolute value of the current around the origin at low voltages. Such an almost ideal behavior  strongly suggests that the oxygen terminated surface of diamond passivates very efficiently the interface defects able to develop leakage currents \cite{Ohmagari}. The determination of $ \alpha A^{**} $ needs temperature variations in such a way that the saturation current density over $ T^2 $, namely $ \alpha A^{**} \exp \left( -\frac{e_0 \Phi_{B}^0}{k_B T} \right) $, obtained from the extrapolated value of the current-voltage characteristic at $ V_j = 0 $, can be drawn. Such Richardson plots are displayed in Fig. \ref{fig2}. The effective Richardson constants $\alpha A^{**} $, ideality factors at room temperature and parameters described in the following are indicated in the Table \ref{Table1}. It must be stressed firstly that the $\alpha A^{**} $ values are at least ten times lower than the standard value $ A^{**} $ and secondly that a careful inspection of the Fig. \ref{fig2} shows that the data points follow systematically a slight upward curvature, indicating a barrier height increase with temperature. The first result confirms indubitably the presence of an insulating interfacial layer, which is able to increase considerably the quantum reflection coefficient and decrease the effective velocity of carriers, mainly because of scattering due to roughness, chemical disorder and differential dielectric response. The second one is the preliminary indication of the presence of barrier inhomogeneities, a conclusion similar to that obtained in ref. \cite{Werner} about silicides-silicon interfaces, also implying additional consequences which have to be analyzed more thoroughly.

\begin{table} 
\begin{tabular}{||c||c|c|c|c||}
\hline
\hline \rule[-2ex]{0pt}{5.5ex} Sample type and & $\alpha A^{**}$ &$ n $&~~~~~$ \overline{\Phi_{B,\infty}^0} $~~~~~& ~~~~~~$\sigma_{\Phi}^0$~~~~~~ \\ 
\hline \rule[-2ex]{0pt}{5.5ex} annealing temperature (\textcelsius) & A~cm$^{-2}$~K$^{-2}$ & (at 300 K) & V & mV \\
\hline 
\hline \rule[-2ex]{0pt}{5.5ex} (a) $ < $~300~\textcelsius & 11 & 1.52 & 1.93 & 71 \\ 
\hline \rule[-2ex]{0pt}{5.5ex} (b) 350~\textcelsius & 2.8 & 1.29 & 1.57 & 87 \\ 
\hline \rule[-2ex]{0pt}{5.5ex} (c) 450~\textcelsius & 8.5 & ~1.00 to 1.07~ & 0.93 & 54 \\ 
\hline
\hline
\end{tabular}
\caption{Parameters of the Schottky barrier in the three interfaces according to the model of section \ref{secmodel1}.}
\label{Table1}
\end{table} 

\subsection{Richardson constant, ideality factor and potential barrier behaviors} 
\begin{figure}
\includegraphics{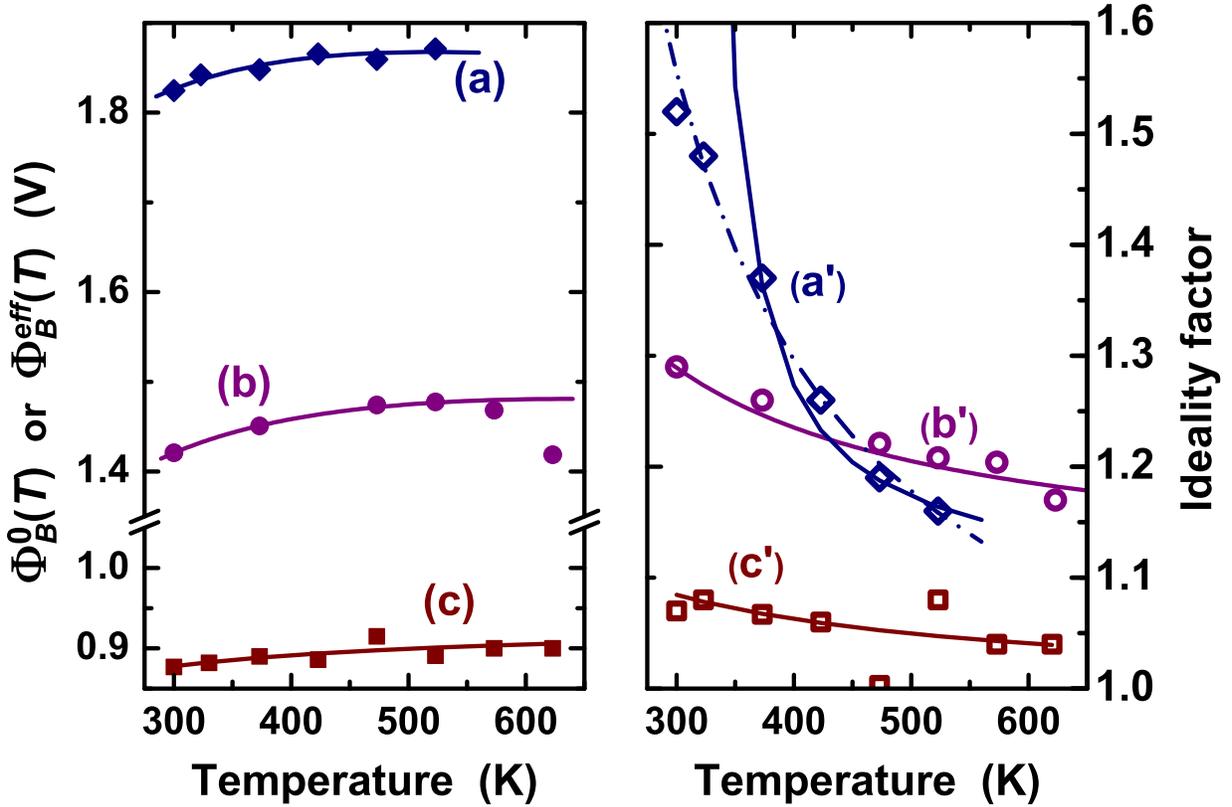}
\caption{\label{fig3} At left, effective barrier heights as a function of temperature in (a) diodes not annealed at temperatures higher than 300 \textcelsius; (b) annealed at 350 \textcelsius; (c) annealed at 450 \textcelsius ; full curves are the best fits according respectively to Eq.~(\ref{barrierinhomog}) or Eq.~(\ref{newbarrierTung}) which results essentially in the same behavior for the effective barrier. The point at the highest temperature in data (b) must be discarded because it has been measured at the same temperature as the annealing one, involving an additional annealing time which has induced a further evolution of the interface towards its final state (c). At right, ideality factors for the same diodes after the same respective post-treatments in (a'), (b') and (c') ; the best fits to the ideality factor data obtained with the help of Eq.~(\ref{idealityn}) are displayed in full curves for (b') and (c') and as a dash-dot curve for (a'), but using a non physical parameter in the last case. The full curve adjusted on (a') data exploits Eq.~(\ref{ideality-global}) and related ones with more meaningful parameters at the expense of a worse agreement with data at the lowest temperature (see text).}
\end{figure}

Zero voltage barrier heights and averaged ideality factors are displayed in Fig. \ref{fig3} as a function of temperature. On the first hand, the former quantities experience indeed an increase with temperature with generally an asymptotic behavior towards highest temperatures, while the latter ones are decreasing, starting from very different values at room temperature (which are reported in Table \ref{Table1}) and displaying very different negative slopes. As noticed previously, such variable ideality factors cannot be ascribed to a voltage drop within an interfacial layer bearing an additional barrier to that built in the semiconductor, because they would not be coherent with oxide interfacial layers which still exist at the three interfaces. On the other hand, experimental data plotted in Fig. \ref{fig1} deviate from the simulated curves drawn in dash-dot lines on the basis of a constant ideality factor when voltage drop in the series resistance and other effects remain negligible. A first attempt to fit more accurately experimental results uses an alternative definition of the current density, which is also more physically meaningful. It consists in using a temperature and voltage dependent barrier height $\Phi_{B}(T,V_j) $, so that the current density can be rewritten more usefully for the homogeneous barrier case:
\begin{equation}
J_{z,\ref{JphiBvoltdep}}= \alpha A^{**} T^2 \exp \left( -\frac{e_0 \Phi_{B}(T,V_j)}{k_B T} \right) \left[ \exp \left(\frac{e_0 V_j}{ k_B T} \right) -1 \right]
\label{JphiBvoltdep}
\end{equation}    
Introducing an inhomogeneous barrier which follows a Gaussian spatial distribution in the $x-y$ plane of the interface, characterized by an average value $ \overline{\Phi_{B}}(T,V_j)$ and a standard deviation $ \sigma_{\Phi}(V_j)$, permits to derive \cite{Monchbook,Werner} a law similar either to Eq. (\ref{Jnonideal}) or (\ref{JphiBvoltdep}), with an effective barrier given in the second case by :  
\begin{equation}
\Phi_{B}(T,V_j)= \overline{\Phi_{B}}(T,V_j) - \frac{e_0 \sigma_{\Phi}^2(V_j)}{2 k_B T} 
\label{barrierinhomog}
\end{equation}    
For each temperature, it is always possible to convert Eq. (\ref{JphiBvoltdep}) into Eq. (\ref{Jnonideal}) through the definition of a voltage independent barrier height, conveniently taken at $ V_j = 0 $, and a voltage dependent ideality factor $n(T,V_j) $, valid in all situations, the inverse of which being evaluated from the derivative of the logarithm of the current density given by Eq. (\ref{Jnonideal}) and (\ref{JphiBvoltdep}) when the reverse saturation current is neglected:
\begin{equation}
\frac{k_B T}{e_0} \frac{d(\ln J_z)}{dV_j} = \frac{1}{n(T,V_j)}= 1 - \frac{d \Phi_{B}(T,V_j)}{d V_j} = 1 + \frac{d \Phi_{B}(T,U)}{dU}
\label{idealityfactor}
\end{equation}      
However, in this first level model, no explicit dependence of the parameters is given as a function of junction voltage $ V_j $ or band bending $ U $. Some cases of potential barrier distributions have been analyzed analytically by R. T. Tung \cite{Tung1992}, who showed that the dependence of the parameters of Eq. (\ref{barrierinhomog}) was not linear with the band bending and hence with the junction voltage, on the basis of a physical representation of the barrier lowering by a fictitious dipole located at interface. An additional source of non ideality may be present if a spreading resistance  effect occurs, involving inhomogeneous current density inside the semiconductor because of current crowding near the pinch-off patches with a lowered barrier and located close to the interface \cite{Tung2001, Gammon}. This effect is normally expected to be relevant at current levels where the series resistance voltage becomes larger than some $ k_B T / e_0 $. But the combination of multiple barriers and incomplete ionization of doping impurities may induce deviation from the ideal current-voltage characteristic at much smaller currents because of severe weakening of the carrier concentration and narrowing of the current paths in the lowest barrier patches which cannot be assessed easily from the comparison to a voltage drop in an averaged series resistance. This kind of non ideality will be discussed in the next section.   

In the simple model presented above, the barrier fluctuations are characterized by their average and standard deviation but are not supported by any physical model, thus preventing the derivation of any relationship between physical properties of the interface and electrical ones. Moreover, the behaviors of the barrier lowering and ideality factor with voltage and temperature cannot be deduced. Therefore, the reasons for these barrier inhomogeneities have to be analyzed more accurately and an analytical model which will generalize that developed by R. Tung \cite{Tung1992} will be implemented in the next section, concurrently with a realistic simulation of some potential fluctuations at interface. But within the framework of the simple approach developed in the present section, it is already possible to derive the two parameters $\overline{\Phi_{B}}(\infty,0) $ and $\sigma_{\Phi}(0)$ at zero voltage from the fit of these quantities according to Eq. (\ref{barrierinhomog}), which are renamed respectively $ \overline{\Phi_{B,\infty}^0} $ and $ \sigma_{\Phi}^0$ for sake of simplicity and displayed in Table \ref{Table1}, for the three types of interface. The average barrier height at infinite temperature $ \overline{\Phi_{B,\infty}^0} $ decreases by about 0.5 V after each anneal while the standard deviation $ \sigma_{\Phi}^0$ of the barrier fluctuations goes through a maximum after annealing the interface at 350~\textcelsius. But the deficiency of the present model is confirmed by the different behaviors of the ideality factor with temperature plotted in Fig. \ref{fig3} which cannot be explained. These discrepancies confirm the need for a more accurate model of the effect of potential fluctuations at interface upon current-voltage characteristics.

Meanwhile, an unusual result which ensues from the calculation of the current densities necessary to draw the fitting curves in Fig. \ref{fig1} can be already pointed out. At some current threshold, flat bands are reached (see the arrows in the same figure) and it is necessary to consider that accumulation occurs at the interface to explain the continuous increase of the current densities beyond this threshold. Such a behavior matches Eq. (\ref{Igeneral}) which states the proportionality of the current to a velocity and carrier concentration at interface. For example in the case (c), flat bands are reached at a voltage of 0.63~V, corresponding to an initial band bending of $0.97 - 0.38 = 0.59 $V (here $ V_{dop} $=0.38~V) plus the barrier increase of 0.04 V demonstrated in the next section. The corresponding current density calculated directly with the hole concentration given by formula (1) in ref.\cite{Volpe1}, assuming a compensation of 20$\%$ (donor concentration $ N_D = 10^{15}$cm$^{-3}$) and Eq. (\ref{Igeneral}), with a collection velocity ten times smaller than that for an intimate contact, yields 0.4 A/cm$^{2}$, very close both to the experimental one and that calculated by Eq. (\ref{JphiBvoltdep}).  In cases (b) and (a), still at 300K, flats bands are reached respectively at lower current densities indicated in Fig.~(\ref{fig1}), which cannot be calculated with the help of the initial model because the apparent barrier heights deduced presently are largely underestimated and would produce irrelevant band bending. Therefore, assessing these values needs to wait for more accurate and physically relevant models which will be developed in the next section. In Eq. (\ref{Jhomogeneous}) and subsequent ones, the barrier height has been introduced for convenience and it was often believed as an impassable limit in Schottky diodes. Such a statement is not true as everyone can be convinced if he realizes that Eq. (\ref{Jhomogeneous}) and subsequent ones all derive from Eq. (\ref{Igeneral}) where the current is controlled by a carrier concentration. Band curvature reversal and occurrence of accumulation regime are possible and even necessary to reach high currents, up to more than thousand times those got at flat bands, in these interfaces made on oxygen-terminated diamond, because firstly of the low values of the effective Richardson constant, secondly of the wide forbidden bandgap of diamond and thirdly of the large ionization energy of acceptors. In such a regime, diffusion flow is inverted close to the interface, the barrier seen by majority carriers is canceled and consequently, the models relying on Eq. (\ref{Jhomogeneous}) and following ones, and on the properties of barrier height inhomogeneities, specially their voltage dependence, become inappropriate. Conversely, current crowding near the pinch-off patches with a lowered barrier may become a major source of non ideality. These mechanisms will be either incorporated into the forthcoming models or else, kept in mind to assess the applicability of the following developments.

\section{Potential barrier inhomogeneities and ideality factor} \label{inhom-models}

\subsection{Potential fluctuations due to randomly distributed discrete charges and dipoles} \label{variousterminations}
The electrostatic potential experienced by mobile carriers originates from fixed charges scattered inside the space charge zone and interface. For modeling purposes, it is often derived from constant bulk charge concentration in the space charge zone and homogeneous charge density at interface. This view is correct for mechanisms characterized by lengths larger than the Debye length $ \lambda_D = \sqrt{\varepsilon_{SC} \varepsilon_0 k_B T /(e_0^2 N_A)} $, where $ \varepsilon_{SC} $ and $ \varepsilon_0 $ are respectively the relative permittivity of the semiconductor and the vacuum permittivity, and when average laws are evaluated, but may fail if some physical property of carriers experiences microscopic fluctuations. Indeed, when the effects of potential inhomogeneities are at stake, the discrete nature of fixed charges accommodated within the metal/semiconductor or metal/oxide/semiconductor (MOS) structure must be taken into account and the average distance between these discrete charges have to be compared to $ \lambda_D $. Here, $ \lambda_D $ amounts to 40 nm at 300 K and 52 nm at 500K. Three locations have to be considered: (i) inside the depletion zone, the average distance between ionized acceptors is 58 nm, very close to $ \lambda_D $. Only those which are at such comparable distances from the interface influence the potential barrier height, as clearly evidenced in a following subsection. (ii) The oxygen terminations bonded to the carbon atoms of the (001) diamond surface are making up the negative part of the interface dipole, really present at the interface up to some annealing temperature, an inverted situation in comparison to what happens on the hydrogenated diamond surface, due to the sign change of the electronegativity difference between carbon and oxygen or hydrogen respectively. The positive counterpart of this dipole lies within the first and second carbon layer just below the surface, resulting in a potential jump through the surface of $ \Delta q \, d_{di} \, N_{di} / (\varepsilon_i \, \varepsilon_0) $ where $ \Delta q $ is the charge transfer, $ d_{di} $ the dipolar distance, $ N_{di} $ the site density and $ \varepsilon_i $ the relative interface permittivity  \cite{Monchbook}. This potential jump is directly reflected by the electron affinity change provided by oxygen terminations. In average, it can be estimated close to 1.7 V because the smaller difference between electron affinities of the free $ (2 \times 1) $ reconstructed surface and the oxygenated surface  \cite{Ristein} is probably due to already electron depleted bonds in the last carbon layer of the former one. Taking into account the 30 percent coverage measured by X-rays photo-electron spectroscopy \cite{Arnault} on our samples treated by UV-ozone, equivalent to $ 5 \times 10^{14} $~sites per cm$ ^{2} $, a whole distance normal to the interface of roughly $ d_{di} = 0.1 + 0.2 = 0.3 $~nm and $ \varepsilon_i = 2 $, the estimated charge transfer turns out to be 2$ \times 10^{-20} $~C, that is $\vartheta'$=0.13 electronic charge. However, it is known firstly that there are several oxygen terminations, at least ether (C\textendash O\textendash C), carbonyl or ketone (C=O), hydroxyl (C\textendash OH) and carboxyl (O=C\textendash OH) groups \cite{Klauser, Ghodbane, Sque, Notsu} as individual adsorbates which can be simultaneously present on the diamond surface, and possibly epoxide (bridged oxygen between two adjacent carbons)\cite{Fink} in case of UV-ozone treatment as done on our samples. The electron affinities calculated theoretically for the three first terminations are different \cite{Robertson1,Sque}. Secondly, all the sites are not occupied by ether or ketone terminations (in our case, about 30 percent \cite{Arnault}, like when oxygen plasma treatment is applied to the diamond surface) but when adjacent sites are provided with an oxygen termination, polymerization along chains may occur \cite{Klauser}. Thirdly, comparison between the experimental electron affinity of the oxygen terminated (100)-oriented diamond surface \cite{Muret3, Ristein}, near 1.6~eV or 1.7~eV, and those calculated theoretically \cite{Robertson1,Sque} (the "bulk" one in the sense of the Sque publication, that is the difference between the bottom of the conduction band and vacuum level at the surface) shows that the latter are about twice the experimental ones for ether and carbonyl terminations and close to zero or negative like for the hydrogen termination in the case of hydroxyl. It is therefore quite clear that a mixing of at least these three types of terminations is necessary to reconcile experimental and theoretical electron affinities as long as those measured by photo-electron spectroscopy are averaged values. Therefore, the dipole strength resulting from these various adsorbates and able to modify the potential barrier height is far from being homogeneous at least over a scale length of few nanometers and also along a much larger scale if the various terminations are arranged in macroscopic domains, resulting in a non-uniform local electron affinity. An other reason which may lead to scattered patches with different barrier heights would be the chemical inhomogeneities like often present when carbide formation occurs \cite{Alvarez}, which might also happen if the chemical composition  of the interfacial oxide layer or its thickness were not homogeneous throughout the interface. And last but not least, extrinsic defects such as emerging dislocations cannot be excluded as sources of locally lowered barrier height like revealed at CoSi$_2$/\textit{n}-Si(001) interfaces by ballistic electron energy microscopy \cite{Sirringhaus}. On the contrary, the intrinsic dipole due to charge transfer across the interface because of a possible misalignment of the charge neutrality levels in the two materials is supposed to be homogeneous because it results from the occupation of gap states which are delocalized in the plane of the interface and will not participate in potential fluctuations. (iii) As evidenced in the following, discrete charged centers exist inside the oxide of MIS structures and are expected to be also present in the thin interfacial oxide of the Schottky contacts. These cases are treated quantitatively in the next subsections and are brought face to face to experimental and simulation results.
  
\subsection{Evidence of charged centers in oxide layers}
\begin{figure}
\includegraphics{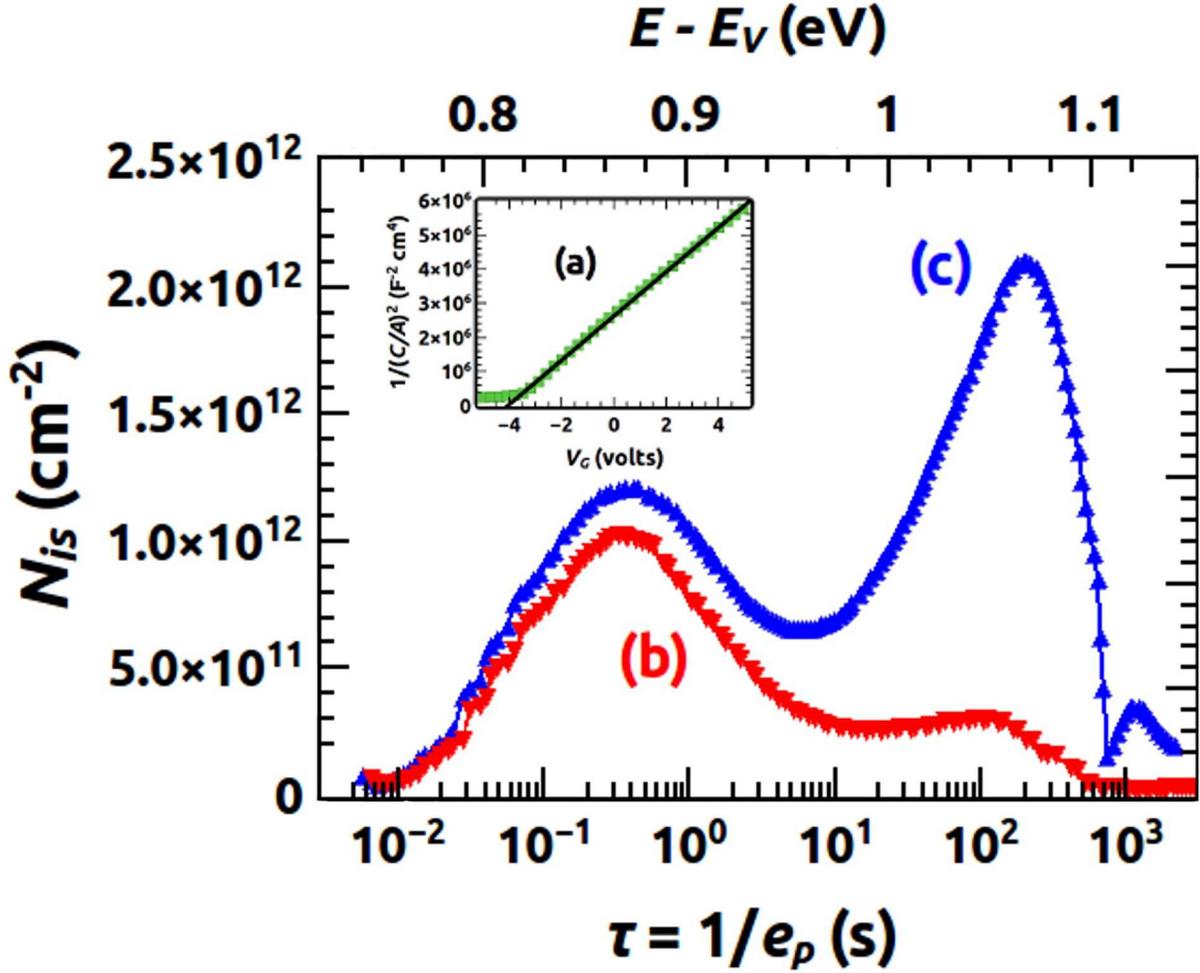}
\caption{\label{fig4} (a) Square of the inverse capacitance density as a function of the gate voltage ; (b) and (c) curves are isothermal transient capacitance spectra respectively after a 1 ms and a 100 ms filling pulse, recorded at a gate voltage of 3 V and temperature of 362 K.}
\end{figure}

To check quantitatively the presence of discrete charged centers inside the oxide, a MIS structure Al/ZrO$_{2}$/p-type (001) oriented diamond has been fabricated. The 25 nm oxide layer has been prepared by atomic layer deposition at 100 \textcelsius. The p-layer has a thickness of 2 $\mu$m, contains some 10$^{17}$ B/cm$^{3}$ and has been epitaxially grown on top of a 200 nm p$^{++}$ diamond layer itself in epitaxial relationship on a (001) oriented Ib substrate. After etching the p-layer and before the deposit of the last stack Al/ZrO$_{2}$, an ohmic contact is fabricated around the sample on the heavily doped (p$^{++}$) layer, ensuring a series resistance close to a few k$\Omega$. In this way, the proper time constant of the MIS structure is inferior to 1 $\mu$s. Such a property allows safe measurements of the static, pulsed and transient capacitances at a frequency of 1 MHz, as described in reference \cite{Muret2}. Results of the reciprocal square capacitance as a function of voltage and the  isothermal spectra of projected interface states as a function of the hole emission time constant deduced from the first real Fourier coefficient of the capacitance transients at two time window durations are displayed respectively in Fig.~\ref{fig4}(a) and Fig.~\ref{fig4}(b) and (c). From the extrapolation of the linear part of the curve in Fig.~\ref{fig4}(a) to zero ordinates, a flat band voltage $ V_{FB} $ of $ -4.1 $~V is deduced  whereas a value equal to $\Phi_{MS}=\Phi_{M}-\chi_{SC}-E_{G}/2e_0-\left(E_{i}-E_{F}\right)/e_0 = -2.4$~V would be calculated with the help of the electron affinity model and $\Phi_{M}$~=~4.28 V for the Al work function \cite{Michaelson}, $\chi_{SC}$~=~+1.6eV for the electron affinity of oxygen-terminated p-type diamond \cite{Ristein, Muret3}, $E_{G}$~=~5.45~eV for the diamond band gap and $\phi_{F}=\left(E_{i}-E_{F}\right)/e_0$~=~2.35~V deduced from the difference between intrinsic and Fermi energy levels in bulk boron-doped diamond, assuming no charge in the oxide.  The discrepancy comes from this last hypothesis which is not verified, implying that the real situation comprises a positive charge inside the parenthesis of the last term in $V_{FB}=\Phi_{MS}-\left(Q_{is}(0)+\overline{Q_{ox}}\right)/C_{ox}$, where $C_{ox}$ is the oxide capacitance density, $Q_{is}(0)$ and $ \overline{Q_{ox}}$ being respectively the charge per unit area in interface states at flat bands and the product of the total charge per unit area within the insulator by the ratio of the barycenter of this charge to the insulator thickness \cite{Muret2}. To reconcile the two values, a total number of positive charges per unit area must amount to 2$\times 10^{13}$~cm$^{-2}$ within the 25 nm thick oxide layer.

The centers which accommodate charge may experience change of their charge state if some mechanisms are occurring, like carrier emission from interface states lying in the common forbidden gap of the two materials, and also transport by hopping and tunneling across some thickness of the insulator. Such effects give rise to transient capacitances in the MIS structure because of the rearrangement of all the individual electric dipoles necessary to comply with the rule of a constant whole electric dipole if the MIS is submitted to a constant voltage \cite{Muret2}. Isothermal spectra are displayed in Fig.~\ref{fig4}(b) and (c) as a function of the characteristic emission time. They are obtained from the Fourier transform of the transients which are recorded after two filling pulses lasting respectively 1 ms and 100 ms. They show a two peaks structure, the slower peak being higher for the longer filling pulse. This response can be viewed as the projected spectrum of states on the interface energy scale inasmuch both states located at interface and some oxide states as well may exchange holes with the valence band of diamond. Differential spectroscopy is here unnecessary just because emission from discrete interface states in the common bandgap largely dominates that from a hypothetical continuum since peaked structures clearly appear in Fig.~\ref{fig4}, and an accurate calibration of the energy scale is not needed. The smaller filling pulse duration $ t_p $~=~1~ms has been chosen to ensure complete filling of interface states even with capture cross sections as low as $ 10^{-19}$~cm$^{2}$, whereas the larger $ t_p $~=~100~ms is able to involve additionally either injection or extraction of carriers trapped in discrete states localized within the insulator at some distance from the interface. The important point consists in the different behaviors followed by the two peaks as a function of $ t_p $: the first occurring at the shorter emission time is very weakly sensitive to $ t_p $ and consequently related to an interface state whereas the second one increases with $ t_p $, involving carrier transport through some insulator slice. This is the demonstration that at least one discrete state exists inside the oxide. An order of magnitude of the distance $ d $ of these centers from the interface can be estimated from the inverse of the attempt-to-escape time checked by Jain and Dahlke for traps in Si/SiO$_2$ interfaces \cite{Jain} and given by $ \nu_0 \exp (-q d) $, with $ \nu_0 \simeq 10^{16} $~s$^{-1}$ and $ \alpha \simeq 1.05 \AA{}^{-1}$, yielding roughly $ d  \simeq $~3~nm. The reciprocal attempt-to-escape time at zero insulator thickness $ \nu_0 $ and the quantum transmission coefficient $q$ are conditioned respectively by the quantum uncertainty about the potential well or barrier height, also close to $ 2\hbar / G $  where $G$ is the line width of the resonance \cite{Collins}, and tunneling transmission coefficient which depends on the square root of the previous quantities. Because the relevant energies are still about 1 eV, the order of magnitude of $ \nu_0 $ and  $q$ are expected not to be largely different in the present case.  The evaluated distance $ d $ is close to one tenth of the whole insulating layer thickness, a ratio similar to that of the center densities deduced by the transient and static capacitance methods in the MIS structure. Consequently, these point defects behaving as repulsive centers for holes and located inside the oxide layer at the Schottky interfaces under study might not be ignored unless other charged centers or dipoles prevail.

\begin{figure}
\includegraphics{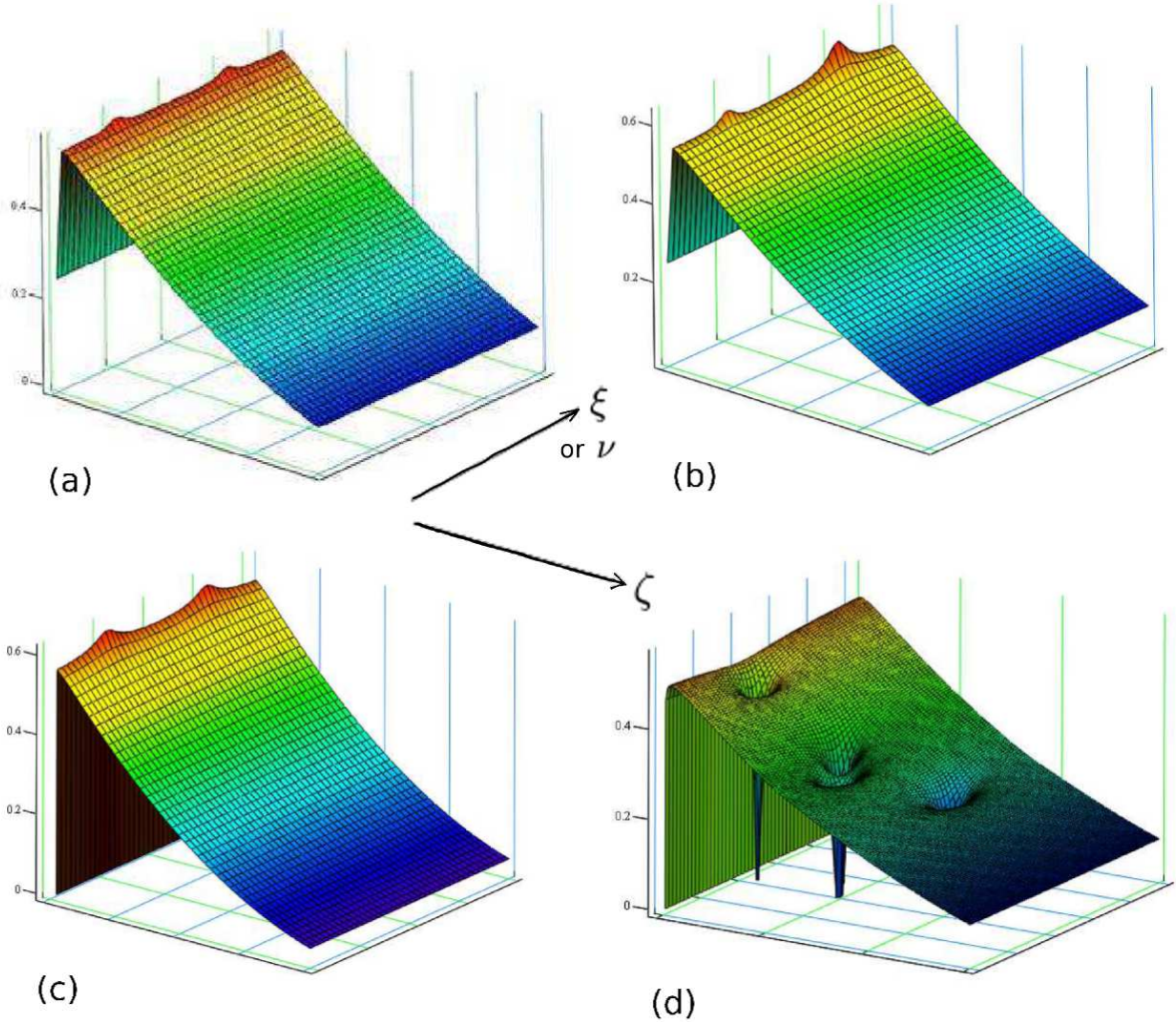}
\caption{\label{fig5} Sketch of the potential barrier experienced by majority carriers with (a) two dipoles across the interface with a repulsive elementary charge on the semiconductor side; (b) one dipole of the same kind across the interface and one repulsive elementary charge beyond it; (c) two repulsive elementary charges beyond the interface; (d) randomly scattered attractive elementary charges inside the depletion zone, inducing a saddle point of the potential energy, visible as a lowered profile of the top of the barrier near the potential well which is the closest to the summit.}
\end{figure}

\subsection{Effects of local barrier enhancement due to random elementary charges and dipoles close to the interface} \label{elementarycharges}
Electrostatic potential along $ x $ and $ y $ directions, parallel to the interface, and $ z $, normal to the interface, can be calculated by adding three terms as a function of normalized distances $ \xi = x/\lambda_D $, $ \nu = y/\lambda_D $ and  $ \zeta = z/\lambda_D $, and band curvature $ U $ in volt: the average potential of the depletion zone $ \left(\sqrt{U - k_B T /e_0} - \zeta \sqrt{k_B T /(2 e_0)} \right)^2 $; the image force potential $ -u_{im}/ \zeta $ with $ u_{im} = e_0 /(16 \pi \varepsilon_{SC} \varepsilon_0 \lambda_D) $ and the potential due to either repulsive elementary charges positioned symmetrically at $ \xi = \pm \xi_0 $, $ \nu = 0 $, and beyond the interface plane by a distance $ \zeta_0 $, that is  
\begin{equation*}
 \frac{4 u_{im}}{\sqrt{(\xi - \xi_0)^2 + \nu ^2 + (\zeta + \zeta_0)^2}} + \frac{4 u_{im}}{\sqrt{(\xi + \xi_0)^2 + \nu ^2 + (\zeta + \zeta_0)^2}}
\label{elemenqpot}
\end{equation*} or replacing either one or two of the previous terms by
\begin{equation*}
 -\frac{4 u_{im}}{\sqrt{(\xi \mp \xi_0)^2 + \nu ^2 + (\zeta + \zeta_0)^2}} + \frac{4 u_{im}}{\sqrt{(\xi \mp \xi_0)^2 + \nu ^2 + (\zeta - \zeta_0)^2}}
\end{equation*} at either one location $ + \xi_0 $ or both locations $ \pm \xi_0 $  to simulate either one or two elementary dipole due to negative oxygen termination beyond the interface and its positive counterpart on the diamond side.
A 3-dimensional view of potential barriers in several cases is given in Fig.~\ref{fig5}, the last one being obtained with an inverted sign before $ \zeta_0 $ and four random locations of elementary charges within the depletion zone. The three first cases are respectively devised to mimic the mechanisms assumed to be at stake in the transport of carriers at interface, resulting from the repulsive effect of charges located just at interface or close to it inside diamond: (a) with negative oxygen terminations inducing interface dipoles which are supposed to dominate; (b) with a mixed influence of repulsive oxide charges and dipoles similar to those of the previous case; (c) with only repulsive oxide charges. In these three cases, elementary charges or dipoles enhance the barrier locally. On the contrary, in the fourth case (d), barrier lowering $ \delta \Phi_{B,1}^{sad} $ may occur with a saddle point at normalized distance $\zeta_{sad}$ from the interface which can be calculated exactly from the solutions of a quartic equation resulting from the cancellation of the potential derivative along the $ \zeta $ direction. However, a rough estimate, easier to derive, relies on the geometric mean of the two extreme solutions of the quartic equation, and can be expressed as $ \zeta_{sad} \approx \sqrt{\chi(U) \left[ \zeta_0-2\chi(U) \right]}$ with $\chi(U) = u_{im}^{1/2} \left (\frac{2k_B T}{e_0} \right )^{-1/4} \left (U-\frac{k_B T}{e_0} \right )^{-1/4}$,  and results in $ \delta \Phi_{B,1}^{sad} \approx 4\:u_{im}/ \left (\zeta_0-\zeta_{sad}\right )$. The resulting effects of this barrier height lowering will not be treated in detail here.  In the three first cases, the distribution function of potential barrier heights is then calculated with a fixed value of $ \zeta_0 $ and random values of $ \xi_0 $ around a mean $ \overline{ \xi_{0}} $ according to a normal law with a standard deviation $ \sigma_{\xi} $, to assess the possible values of $ \overline{\Phi_{B}}(T,V_j)$ and $ \sigma_{\Phi}(V_j)$, the characteristic parameters of the inhomogeneous barrier.

\begin{figure}
\includegraphics{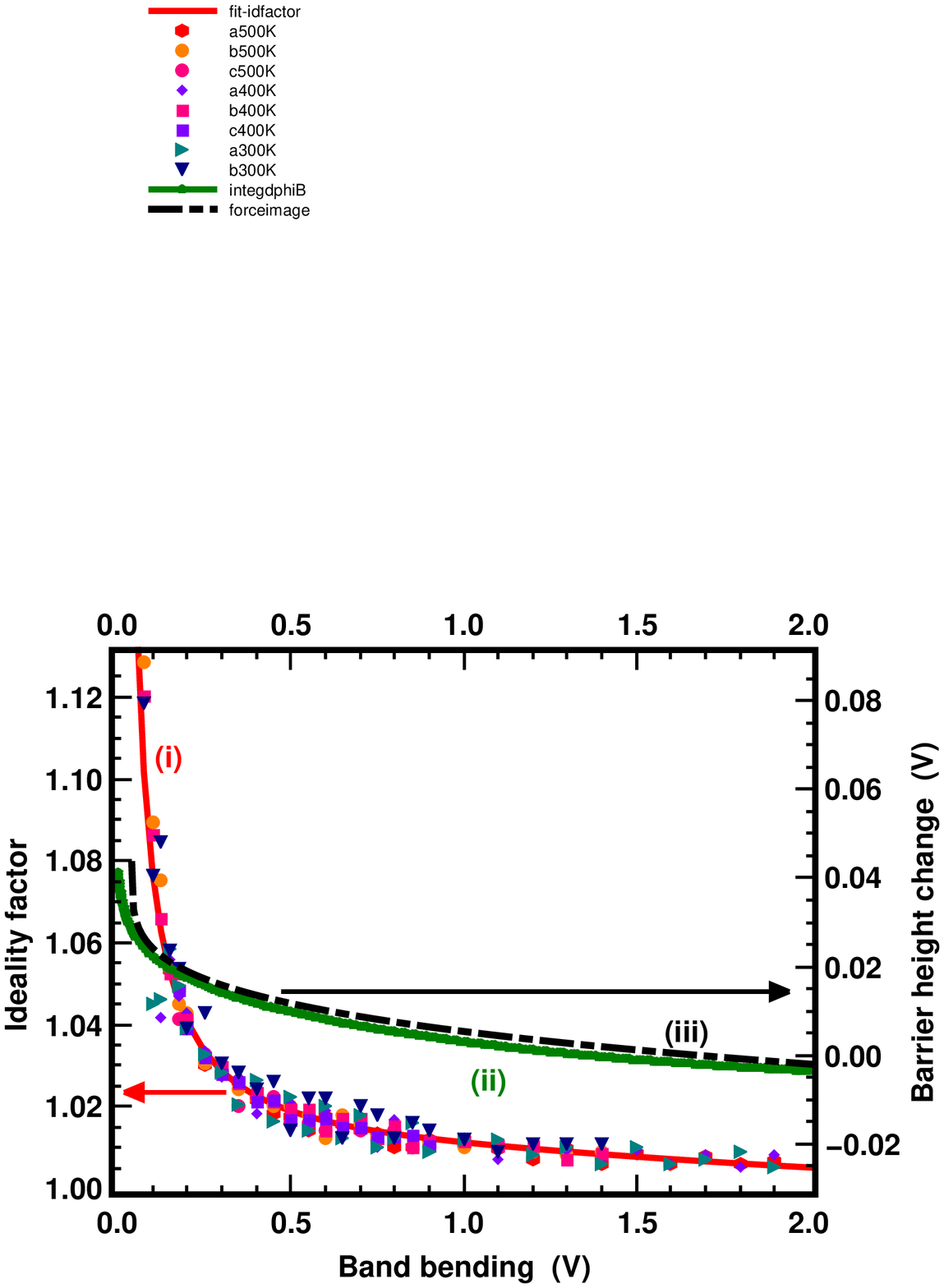}
\caption{\label{fig6} Ideality factors deduced from the microscopic barrier inhomogeneities model for the three types of interface at 300 K (triangles), 400 K (squares) and 500 K (circles) as a function of the band bending $ U $. Curve (i) is the best fit of data points by $ 1+ 1/P(U) $ with $ P(U)=136.8\,U-76.4\,U^2+28.4\,U^3 $. The curve (ii) displays the barrier height variation deduced from (i) with help of Eq. (\ref{idealityfactor}) which implies that $ d\Phi_B\,/\,dU = -1/\left[1+P(U) \right] $. Dash-dot curve (iii) shows the image force lowering variations. }
\end{figure}

In a first step, the parameter $ \sigma_{\Phi}(0)$ is adjusted to match the values displayed in Table~\ref{Table1} at a medium temperature, typically 500~K with realistic values of  $ \zeta_0 $, $ \overline{ \xi_{0}} $ and $ \sigma_{\xi} $. More specifically, $ \zeta_0 \lambda_D $ spans the range 0.2 to 0.46~nm and $ (\overline{ \xi_{0}} \lambda_D)^{-2} $ the range $ 1.5 \times 10^{12} $~cm$ ^{-2} $ to $ 3 \times 10^{14} $~cm$ ^{-2} $, the two later limits featuring respectively a positive charge density inside the oxide layer and the dipole density at the oxygen terminated surface. Accordingly, the average distances $ \overline{ \xi_{0}} $ between point charges or dipoles span a range limited by 0.08 and 0.007 (in  $ \lambda_D $ units) respectively. Therefore, the barrier height fluctuations are occurring along a microscopic scale, well below the Debye length.
In a second step, the band bending is changed and the values of $ \overline{\Phi_{B}}(T,U)$ and $ \sigma_{\Phi}(U)$ are evaluated as a function of the band bending $ U $. In all the cases and whatever the temperature, as suggested in \cite{Werner}, the increase of the average barrier height $ \overline{\Phi_{B}}(T,V_j)$ and the decrease of the standard deviation $ \sigma_{\Phi}(V_j)$ with increasing forward voltage and correlative decreasing band bending is fully confirmed, although the barrier height distribution is not symmetric and cannot be strictly depicted by a normal law. The trends followed by the two terms of Eq. (\ref{barrierinhomog}) both contribute to an increase of the barrier height with forward voltage which can be also interpreted as an increase of the ideality factor, thus inducing a lower current density in comparison to that anticipated with the help of a constant ideality factor like sketched with dash-dot lines in Fig. \ref{fig1}. At variance with the hypothesis used in ref. \cite{Werner}, neither the variation of $ \overline{\Phi_{B}}(T,V_j)$ nor that of $ \sigma_{\Phi}(V_j)^2$ is linear with $ V_j $ or $ U $, a law which would have resulted in a constant ideality factor, in contradiction both with experimental and simulation results. An other major feature is the relative insensitivity of the ideality factor to the density of point charges or dipoles $ (\overline{ \xi_{0}} \lambda_D)^{-2} $, the former starting from 1.01-1.02 for band bending corresponding to zero voltage bias and growing up to 1.05-1.15 when the band bending $ U $ becomes close to a few $ k_B T /e_0  $ like shown in Fig.~\ref{fig6}, irrespective of the temperature. The barrier height variations which are deduced from the ideality factor match very well those of the image force lowering as clear in Fig.~\ref{fig6} too. This result coming from the simulated inhomogeneous barrier height with repulsive elementary charges or dipoles demonstrates that the image force lowering and its changes with the band bending is the main source of the variation of the average barrier height with the applied voltage, because barrier height enhancement does not contribute significantly to the whole barrier experienced by carriers responsible for the current, except for a reduced effective area. Such a result validates this hypothesis already done in ref. \cite{Tung1992}. The whole variation amounts to about 40 mV like inferred in ref. \cite{Monchbook} for the image force lowering alone. It is thus possible to derive a universal law featured by a solid line in Fig.~\ref{fig6} that may be used to derive current density-voltage characteristics with the help of Eq. (\ref{JphiBvoltdep}) and (\ref{idealityfactor}). The first important result coming out from this simulation turns out to be the better agreement of the simulated current density sketched in solid line in the inset of Fig.~\ref{fig1} in the case (c) where the barrier height correction has been taken into account. The second one is the insensitivity of the ideality factor to temperature, a property which matches rather well the behavior of the interface (c) in Fig. \ref{fig3} (c'), in contrast to the absolute variations in data (a') and (b'), although the ideality factor (c') remains slightly higher than the values evaluated from the image force lowering alone. Taking all the arguments evidenced and discussed in the present subsection, one must conclude that microscopic barrier inhomogeneities resulting from repulsive elementary point charges or dipoles which have been however, experimentally evidenced, cannot account for most of the properties of the interfaces presently studied, and that an other type of barrier inhomogeneity must be present.

\subsection{Barrier height and ideality factor of interfaces with lowered barrier patches of size either comparable to or smaller than the Debye length} \label{secmodel2}
\begin{figure}
\includegraphics{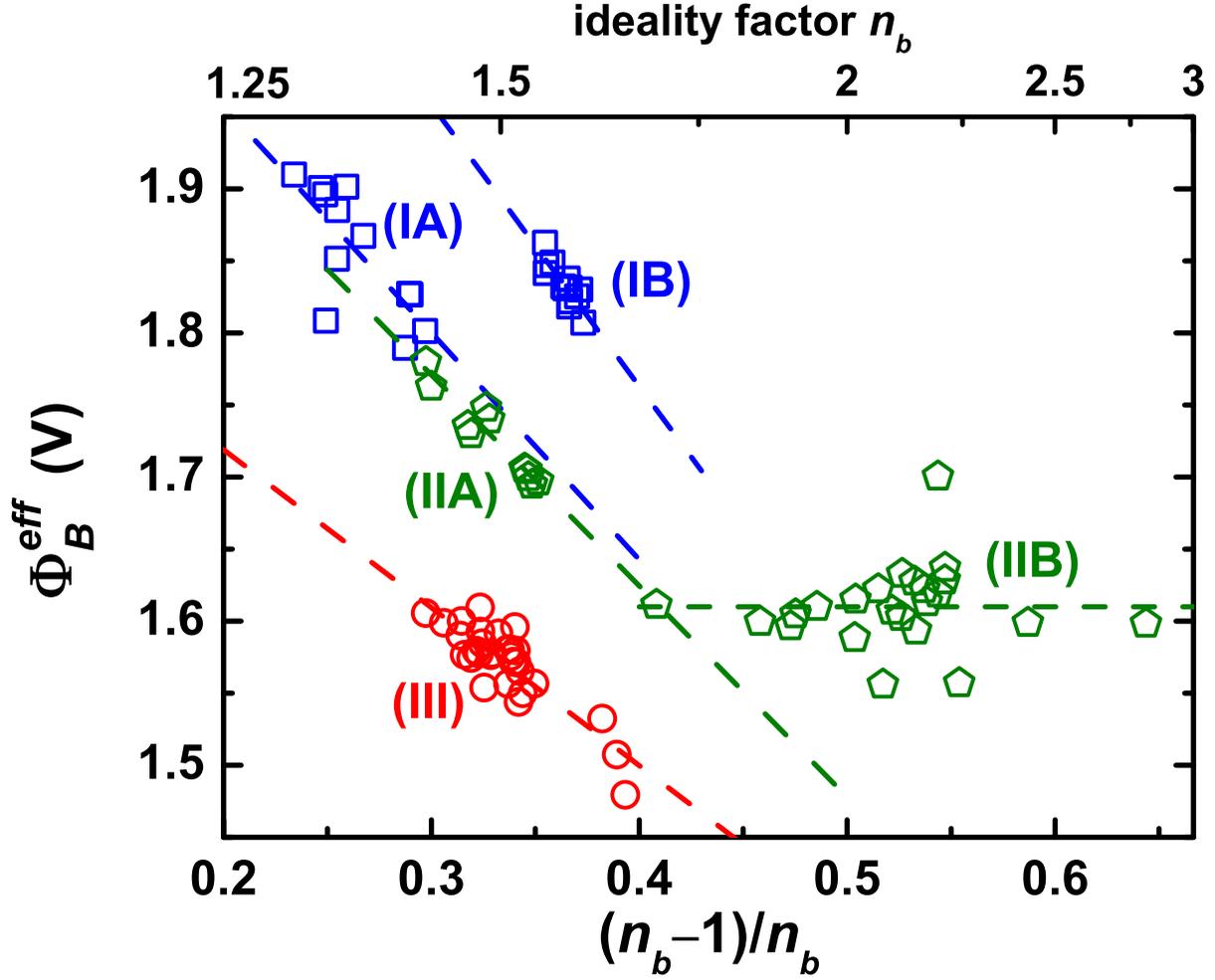}
\caption{\label{fig7} Effective barrier height as a function of $(n_b-1)/n_b$, where $ n_b $ is the ideality factor at band bending $ U_b $ for 95 diodes fabricated on three distinct diamond stacks elaborated like described in section \ref{experimental}, respectively identified by squares (I), pentagons (II) and circles (III). Samples I and II have not been intentionaly annealed while sample III has been annealed at 350 \textcelsius. Dash lines are best linear fits of grouped data IA, IB, IIA, IIB and III according to Eq. (\ref{nfonctiondephiB}).}
\end{figure}

The calculation of barrier height fluctuations due to barrier height lowering $\Delta_p$ at circular patches with radius $ R_0 $ or linear ones with width $ L_0 $ and their effect on the current-voltage characteristics relies on the same principle as previously, except for the weight of each dipole which has to be $ 2 \pi \varepsilon_{SC} \varepsilon_0 \Delta_p R_0^{2} $ to induce the desired barrier change at circular patches which will be only considered in the following because close to those happening naturally in inhomogeneous interfaces. Additionally, only patches with a lowered barrier are considered as demonstrated in the previous subsection. This method, known as the dipole-layer approach, has been implemented analytically by R. T. Tung \cite{Tung1992} and summarized later by W. M\"{o}nch \cite{Monchbook}. For a single circular patch with radius $ R_0 $ and barrier lowering $\Delta_p$ embedded into a homogeneous area characterized by a constant barrier height $ \Phi_{B}^{hom} $, a single parameter containing the two characteristic quantities $ R_0 $ and $\Delta_p$, which is called the patch parameter $ \gamma_p = 3 (\Delta_p~R_0^{2}/4)^{1/3} $ can be defined. Integrating Eq.(\ref{Igeneral}) over the whole area $ A $ with the hypothesis of a parabolic shape of the potential around the saddle point, which is the lowest potential of the patch, restores the form of Eq. (\ref{Jnonideal}). But the new barrier height is $ \Phi_{B}^{hom} $ minus an effective barrier lowering $ \delta \Phi_p^{sad} $ at the potential saddle point, the ideality factor $ n $ is replaced by 1 and an effective area $ A_{eff} = 4 \pi \lambda_D^{2} \Gamma/3 $ affects the current expression. The weight of the patch is proportional to the dimensionless parameter $ \Gamma = \frac{\gamma_p}{3} \frac{1}{\eta^{1/3} U^{2/3}} $ where $ \eta = \varepsilon_{SC}\,\varepsilon_0\,/\,(e_0\,N_A) $. Both the effective barrier lowering $ 3 \Gamma\,U =  \gamma_p \left[ U\,/\,\eta \right] ^{1/3} $ and area $ A_{eff} = \frac{4 \pi \gamma_p k_B T}{9 e_0} \left[ \eta\,/\,U \right] ^{2/3}  $ are dependent on the band bending $ U = \Phi_{B}^{hom} - V_j - V_{dop} $, while the ideality factor is approximately $ 1+\Gamma $, therefore increasing with the forward voltage. The final step consists in introducing a distribution of a variable patch parameter. In the Tung's original work \cite{Tung1992}, two cases were worked out : (i) a sharp distribution of lowered barrier patches approximated by a single patch parameter $ \gamma_0 $ and (ii) a broad distribution featured by half a Gaussian law which limited the $ \gamma_p $ range to the values pertaining to barrier height lowering alone, that are $ \gamma_p \geqslant 0$. This choice was justified by the negligible contribution of patches characterized by a barrier height enhancement to the whole current as we checked in the previous subsection. In order to generalize such an approach, allowing to match continuously any intermediate behavior between the two previous cases, the patch parameter distribution function will be chosen as $ N(\gamma_p) = \frac{ D_p}{\sqrt{2 \pi} \sigma_p} \exp \left( -\frac{(\gamma_p - \gamma_0)^2}{2 \sigma_p^2}\right) $, characterized by a standard deviation $ \sigma_p $ of the patch parameter around the central value  $\gamma_0$ and an areal density $ D_p $. Depending on $ \gamma_0 $, the function $ N(\gamma_p)$ depicts either a peaked distribution around the central patch parameter, or half a Gaussian law decreasing from its maximum at $\gamma_p =0 $ when $\gamma_0 =0 $ or even a decreasing Gaussian tail with a non zero initial slope if $\gamma_0 < 0 $. After summation of all the currents dependent on any positive value of $\gamma_p $ from zero to infinite, this procedure further allows to derive the following expression where only the last factor inside braces contains the effects of inhomogeneities while the others give the ideal thermionic current density of a homogeneous barrier height $\Phi_{B}^{hom} $:

\begin{multline} 
J^{patchy}_{z}= \alpha\:A^{**}\:T^2\:\exp \left( -\frac{e_0 \Phi_{B}^{hom}}{k_B T} \right) \left[ \exp \left(\frac{e_0 V_j}{k_B T} \right) -1 \right] \\
\times \left\lbrace 1 + \frac{2 \sqrt{2\pi}\,k_B T\,D_p\,\sigma_p\,\eta^{2/3}}{9\,e_0\:U^{2/3}} \exp\left( -\frac{\gamma_0^{2}}{2 \sigma_p^2} \right) \left[1\:+ \frac{\sqrt{2\pi}\:e_0 <\!\delta \Phi_{p,0}^{sad}\!>}{k_B T}\exp \left(\frac{1}{2} \left( \frac{e_0}{k_B T} <\!\delta \Phi_{p,0}^{sad}\!> \right)^{2} \right) \right] \right\rbrace
\label{Jpatchy}
\end{multline} 
 
where the standard deviation $ <\!\delta \Phi_{p,0}^{sad}\!> $ of the distribution of the saddle points of the barrier $ \Phi_p^{sad} $ is: 
\begin{equation}
<\!\delta \Phi_{p,0}^{sad}\!> = \sigma_p \left[ \frac{U}{\eta} \right]^{1/3}  + \frac{k_B T\,\gamma_0}{e_0\,\sigma_p}
\label{deltaphisaddle}
\end{equation} 
But the current density deviates notably from the ideal thermionic law only if the factor inside the braces of Eq. (\ref{Jpatchy}) is larger than unity. As one is interested in this case, a reasonable approximation consists in neglecting the unity terms inside the braces, so that the effective barrier height can be deduced from the exponent of the exponential factors, which turns out to be $ \frac{1}{2} \left[ \frac{e_0}{k_B T} <\!\delta \Phi_{p,0}^{sad}\!>^{2} -\frac{\gamma_0^{2}}{\sigma_p^2} \right] $ inside the braces of Eq. (\ref{Jpatchy}) and logarithm of the pre-exponential factor : 
\begin{equation}
\Phi_{B}^{eff}(T,U)= \Phi_{B}^{hom} - \left[\frac{e_0}{2 k_B T} \sigma_p^2 \left( \frac{U}{\eta} \right)^{2/3} + \gamma_0 \left(\frac{U}{\eta}\right)^{1/3} \right] - \frac{k_B T}{e_0} \ln \left[ \frac{4\pi \,D_p\,\sigma_p\,\eta^{2/3}}{9\:U^{2/3}}  <\!\delta \Phi_{p,0}^{sad}\!> \right] 
\label{newbarrierTung}
\end{equation}
This last equation is similar to Eq. (\ref{barrierinhomog}) with the new quadratic average $ <\!\delta \Phi_B\!> $ of barrier fluctuations 
\begin{equation}
 <\!\delta \Phi_{B}\!> = \left[ \sigma_p^2 \left( \frac{U}{\eta} \right)^{2/3} + \frac{2 k_B T}{e_0}\gamma_0 \left(\frac{U}{\eta}\right)^{1/3} \right]^{1/2}
\label{newbarrierlowering}
\end{equation}
and an additional term which has a little influence due to the logarithm and a null value for $ D_p $ in the 10$^{10}$~cm$^{-2}$ range, but may give a non negligible contribution to the ideality factor appearing in next formula when the barrier inhomogeneity level is low. But the main issue  is the dependence of $ <\!\delta \Phi_B\!>$ on temperature which is different from that given by Eq. (\ref{barrierinhomog}) and results in a shift of the high temperature asymptote of the effective barrier in comparison to $ \Phi_{B}^{hom} $ by a quantity proportional to $ \gamma_0 $, leading to an apparent extrapolated barrier $ \Phi_{B}^{app} = \Phi_{B}^{hom} -\gamma_0 \left(\frac{U}{\eta}\right)^{1/3}$. \\
From Eq. (\ref{idealityfactor}), (\ref{deltaphisaddle}) and (\ref{newbarrierTung}), the ideality factor can be calculated as:
\begin{equation}
\begin{split}
\frac{n(T,U)-1}{n(T,U)} & = \frac{e_0}{3 k_B T \:U} \left[ \sigma_p^2 \left( \frac{U}{\eta} \right)^{2/3} + \frac{k_B T}{e_0}\gamma_0 \left(\frac{U}{\eta}\right)^{1/3} \right] - k_\Phi \frac{k_B T }{e_0\,U} \\
& = \frac{e_0}{3 k_B T \:U} <\delta \Phi_B>^2 \; - \: \frac{\gamma_0}{3\eta^{1/3}U^{2/3}}\; - \: k_\Phi \frac{k_B T }{e_0\,U}
\end{split}
\label{idealityn}
\end{equation}
where $ k_\Phi $ is between 1/3 and 2/3 according to the respective dominance of either the first or the second term in $ <\delta \Phi_B> $. This ideality factor increases with decreasing band bending and therefore with increasing applied voltage as long as flat bands are not reached, so that current-voltage characteristics plotted in a semilogarithmic diagram experience a downwards curvature, as in Fig. \ref{fig1} (a) and (b). This model ceases to be valid neat flat bands and beyond because the current calculation relies only on the carrier flow at the saddle points of the potential barrier, neglecting the areas where the barrier is $\Phi_{B}^{hom}$ or higher. It must be also stressed that these ideality factor and effective barrier height must be evaluated around the same band bending $ U=U_b $ if comparison at different temperatures or in different diodes are made. This is often done by assessing the corresponding ideality factor $ n_b $ at the same current density although such a rule is almost true only at vanishing band bending. It is important to notice that various behaviors of the ideality factor as a function of temperature can result from Eq. (\ref{idealityn}), with a trend toward a constant value when the second term inside the brackets dominates at elevated temperatures as one can see in Fig.~(\ref{fig2}b'), the last term in Eq.~(\ref{idealityn}) being negligible except near flat bands and for ideality factors below 1.1. Either neglecting the last term in Eq.~(\ref{newbarrierTung}) or lumping it into $ \Phi_{B}^{hom} $, a relationship between the effective barrier height and ideality factor $ n_b $ at the same band bending $ U_b $ for all the diodes investigated at a given temperature among a set of diodes, comes from Eq. (\ref{newbarrierTung}), (\ref{newbarrierlowering}) and (\ref{idealityn}):
\begin{equation}
\Phi_B^{eff}  \approx \Phi_B^{hom}\;-\:\frac{3 U_b}{2} \frac{n_b-1}{n_b} \;-\:\frac{\gamma_0}{2} \left ( \frac{U_b}{\eta} \right )^{1/3} \;-\: k_\Phi \frac{k_B T }{e_0\,U_b}   
\label{nfonctiondephiB}
\end{equation}
Therefore, the difference between the effective barrier height $ \Phi_B^{eff} $ and a quantity shifted from the homogeneous barrier $ \Phi_B^{hom} $ mainly by a term proportional to the central patch parameter $ \gamma_0 $, follows a linear relationship regarding the ideality function $ (n_b - 1)/n_b $. It does not depend on the standard deviation $ \sigma_p $ of the patch distribution, except for the last term including the factor $ k_\Phi $ which is roughly constant for a given patch distribution. At constant temperature, this last term is indistinguishable from $ \Phi_B^{hom} $  and adds a minor error to it. Conversely, the term proportional to $ \gamma_0 $ may contribute to an important shift of the apparent extrapolated barrier at unity ideality factor in comparison to $ \Phi_B^{hom} $, since this apparent barrier turns out to be $ \Phi_B^{app}= \Phi_B^{hom} \;-\:\frac{\gamma_0}{2} \left ( \frac{U}{\eta} \right )^{1/3} $ and this correction might not be ignored in general. It can also introduce some dispersion of the data when a family of diodes is studied. The effective barrier height $ \Phi_B^{eff} $ is approximately linear with $ n_b $ only for typical values smaller than 1.4 like noticed in ref. \cite{Monchbook}, so that it is better to make plots of the apparent barriers as a function of $ (n_b - 1)/n_b $ for sake of generality, as done in Fig.~(\ref{fig7}), which will be commented in the next subsection. However, it must be stressed that all the expressions derived for the effective barrier height and ideality factor within the framework of the present model, that are Eq.(\ref{newbarrierTung}) to (\ref{nfonctiondephiB}), assumed that the factor inside the braces of Eq. (\ref{Jpatchy}) is largely greater than one. Such a condition implies that the patch density $ D_p $ must be typically much larger than 10$^7$~cm$^{-2}$, so that one has to wonder whether the value deduced from adjustments has really a physical meaning in each situation. 

\begin{table} 
\begin{tabular}{||c||c|c|c|c|c|c||}
\hline
\hline \rule[-2ex]{0pt}{5.5ex} Sample type and & $ \Phi_B^{app} $ & $\sigma_p$ & $\gamma_0$ & $ \Phi_{B,1}^{hom} $ & $ D_p $ & $ \Phi_{B,2}^{hom} $ \\   
\hline \rule[-2ex]{0pt}{5.5ex} annealing temp. & V & V$^{1/3}$~cm$^{2/3}$ & V$^{1/3}$~cm$^{2/3}$ & V & cm$ ^{-2} $ & V \\ 
\hline 
\hline \rule[-2ex]{0pt}{5.5ex} (a) $ < $~300~\textcelsius & 2.04$\pm$0.14 &  (9$\pm$3)$\times$10$^{-5}$ & (1.6$\pm$0.4)$\times$10$^{-4}$  & 2.26$\pm$0.16  & (3$\pm$1)$\times$ 10$^{6}$ & 2.38$\pm$0.02  \\
\hline \rule[-2ex]{0pt}{5.5ex} (b) 350~\textcelsius & 1.68$\pm$0.1 &  (9.0$\pm$1)$\times$10$^{-5} $ &  (1.85$\pm$0.3)$\times$10$^{-4}$ & 1.87$\pm$0.1 & (6$\pm$2)$\times$10$^{9}$ & 1.90$\pm$0.02 \\ 
\hline \rule[-2ex]{0pt}{5.5ex} (c) 450~\textcelsius & 0.93$\pm$0.04 &  (5.5$\pm$1.5)$\times$10$^{-5} $ &  (1.3$\pm$1.2)$\times$10$^{-5}$ & 0.94$\pm$0.04 & $ < $ 10$^{9}$ & 0.99$\pm$0.01 \\  
\hline
\hline
\end{tabular}
\caption{Parameters of the Schottky barrier in the three interfaces according to the models detailed in section \ref{inhom-models}, with $\Phi_{B,1}^{app}$ and $\sigma_p$ deduced from the adjustment of barrier height as a function of temperature according to Eq.~(\ref{newbarrierTung}). Then, $\gamma_0$, and $\Phi_{B,1}^{hom} $ are derived from ideality factors fitted as a function of temperature with respective help of Eq.~(\ref{idealityn}) in the case of interfaces (b) and (c), and Eq.~(\ref{ideality-p-o}) for interface of kind (a), while $ D_p $ and $\Phi_{B,2}^{hom} $ come from the direct adjustment of the current densities shown in Fig.~(\ref{fig1}) respectively to Eq.~(\ref{Jpatchy}) for interfaces of kind (b) and (c) and Eq.~(\ref{Jinhom}) for interface of kind (a) as a function of voltage.}
\label{Table2}
\end{table} 

In one diode alone, the fingerprint of the kind of barrier inhomogeneities breaks out in the behaviors of the ideality factor and to a smaller extent of the effective barrier height as a function of temperature, like shown in Fig. (\ref{fig3}), where evolutions of the ideality factor are markedly different for the three interfaces in Fig. (\ref{fig3}a'), (\ref{fig3}b') and (\ref{fig3}c'). All these features can be reproduced by the laws demonstrated above, specially the ideality factor variations with temperature which are strongly influenced by the central patch parameter $ \gamma_0 $ in Eq.~(\ref{idealityn}). In the case (a), a good agreement with Eq.~(\ref{idealityn}) can be obtained as depicted by the dash-dotted curve in Fig.~(\ref{fig3}a') at the expense of a negative central patch parameter which would induce an apparent barrier $\Phi_{B}^{app} $ larger than the homogeneous one  $\Phi_{B}^{hom} $, obviously leading to an incoherent physical situation. Moreover, the direct adjustment of the current density according to Eq.~(\ref{Jpatchy}) turns out to be impossible because this expression is unable to account for the curvature of the current density-voltage characteristic (a) in Fig.~(\ref{fig1}) below flat band voltage. An alternative model able to involve a more appropriate physical picture is discussed below for interfaces of kind (a). Data collected in 95 diodes prepared on three distinct diamond stacks which underwent the same pre-treatment before metal deposition and displayed in Fig.~(\ref{fig7}), give a first indication: if so different groups of data are possible for diodes whose diameter is 100~$\mu$m, specially the IB against IA and IIA, related to interfaces of kind (a), barrier heights extrapolated at unity ideality factor and averaged on such area must be different. Consequently, the lateral scale of barrier inhomogeneities may no longer be the Debye length but much more, and the approximations of a parabolic potential at the bottom of the lowered barrier patches depicted by the patch parameter $ \gamma_p $ would cease to be valid. Secondly, strong variations of the ideality factor with temperature like in Fig.~(\ref{fig3}a') suggests that diamond resistivity may play a role because acceptors are progressively ionizing in this temperature range. 

\subsection{Barrier domains with multiple sizes and inhomogeneities of the current density} \label{secmodel3}
When the lateral size of barrier inhomogeneities largely exceeds the Debye length, the previous model relying on a parabolic potential at the bottom of the patches with a lowered barrier is no longer relevant. These different areas with distinct barrier height can be treated as independent of the band curvature because the pinch-off of the channels where the current flows becomes negligible. Then, the total current is simply obtained by summation of all the currents like in a parallel circuit of the corresponding paths. But such a model is unable to fit the downward curvature experienced by the logarithm of the current density as a function of voltage in interfaces of kind (a) and yields merely an ideal behavior as soon as the lowest barrier is shifted below the others by some $ k_B T / e_0 $ because the corresponding current flow dominates the others. In fact, this current inhomogeneity due to carrier crowding at the saddle points and lowest barrier domains remains when the $ z $ coordinate move away from the interface and therefore have a dramatic influence on the total voltage applied to the junction for a given current, specially when carrier freeze-in occurs, inducing large variations of the ideality factor with temperature. This effect is relevant both for barrier domains much larger than the Debye length where no pinch-off occurs and for pinched-off patches, and the voltage drops in these shrunk channels must be subtracted from the total applied voltage. In the pinched-off case, the diameter of patches is always much smaller than the lightly doped layer thickness $ t_l $ so that the spreading resistance can be approximated by  $\frac{\rho(T)}{4\sqrt{\mathcal{A}_{eff,i}^{p-o}/\pi}} $ with a good accuracy \cite{Cox}, $ \rho(T) $ being the resistivity of the lightly doped diamond layer. Therefore, the non pinched-off and pinched-off forward currents which flow respectively through individual areas $ \mathcal{A}_i $ and $ \mathcal{A}_{eff,i}^{p-o} $ under an applied voltage $ V_a $ can be derived from the generic thermoionic law but with band bending equal to the homogenous potential barrier minus the relevant barrier lowering and applied voltage $ V_a$ from which the ohmic voltage drop along either non pinched-off or pinched-off channels inside the depletion zone is subtracted: 
\begin{subequations}\label{model3}
\begin{equation} \label{Inonpinched-off}
I_i^{non-p-o}\:=\:\mathcal{A}_i\:\alpha\:A^{**}\:T^2\:\exp\left[-\frac{e_0}{k_B T} \left( \Phi_B^{hom}\:-\:\Delta\Phi\:-\:V_a\:+\:I_i^{non-p-o}\rho(T)/ \lambda_i \right) \right]
\end{equation}

\begin{equation} \label{Ipinched-off}
 I_i^{p-o}\:=\:\mathcal{A}_{eff,i}^{p-o}\:\alpha\:A^{**}\:T^2\:\exp\left[-\frac{e_0}{k_B T} \left( \Phi_B^{hom}\:-\:\gamma_p \left(\frac{U_i}{\eta} \right)^{1/3} \:-\:V_a\:+\:\frac{\rho(T)}{4\sqrt{\mathcal{A}_{eff,i}^{p-o}/\pi}}\:I_i^{p-o}\: \right) \right]
\end{equation} 
with $ \mathcal{A}_{eff,i}^{p-o} = \frac{4\:\pi\:k_B\:T\:\gamma_p}{9\:e_0}\:\left (\frac{\eta}{U_i}\right )^{2/3} $ and $ U_i\:=\:\Phi_B^{hom}\:-\:\gamma_p \left(\frac{U_i}{\eta} \right)^{1/3} \:-\:V_a\:-V_{dop}\:+\:\frac{\rho(T)}{4\sqrt{\mathcal{A}_{eff,i}^{p-o}/\pi}}\:I_i^{p-o} $; \\

the lowest barrier spreading over the areas $ \mathcal{A}_i $ with size much larger than the Debye length being $ \Phi_B^{hom}\:-\:\Delta\Phi $. If the current inhomogeneity remains the same along the whole depth of the lightly doped layer like assumed in ref.~\cite{Tung2001}, the voltage drop along the $ i^{th} $ non pinched-off channel is simply proportional to a resistance $ \rho(T)/ \lambda_i $ where $\lambda_i=\mathcal{A}_i/t_l $. This hypothesis is true in the limit of a diameter of the area with a lowered barrier height well larger than $t_l$, a condition equivalent to $\sqrt{\mathcal{A}_i/\pi} \gg t_l /2 $. But because this diameter can span a large range, from values well below $t_l$ up to much wider ones, the corresponding spreading resistance should take the general form\cite{Cox} $\frac{\rho(T)}{2\sqrt{\pi \mathcal{A}_i}} \arctan(2t_l/\sqrt{\mathcal{A}_i/\pi})  $ and $\lambda_i$ should become $2\sqrt{\pi \mathcal{A}_i}/\arctan(2t_l/\sqrt{\mathcal{A}_i/\pi})$.
It must be noticed that the last term in the exponent of the expression of $ I_i^{p-o} $ and in $ U_i $ is added to $\Phi_B^{hom} $ and therefore plays the same role as a negative central patch parameter $ \gamma_0$ in the effective barrier experienced by the whole current flowing through the pinched-off patches and given in Eq.~(\ref{newbarrierTung}). This fact explains why neglecting the confinement of current in channels which stretch out along some part of the lightly doped layer thickness had to be compensated by the occurrence of a negative figure for $ \gamma_0$ in the simulation results of the previous subsection, leading to an incorrect value because of an inappropriate model applied to interfaces of kind (a). Such a fictitious negative figure for $ \gamma_0$ was necessary to obtain an acceptable, but misleading, agreement of the experimental data with the model developed in the previous subsection since firstly, the resulting $\Phi_{B}^{app} $ value larger than $\Phi_{B}^{hom} $ would contradict the hypothesis which assumed lowered barrier patches and secondly, the same model was unable to match the curvature of the current-voltage characteristics. 
\end{subequations} 
   
Averaging all the currents over the whole diode area $ \mathcal{A}_d $ is not tractable analytically. A reasonable approximation consists in calculating separately the total current and an estimate of the whole conductance within the shrunk channels from the integration of all the parallel conductances $ \lambda_i/\rho(T) $ or $ 4 \sqrt{\mathcal{A}_{eff,i}^{p-o}/\pi}/\rho(T) $ appearing in the last term of the exponent of each expression (\ref{Inonpinched-off}) or (\ref{Ipinched-off}), weighted by the distribution $N(\gamma_p) = \frac{ D_p}{\sqrt{2 \pi} \sigma_p} \exp \left( -\frac{(\gamma_p - \gamma_0)^2}{2 \sigma_p^2}\right)$ in the later case. It is also assumed like previously that only the lowest barrier for the current flow is efficient and that the barrier lowering is approximated by its quadratic average. For non pinched-off channels, the total conductance would result from the integral of $ \lambda_i/\rho(T) $  weighted by an unknown distribution, so that it is more convenient to introduce a fictitious area $\mathcal{A}^{non-p-o}_{\Delta V}$ allowing to write the corresponding voltage drop $ \Delta V^{non-p-o} $. In the case of the pinched-off currents, two integrals must be evaluated numerically as a function of $ \gamma_0/(\sqrt{2}\sigma_p) $. Their ratio appears as a fraction in the next formula and gives the exact value within an error smaller than 10 \% . With $\mathcal{A}^{non-p-o}=\sum_{i} \mathcal{A}_i$ representing the true area of the non pinched-off zones, the average current densities in the whole diode area become respectively for the non pinched-off and pinched-off channels:  
\begin{subequations}\label{Jinhom}
\begin{multline} \label{Jnon-p-o}
J_{z}^{non-p-o}\:\approx\:
\frac{\mathcal{A}^{non-p-o}}{\mathcal{A}_d}\:\alpha A^{**} T^2 \exp \left [ -\frac{e_0}{k_B T} \left (\Phi_B^{hom}\:-\:\Delta\Phi\:-\:V_a\:+ \:\Delta V^{non-p-o} \right ) \right]
\end{multline}
with 
\begin{multline} 
\Delta V^{non-p-o} \:\approx\:\frac{\rho(T)\:\mathcal{A}^{non-p-o}}{(\mathcal{A}^{non-p-o}_{\Delta V})^{1/2}}\;\alpha A^{**} T^2 \exp \left [ -\frac{e_0}{k_B T} \left (\Phi_B^{hom}\:-\:\Delta\Phi\:-\:V_a\:+ \:\Delta V^{non-p-o} \right ) \right]
\end{multline}
and $ J^{p-o}\:=\:J^{patchy}_{z}$ from Eq.~(\ref{Jpatchy}), with $V_j\:=\:V_a-\:\Delta V^{p-o} $ where
\begin{multline} \label{Vj-p-o}
\Delta V^{p-o} \:\approx\:\frac{\pi}{6}\:\rho(T)\:\left(\frac{k_B T}{e_0}\right)^{1/2}\:\left (\frac {\eta}{U}\right )^{1/3}\ \;  \dfrac{1.77\gamma_0\:+\:\sigma_p^2/(\gamma_0\:+\:\sqrt{2}\sigma_p)}{\sqrt{\pi\:\gamma_0\:+\:\sqrt{2}\sigma_p/3}} \\
\times \alpha A^{**} T^2\: \exp \left [ -\frac{e_0}{k_B T} \left (\Phi_B^{hom}\:-\:\frac{e_0}{2k_B T}<\!\delta \Phi_B\!>^{2}\:-\:V_a\:+\:\Delta V^{p-o} \right )  \right ]
\end{multline} 
and
\begin{equation}\label{bandbending-p-o}
U\:\approx\:\Phi_B^{hom}\:-\:\frac{e_0}{2k_B T}<\!\delta \Phi_B\!>^{2}\:-\:V_a-\:V_{dop}
\end{equation}
\end{subequations}   
It must be noticed that the patch density $ D_p $ does not appear in $ \Delta V^{p-o} $ because both the conductance and current are proportional to it, so that it is canceled in the framework of the previous approximation.
From these expressions, it is easy to guess that the lower the temperature, the higher the effect of the smallest barrier, like in previous models. If the non pinched-off current dominates, the apparent barrier lowering would be $ \Delta \Phi\:-\:\frac{k_B T}{e_0} \ln\left( \mathcal{A}_d / \mathcal{A}^{non-p-o}\right) $, inducing a linear decrease of the effective barrier as temperature is lowered, whereas alternatively Eq.~(\ref{newbarrierTung}) would be valid, at least when $\Delta V$ remains small regarding $V_a$. At sufficiently high temperature, the increase of the effective barrier would be linear with temperature when the non pinch-off current dominates, an effect which is not clearly detected in Fig.~(\ref{fig3}a), indicating that the pinch-off current might stay the major one. Such a view is confirmed by the absence of any "S-shaped" current-voltage characteristic in Fig.~(\ref{fig1}a), which otherwise would reveal the existence of two parallel non pinched-off channels like happening in some Ti/4H-SiC diodes\cite{Defives}, each of ones having its own barrier height and series resistance with the corresponding condition $\sqrt{\mathcal{A}_i/\pi} \gg t_l /2 $ indeed achieved. \\

From experimental data displayed in Fig.~(\ref{fig7}), it seems that the general trend involving a linear relationship between effective barrier height and ideality function is still valid. Such a property can be checked from the integration of the central expression in Eq.~(\ref{idealityfactor}) where the total applied voltage $ V_a $ replaces the junction voltage $ V_j $ since the current density is now dependent on $ V_a $ if one relies on the last expressions (\ref{Jinhom}). It results in a linear relationship between the effective barrier and ideality factor function if the ideality factor does not depend on the band bending $ U_b$. Such a condition is only approximately achieved, but better and better when higher and higher band bending are used, like those deduced from these data and reported in Table~\ref{Table3}, because the voltage drop $\Delta V $ remains negligible in this range. It is therefore possible to find a range of voltage and corresponding band bending extending over an interval $ U_b/K$ ($ K$ being a unitless constant) where the ideality factor is constant around $U_b$ and integration of the second and fourth members of Eq.~(\ref{idealityfactor}) gives the apparent barrier height in non ideal interfaces:
\begin{equation}\label{idgeneral}
\Phi_{B}^{app}(T,U_b) = \Phi_{B}^{ideal}\:-\:\frac{U_b}{K}\:\frac{n(T,U_b)-1}{n(T,U_b)}
\end{equation}
where $ n(T,U_b)$ is the ideality factor at temperature $ T$, around the peculiar band bending~$U_b$.  $\Phi_{B}^{ideal} $ is the barrier height at unity ideality factor, which matches accurately $\Phi_B^{hom}\;-\:\frac{\gamma_0}{2} \left ( \frac{U}{\eta} \right )^{1/3}$ for pure pinched-off barrier inhomogeneities as demonstrated in subsection \ref{secmodel2} in formula (\ref{nfonctiondephiB}) where the last term can be neglected, and is only close to $\Phi_B^{hom}$ without a precise knowledge of the deviation in other cases. Therefore, the observed linear correlations in Fig.~\ref{fig7} can still be justified. When the voltage drop inside the lightly doped layer is not negligible, approximate expressions of the ideality function $Y$ can be drawn separately from the previous equations when either $ I^{non-p-o}~=~\mathcal{A}_d J_{z}^{non-p-o} $ or $ I^{p-o}~=~\mathcal{A}_d J_{z}^{p-o} $ dominates, which turns out to be respectively with either the ideality factor $n_a(T,V_a)$ under applied voltage $V_a$ or the ideality factor $n_b(T,V_a)$: 
\begin{subequations}
\begin{equation} \label{ideality-non-p-o}
~~~~~~~~~~Y^{non-p-o}~=~\frac{n_a(T,V_a)-1}{n_a(T,V_a)}~\approx~\frac{e_0}{k_B T} \Delta V^{non-p-o}~~~~~~~~~~~~ 
\end{equation}
\begin{multline}\label{ideality-p-o}
\mathrm{or} ~~~~~~~Y^{p-o}=\frac{n_b(T,V_a)-1}{n_b(T,V_a)} \\
\:\approx\:\left[ \frac{e_0}{3 k_B T \:U}\sigma_p^2 \left( \frac{U}{\eta} \right)^{2/3} + \frac{\gamma_0}{3\:U} \left(\frac{U}{\eta}\right)^{1/3} \right] \left(1\:+\:\frac{e_0}{k_B T}\Delta V^{p-o} \right) \ +\ \frac{e_0}{k_B T}\Delta V^{p-o}
\end{multline}
When the ideality factor has to be studied as a function of temperature, measurements should be done at constant current density, because in such a case the product of $ \alpha A^{**} T^2 $ and exponential factors in Eq.~(\ref{ideality-non-p-o}) and in the term $ \Delta V^{p-o} $ of Eq.~(\ref{ideality-p-o}) can be considered as almost invariant, making any parameter adjustment much easier.
\end{subequations}   
In each case, these expressions shows that the ideality factor may become very sensitive to temperature, specially because it follows the variations of the resistivity $ \rho(T) $ which are very important in the range where freezing out of carrier occurs, typically from 300~K to 550~K in this study. Involvement of the voltage drop due to the flow of current through the narrow bottlenecks which lengthen the lowered barrier regions inwards the lightly doped layer is able to explain qualitatively the strong negative slope of the ideality factor in Fig.~(\ref{fig3}a'). The general case relies on the combination of the two kinds of current with a proportion $ P = I^{p-o}/(I^{p-o}+I^{non-p-o}) $ and on Eq.~(\ref{idealityfactor}), giving the global ideality function:
\begin{equation}\label{ideality-global}
Y\:=\:\frac{n(T,V_a)-1}{n(T,V_a)}=P~Y^{p-o} + (1-P)~Y^{non-p-o}  
\end{equation}
The best match to the four data at highest temperatures is obtained with $ P = 72 \% $ and plotted as a full curve in Fig.~(\ref{fig3}a'), confirming the previous hypothesis about the dominance of the pinched-off current. The discrepancy at the lowest temperatures is commented in the following. But it must be stressed that the first term in Eq.~(\ref{ideality-global}) is unable to account for the ideality factor variations, which are mainly due to the second term despite its smaller weight. It is thus confirmed that the non pinched-off currents contribution is necessary to explain both the current density-voltage curvature and the ideality function shape with temperature in interfaces of kind (a). The more important voltage drop $ \Delta V^{non-p-o} $ in these interfaces is also able to explain why flat bands are reached at a lower current density (see Fig.~\ref{fig1}) in comparison to interfaces of kind (b) and (c) in which the patch and channel densities are so high (see Table~\ref{Table2}) that the voltage drop remains negligible in this bias range.  \\

At sufficiently high forward bias, band curvature can be inverted in some areas or even in the whole diode, as the reader can realized from the comparison of $ \Phi_B^{hom} $ (in Table~\ref{Table2})  minus $ V_{dop}=0.38$~V to the applied voltage in Fig.~(\ref{fig1}), so that the previous expressions can no longer be used. In other words, energy bands bend upward near the interface when the flat band voltage is exceeded. From Eq.~(\ref{newbarrierlowering}), barrier lowering vanishes at flat bands for $ I^{p-o} $ while it stays constant for $ I^{non-p-o} $ and one can think that the effective barrier height remains more or less at the value obtained at flat bands beyond this threshold. But previous models which rely only on barrier lowering are unable to take into account the progressive disappearance of current inhomogeneities in the lightly doped layer due to the general increase of the carrier concentration at the whole interface. Such a behavior would finally lead to a more homogeneous current density and a voltage drop within the lightly doped layer expected to be simply the product of a series resistance and total current. But even this view fails because voltage drop turns out to be lower in Fig.~(\ref{fig1}) as if the resistivity were smaller than the predicted one. This discrepancy is neither an artifact of the adjustment procedure nor a self-heating effect which would hardly be justified because of the low dissipated power (typically one to some tens of mW in the range of interest), but is rather due to the injection of carrier by the heavily doped layer, involving a conductivity modulation due to the increase of carrier concentration above the equilibrium one and occurrence of a diffusion current starting from this heavily doped layer. This effect is evidenced from the comparison of the current-density data and curves simulated with a constant series resistance in Fig.~(\ref{fig1}) and it can also be checked because of the lower value of the current-density voltage derivative compared to the theoretical specific resistance (not shown) at sufficiently high current. A detailed study relying both on experiment and simulation is published elsewhere\cite{Muret4} and demonstrates that the injection effect is more and more important as the temperature is lowered below the freeze out of carriers and the active layer thickness is smaller than 2~$\mu$m, two conditions which are fulfilled here. It had been probably observed by other authors in Ni/SiC diodes\cite{Gammon}, although they did not give such an interpretation. This carrier injection which decreases the resistivity is still efficient at lower current density, so that the hole concentration at interface is underestimated even at the bias voltage where ideality factors are measured, at least when complete ionization of acceptors is not achieved. Correlatively, the effective resistivity is overestimated when its equilibrium value is taken into account as done in the present calculation, increasing thoughtlessly the simulated voltage drop and ideality factor in Fig.~(\ref{fig3}a') at the two lowest temperatures. This injection effect probably occurs because the lightly doped layer is only 1.4~$\mu$m thick and would disappear for the larger thicknesses suited to higher voltage operation. In any cases, it must be noticed that all the previous models discussed and built in this work are applicable to a range of bias voltages and currents limited by the occurrence of flat bands in the junction but that high injection effects may be present near flat bands and beyond.

\begin{table}
\begin{tabular}{||c||c|c|c|c|c||}
\hline
\hline \rule[-2ex]{0pt}{5.5ex} Interface type, sample & $ \Phi_B^{app}~~$ (V) & $U_b$~~(V) & $\Phi_B^{hom}~~$ (V) \\   
\hline \rule[-2ex]{0pt}{5.5ex} and diodes set identifiers &  &  & corrected for $ \gamma_0 $  \\ 
\hline 
\hline \rule[-2ex]{0pt}{5.5ex} (a) I A & 2.27 $\pm$ 0.12 & 1.05 $\pm$ 0.28 &   \\
\hline \rule[-2ex]{0pt}{5.5ex} (a) I B & 2.54 $\pm$ 0.14 & 1.30 $\pm$ 0.24 &   \\ 
\hline \rule[-2ex]{0pt}{5.5ex} (a) II A & 2.21 $\pm$ 0.04 & 0.97 $\pm$ 0.04 &  \\  
\hline \rule[-2ex]{0pt}{5.5ex} (a) II B & 1.61 $\pm$ 0.06 &   &   \\  
\hline \rule[-2ex]{0pt}{5.5ex} (b) III & 1.94 $\pm$ 0.04 & 0.73 $\pm$ 0.08 &  2.04 $\pm$ 0.04  \\  
\hline
\hline
\end{tabular}
\caption{Schottky barriers extrapolated at unity ideality factor at band bending $ U_b $ in different sets of diodes fabricated on three distinct diamond stacks numbered I, II and III, according to best fits of Eq. (\ref{nfonctiondephiB}) and $\Phi_B^{hom}$ after the $ \gamma_0 $ correction for the interface of kind (b).}
\label{Table3}
\end{table} 
 
\subsection{Characteristics and origin of barrier height inhomogeneities in the three types of interface}
The linear relationship between the effective barrier height as a function of the ideality factor function $(n_b-1)/n_b$ is actually checked for diodes prepared with the same procedure on three different substrates either without intentional post-annealing or annealed near 350\textcelsius~in Fig.~(\ref{fig7}). The correction of the constant term in Eq.~(\ref{nfonctiondephiB}) is done only in the case of indisputable relevance of the model of subsection~~\ref{secmodel2} and as it is not certain for interfaces of kind (a), extrapolated barrier heights $ \Phi_B^{app}$ and $\Phi_B^{hom}$ will be considered as the same in table~\ref{Table3}. It is noteworthy that the homogeneous barrier height $\Phi_B^{hom}$ can be different in two families of diodes (IA and IB; IIA and IIB) on the same diamond stack while it turns out to be very similar for two families of diodes (IA and IIA) belonging to two distinct diamond samples, whose the initial surfaces are all oxygen-terminated. Such results confirm that oxygen terminations, which are responsible for the enhanced electron affinity of the oxygen-terminated diamond surfaces and highest barrier at interfaces build on such surfaces, are expected to be organized in domains displaying distinct electron affinities\cite{Arnault,Klauser,Ghodbane,Sque,Fink,Robertson1}, resulting in distinct barrier heights, and that the same kind of oxygen terminations can exist on different samples. This dispersion matches that of the various molecules and bonds presented in subsection \ref{variousterminations} and probably also the lack of full coverage by oxygen related molecules. The linear relationship of Eq.~(\ref{nfonctiondephiB}) or (\ref{idgeneral}) seems to apply irrespective of the scale of the barrier inhomogeneities in comparison to the Debye length provided ideality factor is measured at high enough band bending, typically two third of the barrier height, as justified in the previous subsection. But because the interface annealed at 350\textcelsius~(Fig.~(\ref{fig7},III)) is a matter of the lowered barrier pinched-off patch model detailed in subsection \ref{secmodel2}, the correction which gives the homogeneous barrier from the apparent one is only exerted in this case in table~\ref{Table3} where the parameters are collected and displayed for all the diodes families. It must be noticed that the coherence of the various $\Phi_B^{hom}$ values appearing in Tables \ref{Table2} and \ref{Table3} is reasonable and that an average value of 2.04$\pm$0.04~V can be inferred for the interface of kind (b), only about 0.2-0.5 V below the values at stake for the interfaces of kind (a). The relevance of the model relying for interfaces of kind (b) on lowered barrier patches with sizes comparable or smaller than the Debye length and depicted by a patch parameter distribution centered on a peculiar value $ \gamma_0\,>\,0 $, with a much higher patch density than for unannealed interfaces, demonstrates that the main barrier inhomogeneities take place now along this microscopic scale, in contrast to what happened in unannealed interfaces. One must remind that the patch parameters $\gamma_p$, $\sigma_p$ and $\gamma_0$ depends on the one third power of the product of the square radius $R_0$ and initial barrier lowering $\Delta_p$ so that similar values of the two last quantities will not involve necessarily the same $R_0$ and $\Delta_p$, which can move in opposite directions. From results given in Table~\ref{Table2}, such statements suggest that the 350\textcelsius~ anneal resulted in the partial and random cancellation of the dipoles O$^{- \nu'}-$C$^{+ \nu'}$ whose stability is mainly governed by the enthalpy of the thermodynamical chemistry of these species and temperature, leaving the interface in an intermediate state and inducing new barrier lowering with a magnitude equal to the difference of the homogeneous potential barrier in interfaces of kind (b) and (c), i.e. $\Delta_p\approx1.1$~V at a microscopic scale, and increasing considerably the patch density. On the contrary, the initial inhomogeneities in unannealed interfaces of kind (a) likely resulted from the oxygen terminations dispersion over areas with larger size but smaller barrier lowering and much lighter density, with typical $\Delta_p\approx0.3$~V, similar to the difference in potential barriers reported for various interfaces of type (a) in Table~\ref{Table3}. To comply with the similarity of $\sigma_p$ and $\gamma_0$ for interfaces of kinds (a) and (b) in Table~\ref{Table2}, patches area has to be decreased by a factor of nearly four when going from case (a) to case (b), leading to a very good relevance of the model developed in subsection \ref{secmodel2}. In the case of interfaces of kind (c) annealed at 450~\textcelsius, these dipoles have completely disappeared, involving a barrier height as low as 0.96$\pm$0.04~V (average from Table~\ref{Table2}) and very low barrier fluctuations, as indicated by a central patch parameter $\gamma_0$ one decade smaller than that of other cases, which induced an almost ideal behavior of the current-voltage characteristics. In this last case, residual barrier inhomogeneities could be due to the random influence of the ionized acceptors close enough to the interface,  as it can be evidenced according to the model sketched in Fig.~(\ref{fig5}d) and invoked in subsection \ref{elementarycharges}, but the calculation will not be developed further and the reader interested in this question is invited to refer to ref.\cite{Tung2014} and references therein. An other residual source of barrier inhomogeneities lies in the possible fluctuations of the oxide thickness, because of the link between these two quantities as discussed in the next section. Anyway, indications derived from modeling the electrical properties of the junctions match very well the conclusions of the HRTEM and EELS studies developed in section \ref{experimental}, which showed the sharpening of the interfacial oxide layer after annealing and led to the very probable hypothesis of a chemical mechanism involving bond strengthening between the first oxygen layer and the last carbon one. The main consequence of this bonds transformation shows out through the very large change occurring in the averaged barrier heights of interfaces (a) and (c), with a difference near 1.4~V, resulting from the partial or complete cancellation of the dipoles O$^{- \nu'}-$C$^{+ \nu'}$ and also other dipole changes as discussed in the next section. The case of the diodes set IIB in Fig. (\ref{fig7}) does not comply with any previous models, a fact which suggests that the main part of the current in these diodes is not governed by an exponential law depending on the applied voltage, which was a common hypothesis of all the previous calculations. Therefore, leakage currents like those which might occur in threading dislocations or defects appearing as dark spots in cathodoluminescence images\cite{Ohmagari} could be responsible for a behavior characterized by rather higher ideality factors and constant apparent barrier height in Fig. (\ref{fig7},IIB).  Finally, homogeneous barrier heights $\Phi_{B,2}^{hom}$ appearing in Table \ref{Table2} can be considered as the most representative of the physical situation of each interface, except that unannealed interfaces of type (a) may experience two or three different values like mentioned in Table~\ref{Table3}, the lowest being about 0.2~V below that indicated in Table \ref{Table2} for a homogeneous set of diodes. This local barrier lowering may agree the quantity $ \Delta\Phi $ invoked in subsection~\ref{secmodel3} over areas with lateral size larger than the Debye length to built the last electrical model relying on non pinched-off channels in the previous subsection. It can be concluded that, despite some minor imperfect agreements between experimental data and models of inhomogeneous barriers detailed in this section and their limitations to the depletion regime of Schottky junctions, more informations and deeper physical insight into the situation of these metal-diamond interfaces have been gained and can be reinvested in barrier height models discussed in the next section. The most important quantity which can be extracted from the analysis of electrical characteristics of Schottky diodes is obviously the true potential barrier height at an interface free of inhomogeneities and it has been defined at the beginning of this study as  $\Phi_{B}^{hom}$. It is always underestimated if the barrier height is measured from the slope of a Richardson plot and a relevant value $\Phi_{B}^{hom}$ needs to be deduced either from adjustment of the measured barrier height as a function of temperature, or from the direct adjustment of the current-voltage density with the appropriate model chosen among those described in this section, or from the statistics of barrier heights as a function of ideality factors in a set of diodes, or preferably from several methods.

\section{Barrier height models at metal-oxygenated diamond interfaces}
The barrier heights issue at metal-diamond interfaces belongs to the long time debated topic of how the Fermi level position is determined at semiconductor surfaces and interfaces, discussed as early as in the thirties by Sir Nevill Mott \cite{Mott} concerning the rectification theory and by Walter Schottky \cite{Schottky}. Because this first model, which stated the linear dependence of the barrier height upon the metal work function with a unity slope, did not comply with the huge amount of data gathered since its publication, other explanations have been proposed to justify the weaker dependence experimentally observed, known as Fermi level pinning. New barrier height models have been progressively introduced within the framework of the "metal induced gap states" principle\cite{Tersoff} or its generalization to "interface induced gap states"\cite{Monchbook} in order to link band alignment and either the work function or electronegativity of the metal at the broadest possible panel of metal-semiconductor interfaces. All rely on the existence of an interface state density localized in the first elementary cells of the semiconductor which is neutral when the Fermi level coincides with an energy level $ \Phi_0 $ characteristic of the semiconductor and referred as to the neutral level, defined here as the energy difference between such a level and the valence band edge at interface. Deviations from this position and from neutrality occur if a metal having an electronegativity different from that of the semiconductor is put in intimate contact with the semiconductor surface, assuming that the interface states inside the semiconductor remain invariant. This last hypothesis leads to a negative feedback effect due to the reaction of the charge inside interface states onto the voltage jump due to the double layer of the areal dipole astride the interface, which induces a linear relationship between the barrier height and the difference of the metal and semiconductor electronegativities $ X_m - X_s$, with either a positive slope $ S_X $ on a n-type semiconductor, or a negative one, $ -S_X $, on a p-type semiconductor as here:        
\begin{subequations}
\begin{equation}\label{MIGS}
e_0\,\Phi_{Bp}^{hom}\,=\,\Phi_0 - S_X\,(X_m - X_s)
\end{equation} 
where
\begin{equation}
S_X\,=\,\frac{A_X}{1+(e_0^2\,/\,\varepsilon_i \, \varepsilon_0)\,D_{is}\,d_{di}}
\end{equation} 
with $ D_{is} $ the interface state density and $ A_X $ a coefficient which depends on the electronegativity scale, equal to 0.86 if the Miedema scale is used\cite{Monchbook}.
\end{subequations}
This theory was rather successful for intimate metal-semiconductor contacts, which have been better and better controlled at an atomic scale for the seventies, sometimes with an epitaxial metal layer. However, some dispersion still remained, which was not attributable to experimental uncertainties. This is specially true for diamond, which has not been the subject of extensive reports since the nineties, excepted to evidence the wide range of barrier heights existing for the same metal\cite{Evans} and with the notable exception of the hydrogenated surface\cite{Tsugawa}. Such a dispersion has been ascribed to the various structures, terminations, ad-atoms and natures of bond which can exist at the diamond surfaces and interfaces, like already invoked in the case of oxygen termination.
\begin{figure}
\includegraphics{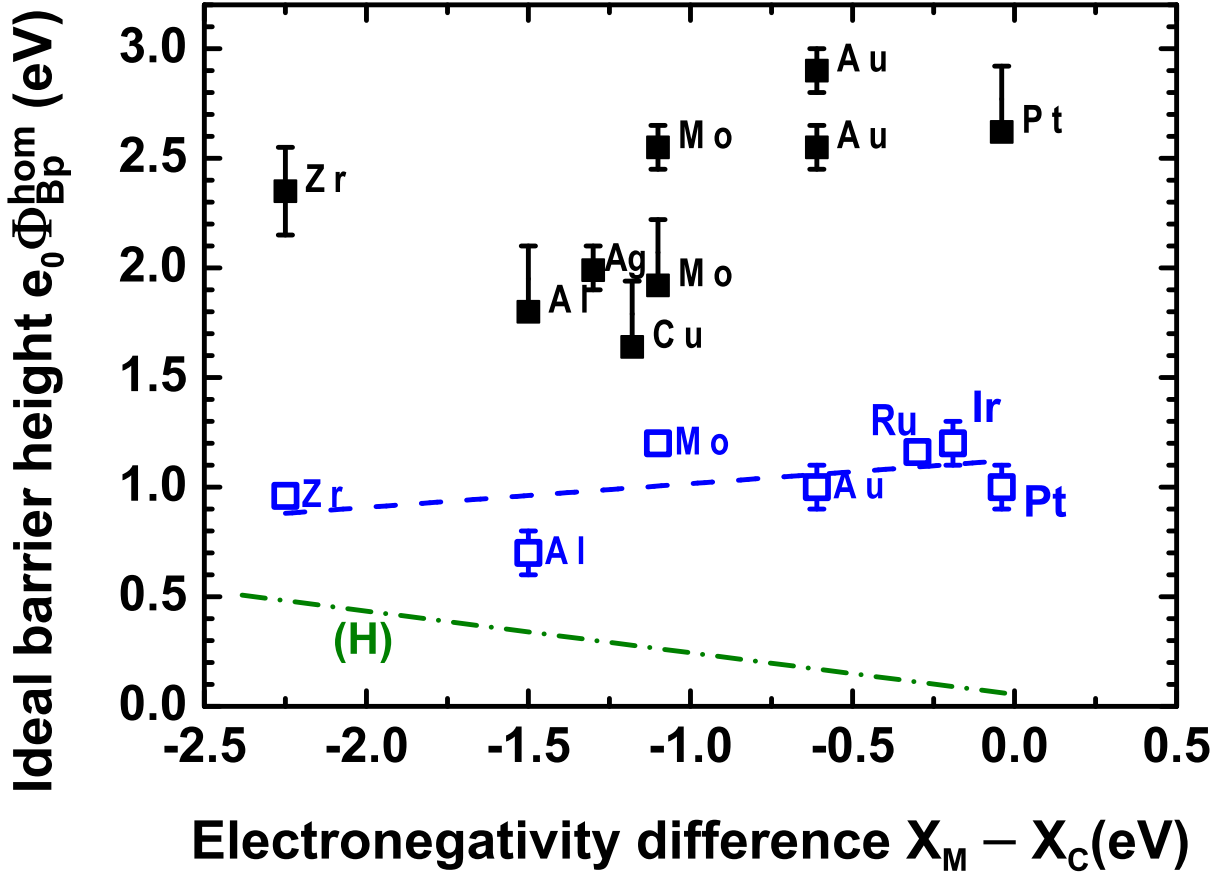}
\caption{\label{fig8} Ideal barrier heights as a function of the Miedama electronegativity difference between metal and carbon (assumed to be 5.7~eV) for oxygenated interfaces annealed below 250~\textcelsius~(full squares) and for those annealed in the range 400-650~\textcelsius~(open squares), together with a linear fit in dashed line in the second case. The dash-dotted line (H) is the best fit of barrier data on the hydrogenated surface by Tsugawa\cite{Tsugawa} \textit{et al.} as a function of the same abscissa. Multiple data for the same metal (Au and Mo) come from different publications.}
\end{figure}     
Recently, measurements of barrier heights have been published at carefully prepared  oxygenated and eventually post-annealed diamond surfaces\cite{Teraji1,Teraji2,Kone,Umezawa,Nawawi,Ueda1,Ueda2}. Together with the present results for Zr-diamond interfaces, ideal barrier heights are plotted in Fig.~(\ref{fig8}) with error bars added according to the rules discussed in the previous section: either only above the published value when the method underestimated the ideal homogeneous barrier $ \Phi_{Bp}^{hom} $, or symmetrical when $ \Phi_{Bp}^{hom} $ is directly indicated by authors or reassessed with the help of an appropriate method among those discussed in the preceding section. Despite some dispersion, illustrated sometimes in Fig.~(\ref{fig8}) and (\ref{fig9}) with different barrier heights reported for the same metal in literature, two groups of data appear well separated, one for the contacts annealed at typical temperatures less than 250~\textcelsius~or as-deposited and an other for contacts annealed at temperatures in the range 400-650~\textcelsius, as a function of the Miedema electronegativity difference between metal and carbon, the last one being left constant at 5.7~eV. Electronegativity was originally used to describe the interaction between two (or more) atoms in a molecule and thereafter in solids. But it is impossible here to describe the two electronic situations regarding barrier heights in unannealed and annealed interfaces with only two electronegativity values, one for the metal and the other for oxygenated diamond,  since changes after annealing occurred while oxygen still remains at the interface as demonstrated in section \ref{experimental}. Even if the diamond electronegativity were shifted by about $-1.5$~eV to take the oxygen terminations into account in the unannealed interfaces, although this is not a very sensible operation, the whole set of data does not show  clearly any indisputable trend, testifying a first shortcoming. A second one appears if the search for a linear relationship is restricted to the second group of data, which shows a reasonable alignment depicted by the dashed line in Fig.~(\ref{fig8}), but characterized by a positive slope, in contradiction with Eq.~(\ref{MIGS}) and what is experimentally found for all other cases including that of a hydrogenated diamond surface\cite{Tsugawa}. This is the second shortcoming. Incidentally, it is difficult to believe that the energy distribution of "interface induced interface states" may remain the same when going from the hydrogenated then air-exposed diamond surfaces to oxygenated ones, as it is well known that the former is provided with adsorbates and conducting whereas the later is not\cite{RisMaier}. An other approach, relying on physico-chemical quantities able to influence interface bonding, is necessary. \\
\begin{figure}
\includegraphics{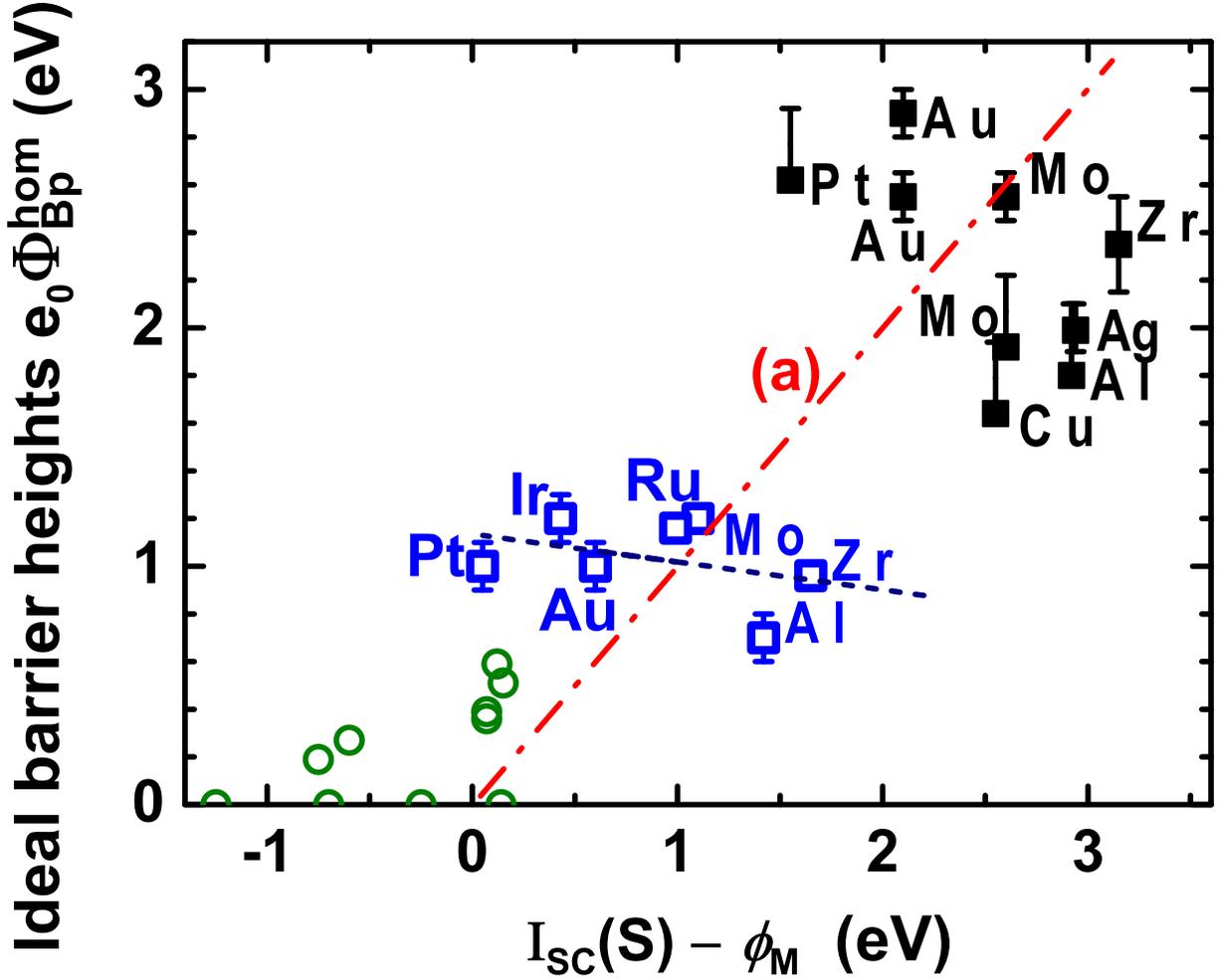}
\caption{\label{fig9} Ideal barrier heights as a function of the difference between the ionization energy of diamond and the work function of metals (multiple data come from different publications) for (i) as-deposited interfaces (full squares) on an oxygenated (100)-oriented surface, (ii) for contacts annealed in the range 400-650~\textcelsius~(open squares) on the same initial surface, together with a linear fit in dotted line, and (iii) for contacts on the hydrogenated (100)-oriented surface (open circles) from ref.\cite{Tsugawa}. Multiple data for the same metal (Au and Mo) come from different publications. The ionization energy includes the electron affinity shift due to surface terminations, yielding (i) $ \chi\,+\,\Delta\chi(S) = 1.7 $~eV; (ii) $\chi\,+\,\Delta\chi(S) = 0.3 $~eV and (iii)~$ \chi\,+\,\Delta\chi(S) = -1.1 $~eV respectively for the oxygenated, bare and hydrogenated (100)-oriented surfaces from ref.\cite{Ristein}. The dash-dot line (a) displays Eq.~(\ref{barriergeneral}) where the last term is ignored, thus featuring the Mott-Schottky relationship.}
\end{figure}  
In the last twenty years, scientists performed modifications of the Schottky barrier heights by means of various ad-atoms and/or physical, chemical and thermal treatments applied to the surface generally before metal deposition excepted for annealing. They realized that these changes are dependent on bonding between interface atoms, and that chemical interactions cannot be ignored between atoms in the neighborhood of the interface. A first improvement of the theory has been brought by R.T. Tung who showed that potential energies in the solids and through the interface were more appropriate to describe band alignment\cite{Tung2001} in such a situation. This idea is applied to data addressed previously which are now plotted in Fig.~(\ref{fig9}). Secondly, the dipole strength of the various double layers occurring at interface is the relevant quantity to take the possible interlayer atoms, the different bonds\cite{Tung2000} and even the possible new states due to transfer doping on the hydrogenated surface\cite{Tsugawa} into account. It is lumped into the last term in the following continuity equation which makes use of the metal work function $ \phi_M $ and electron affinity of the semiconductor as follows:  
\begin{subequations}
\begin{equation}\label{barriergeneral}
e_0\,\Phi_{Bp}^{hom}\,=\,\left (I_{SC}(S)\:-\:\phi_M \right )\:-\:e_0\,\Delta q \, d_{bd} \, N_{bd} / (\varepsilon_i \, \varepsilon_0)
\end{equation} 
where the last term is the potential energy jump due to a surface dipole comprising a charge per bond $ \Delta q $ on the metal side and $ -\Delta q $ on the diamond side separated by the distance $ d_{bd} $, with an areal density of dipoles $ N_{bd}$, while the ionization energy given by:
\begin{equation}
I_{SC}(S)\,=\,E_G\,+\,\chi\,+\,\Delta\chi(S)
\end{equation} 
depends on the diamond band gap $E_G$, electron affinity $\chi$ of the bare surface and possible change in electron affinity $ \Delta\chi(S) $ due to foreign atoms termination or particular reconstruction on each surface $ S $, the $ \chi\,+\,\Delta\chi(S) $ values which are retained being indicated in the caption of Fig.~\ref{fig9}. 
\end{subequations} \\

When the last term is neglected in Eq.~(\ref{barriergeneral}), the Mott-Schottky relationship for contacts on p-type semiconductors is found, with a unity slope, sketched by a dash-dotted line in Fig.~(\ref{fig9}). If the dispersion of data is forgotten in a first approach and the clouds of points related to the three categories of interface in this graph are replaced by their respective barycenters, one can guess that they align roughly on the Mott-Schottky line. This trend has been already evidenced in silicides on n-type silicon interfaces\cite{Tung1993}. The influence of the cloud related to the hydrogenated surfaces should be minored because Pt, Au, Cu and Ag metals contacts are displayed with zero barrier heights although they should be negative but cannot be actually measured and reported. More accurately, if one observes the barrier heights of each metal (Pt, Au, Mo, Al, Zr) whose contact on diamond has been measured in the two groups of oxygenated interfaces, respectively before and after annealing, their difference lies in the range $1.40\pm0.25$~eV for all of them, corresponding to the same change in electron affinity. This common shift indicates that the barrier height for a given metal is primarily controlled by the electron affinity of diamond, with a little influence of the dipole due to interface bonding, reflected in the last term of Eq.~(\ref{barriergeneral}), and testifies the disappearance of the oxygen terminations proper to initial oxygenated surfaces after annealing, or their transformation, irrespective of the presence of an oxide or not. As it was discussed in subsection \ref{variousterminations}, hydroxyl and carboxyl terminations likely exist before annealing and they can participate to bridged hydrogen bonds as suggested in \cite{Sque}, thus allowing adhesion of the atomic layers deposited on the surface by some kind of weak bonding. On the contrary, stronger, and therefore shorter bonds between carbon and atoms of the Zirconium oxide interlayer, that are mainly oxygen ones, would occur after annealing, in close correlation with the shrinkage of the oxide interlayer detected in section \ref{experimental}. But this statement does not mean that "interface induced gap states" have been suppressed or have never existed. They are just accommodating the electric charge which guarantees that the dipole layers, both necessary for setting the electron affinity and the bond dipole counterparts on the diamond side, can be present. Now, at a more subtle level, the general trend in each group of contacts made on the oxygenated surface has an inverted slope in comparison to the Mott-Schottky line, as already noticed in Fig.~\ref{fig8}. The correlation coefficient is better in annealed interfaces, probably because the electron affinity dispersion due to various oxygen terminations has been suppressed. The physical reason for the linear relationship depicted by a dotted line in Fig.~(\ref{fig9}) can be thus discussed with more confidence in the annealed interfaces than in unannealed ones because the exact electron affinity is unknown in the second case for each metal and therefore replaced by the single value measured in vacuum as indicated in the caption of the figure. First, it is useful to notice that neither the first term in Eq.~(\ref{barriergeneral}), which also scales the abscissa in this diagram, nor the second term can account for the observed trend if this later term is supposed to be proportional to the former one because the slope would be increased above one when the ionization energy is held constant. Therefore, one is forced to conclude that hidden physical parameters able to govern this second term might be involved. These are the nature of the bonds and interlayer, the charge transfer across them, their polarizability, and the interface structure with a special attention to the interlayer thickness and relative permittivity. In the case of metals like Pt, Ir and Au, oxide formation is very unlikely since the enthalpy of formation is either positive or close to zero, whereas an oxide is present when Zr, as shown at the beginning of this study, and most probably Al, are used. If the interlayer is supposed to be a dielectric consisting of a ionic solid, its thickness $t_{ox}$ is added to the bond length $d_{bl}$ so that the total dipolar distance is roughly $d_{bd}=t_{ox}+d_{bl}$. Then this distance determines the strength of the last term in Eq.~(\ref{barriergeneral}) since the charge counterpart of the dipole on the metal side is now located at a slightly larger distance than $t_{ox}$ from the last diamond atomic plane and if the charge inside the oxide layer is neglected at first approximation. Moreover, important effects are due to the charge transfer $ \Delta q $ which is governed by the nature of the bonds, electronegativity difference between the metal and semiconductor, the polarizability of bonds and interlayer thickness, as described in the bond polarization model proposed by R.T. Tung\cite{Tung2001}, and also screening which can mainly accounted for by the relative permittivity occurring at the denominator of the last term in Eq.~(\ref{barriergeneral}). To get a deeper insight in the problem at least at a first order level of approximation, let us define the abscissa $ W_{SM}\:=\:\left (I_{SC}(S)\:-\:\phi_M \right)$ and take the derivative of the barrier height $ e_0\,\Phi_{Bp}^{hom} $ when the electron affinity, and hence ionization energy are constant:
\begin{equation} \label{barriertrend}
\frac{d(e_0\,\Phi_{Bp}^{hom})}{d W_{SM}}\,=\,1\,+\,\frac{d (e_0\,\Delta q \, d_{bd} \, N_{bd} / (\varepsilon_i \, \varepsilon_0))}{d \phi_{M}}
\end{equation}      
This slope would stay close to one whenever the second term is negligible but can become negative as observed in Fig.~(\ref{fig9}) if both the second term is negative and its absolute value exceeds unity. Since the main trend occurring when the metal is changed to another one with an increased work function involves a decreased reactivity with oxygen and therefore a thickness reduction or disappearance of the oxide layer, one must wonder what are the resulting effects on the second term of Eq.(\ref{barriertrend}), even if a continuous variation is more artificial than reality. The dipolar distance $ d_{bd} $ is obviously decreased while the interlayer permittivity $ \varepsilon_i $ initially near that of high-k dielectric materials is going down closer to unity, resulting in a smaller screening effect. But because theses two parameters are following the same direction of variation and influence the second term by their ratio, the resulting sign of the derivative turns out to be uncertain. Consequently, the fundamental issue relies mainly on the charge transfer $ \Delta q $ variation, which is dependent on the effective bonding at the metal-semiconductor interfaces\cite{Tung2001, Tung2014}, including not only orbitals of the two original materials but also atoms or molecules belonging to the interlayer. In such a molecular orbital scheme, metals with the lowest work functions (and also more electropositive) would make more ionic bonds, with a polarizability likely augmented by the presence of additional oxygen bonds inside the interlayer, thus inducing a stronger positive charge transfer $ \Delta q $ in the orbitals of the last metal atomic layer. Conversely, noble metals (and also more electronegative ones) are expected to share the electrons of their d states in a more equalized way with carbon orbitals, while carbide forming metals would follow a medium behavior. Such a general trend of the charge transfer $ \Delta q $ which would experience a decrease when the metal work function (and electronegativity) is increased, is able to justify the systematic negative sign of the derivative in the second member of Eq.(\ref{barriertrend}). Consequently, the slope of barrier heights as a function of $ W_{SM}\:=\:\left (I_{SC}(S)\:-\:\phi_M \right)$ can take values smaller than one, as often observed in many semiconductors, and in the case of a magnitude of the absolute value of the last term in the same equation greater than unity, it can become negative as shown in Fig.~(\ref{fig9}), in contrast to what is known in all other semiconductors. Such a model can rely on the electronegativity difference between the semiconductor and metal, which mainly governs the dipole strength induced by interface bonding, but only for the last term in  Eq.(\ref{barriergeneral}), independent of the interface induced gap states at first order. At a second order level of approximation, the bonding nature of atoms inside the oxide and fixed charges in it will be also able to bring changes in $ \Delta q $ and these quantities probably contribute to the small deviations visible in Fig.~(\ref{fig9}) for annealed interfaces, together with the oxide interlayer thickness, permittivity, and changes in the atoms arrangement at interface and in the first diamond atomic layers. The same model also applies both to unannealed oxygenated interfaces and hydrogenated interfaces, however with larger fluctuations in comparison to a linear behavior, and a positive slope in the second case. It covers therefore all the situations, although it is difficult to afford quantitative parameters in the second term of Eq.~(\ref{barriergeneral}) for each metal in an accurate way.

\section{Conclusion}
In this work, interfaces of Zirconium with oxygenated diamond surfaces before and after two kinds of anneal has been characterized on the one hand by HRTEM imaging and EELS spectroscopy, and on the other hand by thorough electrical measurements made on the Schottky junctions. HRTEM and EELS techniques evidenced an oxide interlayer comprising approximately one or two Zr oxide atomic layers, which sharpened after annealing. Electrical measurements were performed to collect and analyze  current-voltage characteristics at several temperatures in many diodes in order to investigate the detailed shape of the current density behavior with applied voltage, the Richardson constant, the apparent barrier height and ideality factor. Models taking the barrier height and current density inhomogeneities into account have been implemented and brought face to face with experimental results. From these comparisons, several physical parameters have been estimated. The Richardson constant turned out to be close to one tenth of the standard value, and even one thirteenth when the barrier height disorder reached its maximum value. The effect of different shapes of barrier fluctuations has been studied either numerically in the case of local barrier enhancements or from two analytical models adapted to local barrier lowering and current crowding in the active layer of the diode. In the former case, which happens when fixed repulsive charges are close to the interface like evidenced in metal-ZrO$ _{2} $ -diamond structures, the image force effect is the only relevant one as far as current-voltage characteristics are concerned. In the case of random patches comprising a smaller barrier than the ideal homogeneous one, it has been shown that barrier inhomogeneities can be characterized by several parameters related to their amplitude, density, and to the presence or absence of both the pinch-off effect in these lowered barrier patches and current crowding in narrowed channels of the depletion zone. Mixing two of the last effects turns out to be necessary to quantitatively explain the curvature of current-voltage characteristics of unannealed interfaces in a semi-logarithmic plot. The ideal homogeneous barrier is the truly representative physical quantity of the interface because it is free from the local and random inhomogeneities which are responsible for the non ideality of the diodes. It has been inferred from matching previous models to electrical data. Rules and uncertainties have been derived for a confident evaluation of this fundamental quantity. All other parameters have been extracted in turn from the comparison of the non ideal behaviors with the quantitative predictions of the models. Main conclusions were the strong reduction of both the barrier inhomogeneities and homogeneous barrier of 1.4 eV after annealing interfaces at 450~\textcelsius. Because of the known inhomogeneity of the oxygen terminations and enhancement of the electron affinity by also 1.4 eV on the initial oxygenated diamond surface, these facts strongly suggested that original oxygen terminations vanished after such an annealing and are replaced by stronger bonds between carbon and atoms of the Zirconium oxide interlayer. Then, the more general issue of barrier height determination and Fermi level pinning has been addressed by means of an analysis of the experimental data collected in the present work and in recent literature, reassessed with the help of the methods presented before when necessary. The same evolution of barrier heights for four other metals after annealing confirms the disappearance of oxygen terminations, irrespective of the presence of an oxide interlayer, and the electron affinity of the diamond surfaces as a relevant quantity for barrier height description. For the whole set of metals in contact with either unannealed or annealed oxygenated surfaces, or hydrogenated surfaces, a rough but clear alignment of barrier heights along the Mott-Schottky straight line, with a unity slope, really emerged regarding the difference between the ionization energy of diamond surfaces and metal work function. However, for each particular interface, the behavior at a smaller scale appeared different. In the case of annealed oxygenated interfaces, the slope turns out to be slightly negative, a result incompatible with the ”interface induced gap states” model which assumes these states as invariant from one metal to another. On the contrary, the bond-polarization scheme, which relies on the interface dipole strength, itself dependent of the molecular orbitals involved in the bonds across the interface, may justify the slopes both at a global level and for a particular surface. All these concerns have been presented and discussed with the help of new experimental results, improved models and recent high quality experimental works in the purpose of fostering new advances in understanding the old problem of potential barrier at metal-covalent semiconductor interfaces.

\bibliography{barriersZrC_rev}

\begin{thebibliography}{59}%
\makeatletter
\providecommand \@ifxundefined [1]{%
 \@ifx{#1\undefined}
}%
\providecommand \@ifnum [1]{%
 \ifnum #1\expandafter \@firstoftwo
 \else \expandafter \@secondoftwo
 \fi
}%
\providecommand \@ifx [1]{%
 \ifx #1\expandafter \@firstoftwo
 \else \expandafter \@secondoftwo
 \fi
}%
\providecommand \natexlab [1]{#1}%
\providecommand \enquote  [1]{``#1''}%
\providecommand \bibnamefont  [1]{#1}%
\providecommand \bibfnamefont [1]{#1}%
\providecommand \citenamefont [1]{#1}%
\providecommand \href@noop [0]{\@secondoftwo}%
\providecommand \href [0]{\begingroup \@sanitize@url \@href}%
\providecommand \@href[1]{\@@startlink{#1}\@@href}%
\providecommand \@@href[1]{\endgroup#1\@@endlink}%
\providecommand \@sanitize@url [0]{\catcode `\\12\catcode `\$12\catcode
  `\&12\catcode `\#12\catcode `\^12\catcode `\_12\catcode `\%12\relax}%
\providecommand \@@startlink[1]{}%
\providecommand \@@endlink[0]{}%
\providecommand \url  [0]{\begingroup\@sanitize@url \@url }%
\providecommand \@url [1]{\endgroup\@href {#1}{\urlprefix }}%
\providecommand \urlprefix  [0]{URL }%
\providecommand \Eprint [0]{\href }%
\providecommand \doibase [0]{http://dx.doi.org/}%
\providecommand \selectlanguage [0]{\@gobble}%
\providecommand \bibinfo  [0]{\@secondoftwo}%
\providecommand \bibfield  [0]{\@secondoftwo}%
\providecommand \translation [1]{[#1]}%
\providecommand \BibitemOpen [0]{}%
\providecommand \bibitemStop [0]{}%
\providecommand \bibitemNoStop [0]{.\EOS\space}%
\providecommand \EOS [0]{\spacefactor3000\relax}%
\providecommand \BibitemShut  [1]{\csname bibitem#1\endcsname}%
\let\auto@bib@innerbib\@empty
%</preamble>
\bibitem [{\citenamefont {Volpe}\ \emph {et~al.}(2009)\citenamefont {Volpe},
  \citenamefont {Pernot}, \citenamefont {Muret},\ and\ \citenamefont
  {Omn\`{e}s}}]{Volpe1}%
  \BibitemOpen
  \bibfield  {author} {\bibinfo {author} {\bibfnamefont {P.-N.}\ \bibnamefont
  {Volpe}}, \bibinfo {author} {\bibfnamefont {J.}~\bibnamefont {Pernot}},
  \bibinfo {author} {\bibfnamefont {P.}~\bibnamefont {Muret}}, \ and\ \bibinfo
  {author} {\bibfnamefont {F.}~\bibnamefont {Omn\`{e}s}},\ }\href {\doibase
  http://dx.doi.org/10.1063/1.3086397} {\bibfield  {journal} {\bibinfo
  {journal} {Applied Physics Letters}\ }\textbf {\bibinfo {volume} {94}},\
  \bibinfo {eid} {092102} (\bibinfo {year} {2009})}\BibitemShut {NoStop}%
\bibitem [{\citenamefont {Volpe}\ \emph {et~al.}(2010)\citenamefont {Volpe},
  \citenamefont {Muret}, \citenamefont {Pernot}, \citenamefont {Omn\`{e}s},
  \citenamefont {Teraji}, \citenamefont {Koide}, \citenamefont {Jomard},
  \citenamefont {Planson}, \citenamefont {Brosselard}, \citenamefont {Dheilly},
  \citenamefont {Vergne},\ and\ \citenamefont {Scharnholz}}]{Volpe2}%
  \BibitemOpen
  \bibfield  {author} {\bibinfo {author} {\bibfnamefont {P.-N.}\ \bibnamefont
  {Volpe}}, \bibinfo {author} {\bibfnamefont {P.}~\bibnamefont {Muret}},
  \bibinfo {author} {\bibfnamefont {J.}~\bibnamefont {Pernot}}, \bibinfo
  {author} {\bibfnamefont {F.}~\bibnamefont {Omn\`{e}s}}, \bibinfo {author}
  {\bibfnamefont {T.}~\bibnamefont {Teraji}}, \bibinfo {author} {\bibfnamefont
  {Y.}~\bibnamefont {Koide}}, \bibinfo {author} {\bibfnamefont
  {F.}~\bibnamefont {Jomard}}, \bibinfo {author} {\bibfnamefont
  {D.}~\bibnamefont {Planson}}, \bibinfo {author} {\bibfnamefont
  {P.}~\bibnamefont {Brosselard}}, \bibinfo {author} {\bibfnamefont
  {N.}~\bibnamefont {Dheilly}}, \bibinfo {author} {\bibfnamefont
  {B.}~\bibnamefont {Vergne}}, \ and\ \bibinfo {author} {\bibfnamefont
  {S.}~\bibnamefont {Scharnholz}},\ }\href {\doibase
  http://dx.doi.org/10.1063/1.3520140} {\bibfield  {journal} {\bibinfo
  {journal} {Applied Physics Letters}\ }\textbf {\bibinfo {volume} {97}},\
  \bibinfo {eid} {223501} (\bibinfo {year} {2010})}\BibitemShut {NoStop}%
\bibitem [{\citenamefont {Vescan}\ \emph {et~al.}(1998)\citenamefont {Vescan},
  \citenamefont {Daumiller}, \citenamefont {Gluche}, \citenamefont {Ebert},\
  and\ \citenamefont {Kohn}}]{Vescan}%
  \BibitemOpen
  \bibfield  {author} {\bibinfo {author} {\bibfnamefont {A.}~\bibnamefont
  {Vescan}}, \bibinfo {author} {\bibfnamefont {I.}~\bibnamefont {Daumiller}},
  \bibinfo {author} {\bibfnamefont {P.}~\bibnamefont {Gluche}}, \bibinfo
  {author} {\bibfnamefont {W.}~\bibnamefont {Ebert}}, \ and\ \bibinfo {author}
  {\bibfnamefont {E.}~\bibnamefont {Kohn}},\ }\href {\doibase
  http://dx.doi.org/10.1016/S0925-9635(97)00200-8} {\bibfield  {journal}
  {\bibinfo  {journal} {Diamond and Related Materials}\ }\textbf {\bibinfo
  {volume} {7}},\ \bibinfo {pages} {581 } (\bibinfo {year} {1998})}\BibitemShut
  {NoStop}%
\bibitem [{\citenamefont {Muret}\ \emph {et~al.}(1999)\citenamefont {Muret},
  \citenamefont {Pruvost}, \citenamefont {Saby}, \citenamefont {Lucazeau},
  \citenamefont {Tan}, \citenamefont {Gheeraert},\ and\ \citenamefont
  {Deneuville}}]{Muret1}%
  \BibitemOpen
  \bibfield  {author} {\bibinfo {author} {\bibfnamefont {P.}~\bibnamefont
  {Muret}}, \bibinfo {author} {\bibfnamefont {F.}~\bibnamefont {Pruvost}},
  \bibinfo {author} {\bibfnamefont {C.}~\bibnamefont {Saby}}, \bibinfo {author}
  {\bibfnamefont {E.}~\bibnamefont {Lucazeau}}, \bibinfo {author}
  {\bibfnamefont {T.~N.}\ \bibnamefont {Tan}}, \bibinfo {author} {\bibfnamefont
  {E.}~\bibnamefont {Gheeraert}}, \ and\ \bibinfo {author} {\bibfnamefont
  {A.}~\bibnamefont {Deneuville}},\ }\href {\doibase
  http://dx.doi.org/10.1016/S0925-9635(98)00380-X} {\bibfield  {journal}
  {\bibinfo  {journal} {Diamond and Related Materials}\ }\textbf {\bibinfo
  {volume} {8}},\ \bibinfo {pages} {961 } (\bibinfo {year} {1999})}\BibitemShut
  {NoStop}%
\bibitem [{\citenamefont {Liao}\ \emph {et~al.}(2005)\citenamefont {Liao},
  \citenamefont {Alvarez},\ and\ \citenamefont {Koide}}]{Liao}%
  \BibitemOpen
  \bibfield  {author} {\bibinfo {author} {\bibfnamefont {M.}~\bibnamefont
  {Liao}}, \bibinfo {author} {\bibfnamefont {J.}~\bibnamefont {Alvarez}}, \
  and\ \bibinfo {author} {\bibfnamefont {Y.}~\bibnamefont {Koide}},\ }\href
  {\doibase 10.1143/JJAP.44.7832} {\bibfield  {journal} {\bibinfo  {journal}
  {Japanese Journal of Applied Physics}\ }\textbf {\bibinfo {volume} {44}},\
  \bibinfo {pages} {7832} (\bibinfo {year} {2005})}\BibitemShut {NoStop}%
\bibitem [{\citenamefont {Craciun}\ \emph {et~al.}(2004)\citenamefont
  {Craciun}, \citenamefont {Saby}, \citenamefont {Muret},\ and\ \citenamefont
  {Deneuville}}]{Craciun}%
  \BibitemOpen
  \bibfield  {author} {\bibinfo {author} {\bibfnamefont {M.}~\bibnamefont
  {Craciun}}, \bibinfo {author} {\bibfnamefont {C.}~\bibnamefont {Saby}},
  \bibinfo {author} {\bibfnamefont {P.}~\bibnamefont {Muret}}, \ and\ \bibinfo
  {author} {\bibfnamefont {A.}~\bibnamefont {Deneuville}},\ }\href {\doibase
  http://dx.doi.org/10.1016/j.diamond.2003.10.012} {\bibfield  {journal}
  {\bibinfo  {journal} {Diamond and Related Materials}\ }\textbf {\bibinfo
  {volume} {13}},\ \bibinfo {pages} {292 } (\bibinfo {year} {2004})},\ \bibinfo
  {note} {carbon Materials for Active Electronics. Proceedings of Symposium L,
  E-MRS Spring Meeting 2003}\BibitemShut {NoStop}%
\bibitem [{\citenamefont {Saby}\ \emph {et~al.}(2002)\citenamefont {Saby},
  \citenamefont {Muret}, \citenamefont {Pruvost},\ and\ \citenamefont
  {Patrat}}]{Saby1}%
  \BibitemOpen
  \bibfield  {author} {\bibinfo {author} {\bibfnamefont {C.}~\bibnamefont
  {Saby}}, \bibinfo {author} {\bibfnamefont {P.}~\bibnamefont {Muret}},
  \bibinfo {author} {\bibfnamefont {F.}~\bibnamefont {Pruvost}}, \ and\
  \bibinfo {author} {\bibfnamefont {G.}~\bibnamefont {Patrat}},\ }\href
  {\doibase http://dx.doi.org/10.1016/S0925-9635(01)00652-5} {\bibfield
  {journal} {\bibinfo  {journal} {Diamond and Related Materials}\ }\textbf
  {\bibinfo {volume} {11}},\ \bibinfo {pages} {1332 } (\bibinfo {year}
  {2002})}\BibitemShut {NoStop}%
\bibitem [{\citenamefont {Evans}\ \emph {et~al.}(2009)\citenamefont {Evans},
  \citenamefont {Roberts}, \citenamefont {Williams}, \citenamefont
  {Vearey-Roberts}, \citenamefont {Bain}, \citenamefont {Evans}, \citenamefont
  {Langstaff},\ and\ \citenamefont {Twitchen}}]{Evans}%
  \BibitemOpen
  \bibfield  {author} {\bibinfo {author} {\bibfnamefont {D.~A.}\ \bibnamefont
  {Evans}}, \bibinfo {author} {\bibfnamefont {O.~R.}\ \bibnamefont {Roberts}},
  \bibinfo {author} {\bibfnamefont {G.~T.}\ \bibnamefont {Williams}}, \bibinfo
  {author} {\bibfnamefont {A.~R.}\ \bibnamefont {Vearey-Roberts}}, \bibinfo
  {author} {\bibfnamefont {F.}~\bibnamefont {Bain}}, \bibinfo {author}
  {\bibfnamefont {S.}~\bibnamefont {Evans}}, \bibinfo {author} {\bibfnamefont
  {D.~P.}\ \bibnamefont {Langstaff}}, \ and\ \bibinfo {author} {\bibfnamefont
  {D.~J.}\ \bibnamefont {Twitchen}},\ }\href {\doibase
  10.1088/0953-8984/21/36/364223} {\bibfield  {journal} {\bibinfo  {journal}
  {Journal of Physics: Condensed Matter}\ }\textbf {\bibinfo {volume} {21}},\
  \bibinfo {pages} {364223} (\bibinfo {year} {2009})}\BibitemShut {NoStop}%
\bibitem [{\citenamefont {Teraji}\ \emph
  {et~al.}(2009{\natexlab{a}})\citenamefont {Teraji}, \citenamefont {Garino},
  \citenamefont {Koide},\ and\ \citenamefont {Ito}}]{Teraji1}%
  \BibitemOpen
  \bibfield  {author} {\bibinfo {author} {\bibfnamefont {T.}~\bibnamefont
  {Teraji}}, \bibinfo {author} {\bibfnamefont {Y.}~\bibnamefont {Garino}},
  \bibinfo {author} {\bibfnamefont {Y.}~\bibnamefont {Koide}}, \ and\ \bibinfo
  {author} {\bibfnamefont {T.}~\bibnamefont {Ito}},\ }\href {\doibase
  http://dx.doi.org/10.1063/1.3153986} {\bibfield  {journal} {\bibinfo
  {journal} {Journal of Applied Physics}\ }\textbf {\bibinfo {volume} {105}},\
  \bibinfo {eid} {126109} (\bibinfo {year} {2009}{\natexlab{a}})}\BibitemShut
  {NoStop}%
\bibitem [{\citenamefont {Baumann}\ and\ \citenamefont
  {Nemanich}(1998)}]{Baumann}%
  \BibitemOpen
  \bibfield  {author} {\bibinfo {author} {\bibfnamefont {P.~K.}\ \bibnamefont
  {Baumann}}\ and\ \bibinfo {author} {\bibfnamefont {R.~J.}\ \bibnamefont
  {Nemanich}},\ }\href {\doibase http://dx.doi.org/10.1063/1.366940} {\bibfield
   {journal} {\bibinfo  {journal} {Journal of Applied Physics}\ }\textbf
  {\bibinfo {volume} {83}},\ \bibinfo {pages} {2072} (\bibinfo {year}
  {1998})}\BibitemShut {NoStop}%
\bibitem [{\citenamefont {Kawarada}(1996)}]{Kawarada}%
  \BibitemOpen
  \bibfield  {author} {\bibinfo {author} {\bibfnamefont {H.}~\bibnamefont
  {Kawarada}},\ }\href {\doibase
  http://dx.doi.org/10.1016/S0167-5729(97)80002-7} {\bibfield  {journal}
  {\bibinfo  {journal} {Surface Science Reports}\ }\textbf {\bibinfo {volume}
  {26}},\ \bibinfo {pages} {205 } (\bibinfo {year} {1996})}\BibitemShut
  {NoStop}%
\bibitem [{\citenamefont {Aoki}\ and\ \citenamefont {Kawarada}(1994)}]{Aoki}%
  \BibitemOpen
  \bibfield  {author} {\bibinfo {author} {\bibfnamefont {M.}~\bibnamefont
  {Aoki}}\ and\ \bibinfo {author} {\bibfnamefont {H.}~\bibnamefont
  {Kawarada}},\ }\href {\doibase 10.1143/JJAP.33.L708} {\bibfield  {journal}
  {\bibinfo  {journal} {Japanese Journal of Applied Physics}\ }\textbf
  {\bibinfo {volume} {33}},\ \bibinfo {pages} {L708} (\bibinfo {year}
  {1994})}\BibitemShut {NoStop}%
\bibitem [{\citenamefont {Alvarez}\ \emph {et~al.}(2006)\citenamefont
  {Alvarez}, \citenamefont {Houz\'{e}}, \citenamefont {Kleider}, \citenamefont
  {Liao},\ and\ \citenamefont {Koide}}]{Alvarez}%
  \BibitemOpen
  \bibfield  {author} {\bibinfo {author} {\bibfnamefont {J.}~\bibnamefont
  {Alvarez}}, \bibinfo {author} {\bibfnamefont {F.}~\bibnamefont {Houz\'{e}}},
  \bibinfo {author} {\bibfnamefont {J.}~\bibnamefont {Kleider}}, \bibinfo
  {author} {\bibfnamefont {M.}~\bibnamefont {Liao}}, \ and\ \bibinfo {author}
  {\bibfnamefont {Y.}~\bibnamefont {Koide}},\ }\href {\doibase
  http://dx.doi.org/10.1016/j.spmi.2006.07.027} {\bibfield  {journal} {\bibinfo
   {journal} {Superlattices and Microstructures}\ }\textbf {\bibinfo {volume}
  {40}},\ \bibinfo {pages} {343 } (\bibinfo {year} {2006})},\ \bibinfo {note}
  {e-MRS 2006 Symposium S: Material Science and Technology of Wide Bandgap
  Semiconductors 2006 Spring Meeting of the European Materials Research
  Society}\BibitemShut {NoStop}%
\bibitem [{\citenamefont {Daicho}\ \emph {et~al.}(2014)\citenamefont {Daicho},
  \citenamefont {Saito}, \citenamefont {Kurihara}, \citenamefont {Hiraiwa},\
  and\ \citenamefont {Kawarada}}]{Daicho}%
  \BibitemOpen
  \bibfield  {author} {\bibinfo {author} {\bibfnamefont {A.}~\bibnamefont
  {Daicho}}, \bibinfo {author} {\bibfnamefont {T.}~\bibnamefont {Saito}},
  \bibinfo {author} {\bibfnamefont {S.}~\bibnamefont {Kurihara}}, \bibinfo
  {author} {\bibfnamefont {A.}~\bibnamefont {Hiraiwa}}, \ and\ \bibinfo
  {author} {\bibfnamefont {H.}~\bibnamefont {Kawarada}},\ }\href {\doibase
  http://dx.doi.org/10.1063/1.4881524} {\bibfield  {journal} {\bibinfo
  {journal} {Journal of Applied Physics}\ }\textbf {\bibinfo {volume} {115}},\
  \bibinfo {eid} {223711} (\bibinfo {year} {2014})}\BibitemShut {NoStop}%
\bibitem [{\citenamefont {Kawarada}\ \emph {et~al.}(2014)\citenamefont
  {Kawarada}, \citenamefont {Tsuboi}, \citenamefont {Naruo}, \citenamefont
  {Yamada}, \citenamefont {Xu}, \citenamefont {Daicho}, \citenamefont {Saito},\
  and\ \citenamefont {Hiraiwa}}]{KawaradaFET}%
  \BibitemOpen
  \bibfield  {author} {\bibinfo {author} {\bibfnamefont {H.}~\bibnamefont
  {Kawarada}}, \bibinfo {author} {\bibfnamefont {H.}~\bibnamefont {Tsuboi}},
  \bibinfo {author} {\bibfnamefont {T.}~\bibnamefont {Naruo}}, \bibinfo
  {author} {\bibfnamefont {T.}~\bibnamefont {Yamada}}, \bibinfo {author}
  {\bibfnamefont {D.}~\bibnamefont {Xu}}, \bibinfo {author} {\bibfnamefont
  {A.}~\bibnamefont {Daicho}}, \bibinfo {author} {\bibfnamefont
  {T.}~\bibnamefont {Saito}}, \ and\ \bibinfo {author} {\bibfnamefont
  {A.}~\bibnamefont {Hiraiwa}},\ }\href {\doibase
  http://dx.doi.org/10.1063/1.4884828} {\bibfield  {journal} {\bibinfo
  {journal} {Applied Physics Letters}\ }\textbf {\bibinfo {volume} {105}},\
  \bibinfo {eid} {013510} (\bibinfo {year} {2014})}\BibitemShut {NoStop}%
\bibitem [{\citenamefont {Tsugawa}\ \emph {et~al.}(2010)\citenamefont
  {Tsugawa}, \citenamefont {Noda}, \citenamefont {Hirose},\ and\ \citenamefont
  {Kawarada}}]{Tsugawa}%
  \BibitemOpen
  \bibfield  {author} {\bibinfo {author} {\bibfnamefont {K.}~\bibnamefont
  {Tsugawa}}, \bibinfo {author} {\bibfnamefont {H.}~\bibnamefont {Noda}},
  \bibinfo {author} {\bibfnamefont {K.}~\bibnamefont {Hirose}}, \ and\ \bibinfo
  {author} {\bibfnamefont {H.}~\bibnamefont {Kawarada}},\ }\href {\doibase
  10.1103/PhysRevB.81.045303} {\bibfield  {journal} {\bibinfo  {journal} {Phys.
  Rev. B}\ }\textbf {\bibinfo {volume} {81}},\ \bibinfo {pages} {045303}
  (\bibinfo {year} {2010})}\BibitemShut {NoStop}%
\bibitem [{\citenamefont {Ikeda}\ \emph {et~al.}(2009)\citenamefont {Ikeda},
  \citenamefont {Umezawa}, \citenamefont {Ramanujam},\ and\ \citenamefont {ichi
  Shikata}}]{Ikeda}%
  \BibitemOpen
  \bibfield  {author} {\bibinfo {author} {\bibfnamefont {K.}~\bibnamefont
  {Ikeda}}, \bibinfo {author} {\bibfnamefont {H.}~\bibnamefont {Umezawa}},
  \bibinfo {author} {\bibfnamefont {K.}~\bibnamefont {Ramanujam}}, \ and\
  \bibinfo {author} {\bibfnamefont {S.}~\bibnamefont {ichi Shikata}},\ }\href
  {\doibase 10.1143/APEX.2.011202} {\bibfield  {journal} {\bibinfo  {journal}
  {Applied Physics Express}\ }\textbf {\bibinfo {volume} {2}},\ \bibinfo
  {pages} {011202} (\bibinfo {year} {2009})}\BibitemShut {NoStop}%
\bibitem [{\citenamefont {Teraji}\ \emph
  {et~al.}(2009{\natexlab{b}})\citenamefont {Teraji}, \citenamefont {Koide},\
  and\ \citenamefont {Ito}}]{Teraji2}%
  \BibitemOpen
  \bibfield  {author} {\bibinfo {author} {\bibfnamefont {T.}~\bibnamefont
  {Teraji}}, \bibinfo {author} {\bibfnamefont {Y.}~\bibnamefont {Koide}}, \
  and\ \bibinfo {author} {\bibfnamefont {T.}~\bibnamefont {Ito}},\ }\href
  {\doibase 10.1002/pssr.200903151} {\bibfield  {journal} {\bibinfo  {journal}
  {physica status solidi (RRL) – Rapid Research Letters}\ }\textbf {\bibinfo
  {volume} {3}},\ \bibinfo {pages} {211} (\bibinfo {year}
  {2009}{\natexlab{b}})}\BibitemShut {NoStop}%
\bibitem [{\citenamefont {Traor\'{e}}\ \emph {et~al.}(2014)\citenamefont
  {Traor\'{e}}, \citenamefont {Muret}, \citenamefont {Fiori}, \citenamefont
  {Eon}, \citenamefont {Gheeraert},\ and\ \citenamefont {Pernot}}]{Traore}%
  \BibitemOpen
  \bibfield  {author} {\bibinfo {author} {\bibfnamefont {A.}~\bibnamefont
  {Traor\'{e}}}, \bibinfo {author} {\bibfnamefont {P.}~\bibnamefont {Muret}},
  \bibinfo {author} {\bibfnamefont {A.}~\bibnamefont {Fiori}}, \bibinfo
  {author} {\bibfnamefont {D.}~\bibnamefont {Eon}}, \bibinfo {author}
  {\bibfnamefont {E.}~\bibnamefont {Gheeraert}}, \ and\ \bibinfo {author}
  {\bibfnamefont {J.}~\bibnamefont {Pernot}},\ }\href {\doibase
  http://dx.doi.org/10.1063/1.4864060} {\bibfield  {journal} {\bibinfo
  {journal} {Applied Physics Letters}\ }\textbf {\bibinfo {volume} {104}},\
  \bibinfo {eid} {052105} (\bibinfo {year} {2014})}\BibitemShut {NoStop}%
\bibitem [{\citenamefont {Tung}(2014)}]{Tung2014}%
  \BibitemOpen
  \bibfield  {author} {\bibinfo {author} {\bibfnamefont {R.~T.~è.}\
  \bibnamefont {Tung}},\ }\href {\doibase http://dx.doi.org/10.1063/1.4858400}
  {\bibfield  {journal} {\bibinfo  {journal} {Applied Physics Reviews}\
  }\textbf {\bibinfo {volume} {1}},\ \bibinfo {eid} {011304} (\bibinfo {year}
  {2014}),\ http://dx.doi.org/10.1063/1.4858400}\BibitemShut {NoStop}%
\bibitem [{\citenamefont {Arnault}()}]{Arnault}%
  \BibitemOpen
  \bibfield  {author} {\bibinfo {author} {\bibfnamefont {J.-C.}\ \bibnamefont
  {Arnault}},\ }\href@noop {} {}\bibinfo {note} {Unpublished}\BibitemShut
  {NoStop}%
\bibitem [{\citenamefont {Rhoderick}(1972)}]{Rhoderick1}%
  \BibitemOpen
  \bibfield  {author} {\bibinfo {author} {\bibfnamefont {E.~H.}\ \bibnamefont
  {Rhoderick}},\ }\href {\doibase 10.1088/0022-3727/5/10/324} {\bibfield
  {journal} {\bibinfo  {journal} {Journal of Physics D: Applied Physics}\
  }\textbf {\bibinfo {volume} {5}},\ \bibinfo {pages} {1920} (\bibinfo {year}
  {1972})}\BibitemShut {NoStop}%
\bibitem [{\citenamefont {Rhoderick}\ and\ \citenamefont
  {Williams}(1988)}]{Rhoderick2}%
  \BibitemOpen
  \bibfield  {author} {\bibinfo {author} {\bibfnamefont {E.~H.}\ \bibnamefont
  {Rhoderick}}\ and\ \bibinfo {author} {\bibfnamefont {R.~H.}\ \bibnamefont
  {Williams}},\ }\href@noop {} {\emph {\bibinfo {title} {Metal-Semiconductor
  Contacts}}},\ \bibinfo {edition} {2nd}\ ed.\ (\bibinfo  {publisher}
  {Clarendon Press},\ \bibinfo {address} {Oxford},\ \bibinfo {year}
  {1988})\BibitemShut {NoStop}%
\bibitem [{\citenamefont {M\"{o}nch}(2004)}]{Monchbook}%
  \BibitemOpen
  \bibfield  {author} {\bibinfo {author} {\bibfnamefont {W.}~\bibnamefont
  {M\"{o}nch}},\ }\href@noop {} {\emph {\bibinfo {title} {Electronic Properties
  of Semiconductor Interfaces}}},\ Surface Sciences\ (\bibinfo  {publisher}
  {Springer-Verlag},\ \bibinfo {address} {Berlin Heidelberg},\ \bibinfo {year}
  {2004})\BibitemShut {NoStop}%
\bibitem [{\citenamefont {Crowell}\ and\ \citenamefont
  {Beguwala}(1971)}]{Crowell}%
  \BibitemOpen
  \bibfield  {author} {\bibinfo {author} {\bibfnamefont {C.~R.}\ \bibnamefont
  {Crowell}}\ and\ \bibinfo {author} {\bibfnamefont {M.}~\bibnamefont
  {Beguwala}},\ }\href {\doibase 10.1016/0038-1101(71)90027-X} {\bibfield
  {journal} {\bibinfo  {journal} {Solid-State Electronics}\ }\textbf {\bibinfo
  {volume} {14}},\ \bibinfo {pages} {1149} (\bibinfo {year}
  {1971})}\BibitemShut {NoStop}%
\bibitem [{\citenamefont {Card}\ and\ \citenamefont {Rhoderick}(1971)}]{Card}%
  \BibitemOpen
  \bibfield  {author} {\bibinfo {author} {\bibfnamefont {H.}~\bibnamefont
  {Card}}\ and\ \bibinfo {author} {\bibfnamefont {E.}~\bibnamefont
  {Rhoderick}},\ }\href {\doibase 10.1088/0022-3727/4/10/319} {\bibfield
  {journal} {\bibinfo  {journal} {Journal of Physics D: Applied Physics}\
  }\textbf {\bibinfo {volume} {4}},\ \bibinfo {pages} {1589} (\bibinfo {year}
  {1971})}\BibitemShut {NoStop}%
\bibitem [{\citenamefont {Andrews}\ and\ \citenamefont
  {Lepselter}(1970)}]{Andrews}%
  \BibitemOpen
  \bibfield  {author} {\bibinfo {author} {\bibfnamefont {J.}~\bibnamefont
  {Andrews}}\ and\ \bibinfo {author} {\bibfnamefont {M.}~\bibnamefont
  {Lepselter}},\ }\href {\doibase
  http://dx.doi.org/10.1016/0038-1101(70)90098-5} {\bibfield  {journal}
  {\bibinfo  {journal} {Solid-State Electronics}\ }\textbf {\bibinfo {volume}
  {13}},\ \bibinfo {pages} {1011 } (\bibinfo {year} {1970})}\BibitemShut
  {NoStop}%
\bibitem [{\citenamefont {Ohmagari}\ \emph {et~al.}(2011)\citenamefont
  {Ohmagari}, \citenamefont {Teraji},\ and\ \citenamefont {Koide}}]{Ohmagari}%
  \BibitemOpen
  \bibfield  {author} {\bibinfo {author} {\bibfnamefont {S.}~\bibnamefont
  {Ohmagari}}, \bibinfo {author} {\bibfnamefont {T.}~\bibnamefont {Teraji}}, \
  and\ \bibinfo {author} {\bibfnamefont {Y.}~\bibnamefont {Koide}},\ }\href
  {\doibase 10.1063/1.3626791} {\bibfield  {journal} {\bibinfo  {journal}
  {Journal of Applied Physics}\ }\textbf {\bibinfo {volume} {110}},\ \bibinfo
  {pages} {056105} (\bibinfo {year} {2011})}\BibitemShut {NoStop}%
\bibitem [{\citenamefont {Werner}\ and\ \citenamefont
  {G\"{u}ttler}(1991)}]{Werner}%
  \BibitemOpen
  \bibfield  {author} {\bibinfo {author} {\bibfnamefont {J.~H.}\ \bibnamefont
  {Werner}}\ and\ \bibinfo {author} {\bibfnamefont {H.~H.}\ \bibnamefont
  {G\"{u}ttler}},\ }\href {\doibase http://dx.doi.org/10.1063/1.347243}
  {\bibfield  {journal} {\bibinfo  {journal} {Journal of Applied Physics}\
  }\textbf {\bibinfo {volume} {69}},\ \bibinfo {pages} {1522} (\bibinfo {year}
  {1991})}\BibitemShut {NoStop}%
\bibitem [{\citenamefont {Tung}(1992)}]{Tung1992}%
  \BibitemOpen
  \bibfield  {author} {\bibinfo {author} {\bibfnamefont {R.~T.}\ \bibnamefont
  {Tung}},\ }\href {\doibase 10.1103/PhysRevB.45.13509} {\bibfield  {journal}
  {\bibinfo  {journal} {Physical Review B}\ }\textbf {\bibinfo {volume} {45}},\
  \bibinfo {pages} {13509} (\bibinfo {year} {1992})}\BibitemShut {NoStop}%
\bibitem [{\citenamefont {Tung}(2001)}]{Tung2001}%
  \BibitemOpen
  \bibfield  {author} {\bibinfo {author} {\bibfnamefont {R.~T.}\ \bibnamefont
  {Tung}},\ }\href {\doibase http://dx.doi.org/10.1016/S0927-796X(01)00037-7}
  {\bibfield  {journal} {\bibinfo  {journal} {Materials Science and
  Engineering: R: Reports}\ }\textbf {\bibinfo {volume} {35}},\ \bibinfo
  {pages} {1 } (\bibinfo {year} {2001})}\BibitemShut {NoStop}%
\bibitem [{\citenamefont {Gammon}\ \emph {et~al.}(2013)\citenamefont {Gammon},
  \citenamefont {P\'{e}rez-Tom\`{a}s}, \citenamefont {Shah}, \citenamefont
  {Vavasour}, \citenamefont {Donchev}, \citenamefont {Pang}, \citenamefont
  {Myronov}, \citenamefont {Fisher}, \citenamefont {Jennings}, \citenamefont
  {Leadley},\ and\ \citenamefont {Mawby}}]{Gammon}%
  \BibitemOpen
  \bibfield  {author} {\bibinfo {author} {\bibfnamefont {P.~M.}\ \bibnamefont
  {Gammon}}, \bibinfo {author} {\bibfnamefont {A.}~\bibnamefont
  {P\'{e}rez-Tom\`{a}s}}, \bibinfo {author} {\bibfnamefont {V.~A.}\
  \bibnamefont {Shah}}, \bibinfo {author} {\bibfnamefont {O.}~\bibnamefont
  {Vavasour}}, \bibinfo {author} {\bibfnamefont {E.}~\bibnamefont {Donchev}},
  \bibinfo {author} {\bibfnamefont {J.~S.}\ \bibnamefont {Pang}}, \bibinfo
  {author} {\bibfnamefont {M.}~\bibnamefont {Myronov}}, \bibinfo {author}
  {\bibfnamefont {C.~A.}\ \bibnamefont {Fisher}}, \bibinfo {author}
  {\bibfnamefont {M.~R.}\ \bibnamefont {Jennings}}, \bibinfo {author}
  {\bibfnamefont {D.~R.}\ \bibnamefont {Leadley}}, \ and\ \bibinfo {author}
  {\bibfnamefont {P.~A.}\ \bibnamefont {Mawby}},\ }\href {\doibase
  http://dx.doi.org/10.1063/1.4842096} {\bibfield  {journal} {\bibinfo
  {journal} {Journal of Applied Physics}\ }\textbf {\bibinfo {volume} {114}},\
  \bibinfo {eid} {223704} (\bibinfo {year} {2013}),\
  http://dx.doi.org/10.1063/1.4842096}\BibitemShut {NoStop}%
\bibitem [{\citenamefont {Ristein}\ \emph {et~al.}(2004)\citenamefont
  {Ristein}, \citenamefont {Riedel},\ and\ \citenamefont {Ley}}]{Ristein}%
  \BibitemOpen
  \bibfield  {author} {\bibinfo {author} {\bibfnamefont {J.}~\bibnamefont
  {Ristein}}, \bibinfo {author} {\bibfnamefont {M.}~\bibnamefont {Riedel}}, \
  and\ \bibinfo {author} {\bibfnamefont {L.}~\bibnamefont {Ley}},\ }\href
  {\doibase 10.1149/1.1785797} {\bibfield  {journal} {\bibinfo  {journal}
  {Journal of The Electrochemical Society}\ }\textbf {\bibinfo {volume}
  {151}},\ \bibinfo {pages} {E315} (\bibinfo {year} {2004})}\BibitemShut
  {NoStop}%
\bibitem [{\citenamefont {Klauser}\ \emph {et~al.}(2010)\citenamefont
  {Klauser}, \citenamefont {Ghodbane}, \citenamefont {Boukherroub},
  \citenamefont {Szunerits}, \citenamefont {Steinmüller-Nethl}, \citenamefont
  {Bertel},\ and\ \citenamefont {Memmel}}]{Klauser}%
  \BibitemOpen
  \bibfield  {author} {\bibinfo {author} {\bibfnamefont {F.}~\bibnamefont
  {Klauser}}, \bibinfo {author} {\bibfnamefont {S.}~\bibnamefont {Ghodbane}},
  \bibinfo {author} {\bibfnamefont {R.}~\bibnamefont {Boukherroub}}, \bibinfo
  {author} {\bibfnamefont {S.}~\bibnamefont {Szunerits}}, \bibinfo {author}
  {\bibfnamefont {D.}~\bibnamefont {Steinmüller-Nethl}}, \bibinfo {author}
  {\bibfnamefont {E.}~\bibnamefont {Bertel}}, \ and\ \bibinfo {author}
  {\bibfnamefont {N.}~\bibnamefont {Memmel}},\ }\href {\doibase
  http://dx.doi.org/10.1016/j.diamond.2009.11.013} {\bibfield  {journal}
  {\bibinfo  {journal} {Diamond and Related Materials}\ }\textbf {\bibinfo
  {volume} {19}},\ \bibinfo {pages} {474 } (\bibinfo {year}
  {2010})}\BibitemShut {NoStop}%
\bibitem [{\citenamefont {Ghodbane}\ \emph {et~al.}(2010)\citenamefont
  {Ghodbane}, \citenamefont {Ballutaud}, \citenamefont {Omn\`{e}s},\ and\
  \citenamefont {Agn\`{e}s}}]{Ghodbane}%
  \BibitemOpen
  \bibfield  {author} {\bibinfo {author} {\bibfnamefont {S.}~\bibnamefont
  {Ghodbane}}, \bibinfo {author} {\bibfnamefont {D.}~\bibnamefont {Ballutaud}},
  \bibinfo {author} {\bibfnamefont {F.}~\bibnamefont {Omn\`{e}s}}, \ and\
  \bibinfo {author} {\bibfnamefont {C.}~\bibnamefont {Agn\`{e}s}},\ }\href
  {\doibase http://dx.doi.org/10.1016/j.diamond.2010.01.014} {\bibfield
  {journal} {\bibinfo  {journal} {Diamond and Related Materials}\ }\textbf
  {\bibinfo {volume} {19}},\ \bibinfo {pages} {630 } (\bibinfo {year}
  {2010})}\BibitemShut {NoStop}%
\bibitem [{\citenamefont {Sque}\ \emph {et~al.}(2006)\citenamefont {Sque},
  \citenamefont {Jones},\ and\ \citenamefont {Briddon}}]{Sque}%
  \BibitemOpen
  \bibfield  {author} {\bibinfo {author} {\bibfnamefont {S.~J.}\ \bibnamefont
  {Sque}}, \bibinfo {author} {\bibfnamefont {R.}~\bibnamefont {Jones}}, \ and\
  \bibinfo {author} {\bibfnamefont {P.~R.}\ \bibnamefont {Briddon}},\ }\href
  {\doibase 10.1103/PhysRevB.73.085313} {\bibfield  {journal} {\bibinfo
  {journal} {Phys. Rev. B}\ }\textbf {\bibinfo {volume} {73}},\ \bibinfo
  {pages} {085313} (\bibinfo {year} {2006})}\BibitemShut {NoStop}%
\bibitem [{\citenamefont {Notsu}\ \emph {et~al.}(2000)\citenamefont {Notsu},
  \citenamefont {Yagi}, \citenamefont {Tatsuma}, \citenamefont {Tryk},\ and\
  \citenamefont {Fujishima}}]{Notsu}%
  \BibitemOpen
  \bibfield  {author} {\bibinfo {author} {\bibfnamefont {H.}~\bibnamefont
  {Notsu}}, \bibinfo {author} {\bibfnamefont {I.}~\bibnamefont {Yagi}},
  \bibinfo {author} {\bibfnamefont {T.}~\bibnamefont {Tatsuma}}, \bibinfo
  {author} {\bibfnamefont {D.~A.}\ \bibnamefont {Tryk}}, \ and\ \bibinfo
  {author} {\bibfnamefont {A.}~\bibnamefont {Fujishima}},\ }\href {\doibase
  http://dx.doi.org/10.1016/S0022-0728(00)00254-0} {\bibfield  {journal}
  {\bibinfo  {journal} {Journal of Electroanalytical Chemistry}\ }\textbf
  {\bibinfo {volume} {492}},\ \bibinfo {pages} {31 } (\bibinfo {year}
  {2000})}\BibitemShut {NoStop}%
\bibitem [{\citenamefont {Fink}\ and\ \citenamefont {Jenkins}(2009)}]{Fink}%
  \BibitemOpen
  \bibfield  {author} {\bibinfo {author} {\bibfnamefont {C.~K.}\ \bibnamefont
  {Fink}}\ and\ \bibinfo {author} {\bibfnamefont {S.~J.}\ \bibnamefont
  {Jenkins}},\ }\href {\doibase 10.1088/0953-8984/21/26/264010} {\bibfield
  {journal} {\bibinfo  {journal} {Journal of Physics: Condensed Matter}\
  }\textbf {\bibinfo {volume} {21}},\ \bibinfo {pages} {264010} (\bibinfo
  {year} {2009})}\BibitemShut {NoStop}%
\bibitem [{\citenamefont {Robertson}\ and\ \citenamefont
  {Rutter}(1998)}]{Robertson1}%
  \BibitemOpen
  \bibfield  {author} {\bibinfo {author} {\bibfnamefont {J.}~\bibnamefont
  {Robertson}}\ and\ \bibinfo {author} {\bibfnamefont {M.}~\bibnamefont
  {Rutter}},\ }\href {\doibase http://dx.doi.org/10.1016/S0925-9635(97)00257-4}
  {\bibfield  {journal} {\bibinfo  {journal} {Diamond and Related Materials}\
  }\textbf {\bibinfo {volume} {7}},\ \bibinfo {pages} {620 } (\bibinfo {year}
  {1998})}\BibitemShut {NoStop}%
\bibitem [{\citenamefont {Muret}\ and\ \citenamefont {Saby}(2004)}]{Muret3}%
  \BibitemOpen
  \bibfield  {author} {\bibinfo {author} {\bibfnamefont {P.}~\bibnamefont
  {Muret}}\ and\ \bibinfo {author} {\bibfnamefont {C.}~\bibnamefont {Saby}},\
  }\href {\doibase 10.1088/0268-1242/19/1/001} {\bibfield  {journal} {\bibinfo
  {journal} {Semiconductor Science and Technology}\ }\textbf {\bibinfo {volume}
  {19}},\ \bibinfo {pages} {1} (\bibinfo {year} {2004})},\ \bibinfo {note} {; a
  typographic error in table 2 must be corrected by changing the minus sign
  into plus sign in each of the three last columns at the last line which
  recapitulates the electron affinities, only that of the hydrogenated surface
  being really negative}\BibitemShut {NoStop}%
\bibitem [{\citenamefont {Sirringhaus}\ \emph {et~al.}(1996)\citenamefont
  {Sirringhaus}, \citenamefont {Meyer}, \citenamefont {Lee},\ and\
  \citenamefont {von K\"anel}}]{Sirringhaus}%
  \BibitemOpen
  \bibfield  {author} {\bibinfo {author} {\bibfnamefont {H.}~\bibnamefont
  {Sirringhaus}}, \bibinfo {author} {\bibfnamefont {T.}~\bibnamefont {Meyer}},
  \bibinfo {author} {\bibfnamefont {E.~Y.}\ \bibnamefont {Lee}}, \ and\
  \bibinfo {author} {\bibfnamefont {H.}~\bibnamefont {von K\"anel}},\ }\href
  {\doibase 10.1103/PhysRevB.53.15944} {\bibfield  {journal} {\bibinfo
  {journal} {Phys. Rev. B}\ }\textbf {\bibinfo {volume} {53}},\ \bibinfo
  {pages} {15944} (\bibinfo {year} {1996})}\BibitemShut {NoStop}%
\bibitem [{\citenamefont {Muret}(2014)}]{Muret2}%
  \BibitemOpen
  \bibfield  {author} {\bibinfo {author} {\bibfnamefont {P.~R.}\ \bibnamefont
  {Muret}},\ }\href {\doibase 10.1116/1.4865912} {\bibfield  {journal}
  {\bibinfo  {journal} {Journal of Vacuum Science and Technology B}\ }\textbf
  {\bibinfo {volume} {32}},\ \bibinfo {pages} {03D114} (\bibinfo {year}
  {2014})}\BibitemShut {NoStop}%
\bibitem [{\citenamefont {Michaelson}(1977)}]{Michaelson}%
  \BibitemOpen
  \bibfield  {author} {\bibinfo {author} {\bibfnamefont {H.~B.}\ \bibnamefont
  {Michaelson}},\ }\href {\doibase http://dx.doi.org/10.1063/1.323539}
  {\bibfield  {journal} {\bibinfo  {journal} {Journal of Applied Physics}\
  }\textbf {\bibinfo {volume} {48}},\ \bibinfo {pages} {4729} (\bibinfo {year}
  {1977})}\BibitemShut {NoStop}%
\bibitem [{\citenamefont {Jain}\ and\ \citenamefont {Dahlke}(1986)}]{Jain}%
  \BibitemOpen
  \bibfield  {author} {\bibinfo {author} {\bibfnamefont {S.}~\bibnamefont
  {Jain}}\ and\ \bibinfo {author} {\bibfnamefont {W.~E.}\ \bibnamefont
  {Dahlke}},\ }\href {\doibase http://dx.doi.org/10.1016/0038-1101(86)90140-1}
  {\bibfield  {journal} {\bibinfo  {journal} {Solid-State Electronics}\
  }\textbf {\bibinfo {volume} {29}},\ \bibinfo {pages} {597 } (\bibinfo {year}
  {1986})}\BibitemShut {NoStop}%
\bibitem [{\citenamefont {Collins}\ \emph {et~al.}(1987)\citenamefont
  {Collins}, \citenamefont {Lowe},\ and\ \citenamefont {Barker}}]{Collins}%
  \BibitemOpen
  \bibfield  {author} {\bibinfo {author} {\bibfnamefont {S.}~\bibnamefont
  {Collins}}, \bibinfo {author} {\bibfnamefont {D.}~\bibnamefont {Lowe}}, \
  and\ \bibinfo {author} {\bibfnamefont {J.~R.}\ \bibnamefont {Barker}},\
  }\href {\doibase 10.1088/0022-3719/20/36/022} {\bibfield  {journal} {\bibinfo
   {journal} {Journal of Physics C: Solid State Physics}\ }\textbf {\bibinfo
  {volume} {20}},\ \bibinfo {pages} {6233} (\bibinfo {year}
  {1987})}\BibitemShut {NoStop}%
\bibitem [{\citenamefont {Cox}\ and\ \citenamefont {Strack}(1967)}]{Cox}%
  \BibitemOpen
  \bibfield  {author} {\bibinfo {author} {\bibfnamefont {R.~H.}\ \bibnamefont
  {Cox}}\ and\ \bibinfo {author} {\bibfnamefont {H.}~\bibnamefont {Strack}},\
  }\href {\doibase 10.1016/0038-1101(67)90063-9} {\bibfield  {journal}
  {\bibinfo  {journal} {Solid-State Electronics}\ }\textbf {\bibinfo {volume}
  {10}},\ \bibinfo {pages} {1213} (\bibinfo {year} {1967})}\BibitemShut
  {NoStop}%
\bibitem [{\citenamefont {Defives}\ \emph {et~al.}(1999)\citenamefont
  {Defives}, \citenamefont {Noblanc}, \citenamefont {Dua}, \citenamefont
  {Brylinski}, \citenamefont {Barthula},\ and\ \citenamefont
  {Meyer}}]{Defives}%
  \BibitemOpen
  \bibfield  {author} {\bibinfo {author} {\bibfnamefont {D.}~\bibnamefont
  {Defives}}, \bibinfo {author} {\bibfnamefont {O.}~\bibnamefont {Noblanc}},
  \bibinfo {author} {\bibfnamefont {C.}~\bibnamefont {Dua}}, \bibinfo {author}
  {\bibfnamefont {C.}~\bibnamefont {Brylinski}}, \bibinfo {author}
  {\bibfnamefont {M.}~\bibnamefont {Barthula}}, \ and\ \bibinfo {author}
  {\bibfnamefont {F.}~\bibnamefont {Meyer}},\ }\href {\doibase
  http://dx.doi.org/10.1016/S0921-5107(98)00541-8} {\bibfield  {journal}
  {\bibinfo  {journal} {Materials Science and Engineering: B}\ }\textbf
  {\bibinfo {volume} {61–62}},\ \bibinfo {pages} {395 } (\bibinfo {year}
  {1999})}\BibitemShut {NoStop}%
\bibitem [{\citenamefont {Muret}\ \emph {et~al.}(2015)\citenamefont {Muret},
  \citenamefont {Eon}, \citenamefont {Traor\'{e}}, \citenamefont
  {Mar\'{e}chal}, \citenamefont {Pernot},\ and\ \citenamefont
  {Gheeraert}}]{Muret4}%
  \BibitemOpen
  \bibfield  {author} {\bibinfo {author} {\bibfnamefont {P.}~\bibnamefont
  {Muret}}, \bibinfo {author} {\bibfnamefont {D.}~\bibnamefont {Eon}}, \bibinfo
  {author} {\bibfnamefont {A.}~\bibnamefont {Traor\'{e}}}, \bibinfo {author}
  {\bibfnamefont {A.}~\bibnamefont {Mar\'{e}chal}}, \bibinfo {author}
  {\bibfnamefont {J.}~\bibnamefont {Pernot}}, \ and\ \bibinfo {author}
  {\bibfnamefont {E.}~\bibnamefont {Gheeraert}},\ }\href {\doibase
  10.1002/pssa.201532187} {\bibfield  {journal} {\bibinfo  {journal} {physica
  status solidi (a)}\ ,\ \bibinfo {pages} {n/a}} (\bibinfo {year} {2015})},\
  \bibinfo {note} {; in press}\BibitemShut {NoStop}%
\bibitem [{\citenamefont {Mott}(1939)}]{Mott}%
  \BibitemOpen
  \bibfield  {author} {\bibinfo {author} {\bibfnamefont {N.~F.}\ \bibnamefont
  {Mott}},\ }\href {\doibase 10.1098/rspa.1939.0051} {\bibfield  {journal}
  {\bibinfo  {journal} {Proceedings of the Royal Society of London. Series A.
  Mathematical and Physical Sciences}\ }\textbf {\bibinfo {volume} {171}},\
  \bibinfo {pages} {27} (\bibinfo {year} {1939})}\BibitemShut {NoStop}%
\bibitem [{\citenamefont {Schottky}(1939)}]{Schottky}%
  \BibitemOpen
  \bibfield  {author} {\bibinfo {author} {\bibfnamefont {W.}~\bibnamefont
  {Schottky}},\ }\href {\doibase 10.1007/BF01340116} {\bibfield  {journal}
  {\bibinfo  {journal} {Zeitschrift f{\"u}r Physik}\ }\textbf {\bibinfo
  {volume} {113}},\ \bibinfo {pages} {367} (\bibinfo {year}
  {1939})}\BibitemShut {NoStop}%
\bibitem [{\citenamefont {Tersoff}(1984)}]{Tersoff}%
  \BibitemOpen
  \bibfield  {author} {\bibinfo {author} {\bibfnamefont {J.}~\bibnamefont
  {Tersoff}},\ }\href {\doibase 10.1103/PhysRevLett.52.465} {\bibfield
  {journal} {\bibinfo  {journal} {Phys. Rev. Lett.}\ }\textbf {\bibinfo
  {volume} {52}},\ \bibinfo {pages} {465} (\bibinfo {year} {1984})}\BibitemShut
  {NoStop}%
\bibitem [{\citenamefont {Kon\'{e}}\ \emph {et~al.}(2010)\citenamefont
  {Kon\'{e}}, \citenamefont {Civrac}, \citenamefont {Schneider}, \citenamefont
  {Isoird}, \citenamefont {Issaoui}, \citenamefont {Achard},\ and\
  \citenamefont {Gicquel}}]{Kone}%
  \BibitemOpen
  \bibfield  {author} {\bibinfo {author} {\bibfnamefont {S.}~\bibnamefont
  {Kon\'{e}}}, \bibinfo {author} {\bibfnamefont {G.}~\bibnamefont {Civrac}},
  \bibinfo {author} {\bibfnamefont {H.}~\bibnamefont {Schneider}}, \bibinfo
  {author} {\bibfnamefont {K.}~\bibnamefont {Isoird}}, \bibinfo {author}
  {\bibfnamefont {R.}~\bibnamefont {Issaoui}}, \bibinfo {author} {\bibfnamefont
  {J.}~\bibnamefont {Achard}}, \ and\ \bibinfo {author} {\bibfnamefont
  {A.}~\bibnamefont {Gicquel}},\ }\href {\doibase
  http://dx.doi.org/10.1016/j.diamond.2010.01.036} {\bibfield  {journal}
  {\bibinfo  {journal} {Diamond and Related Materials}\ }\textbf {\bibinfo
  {volume} {19}},\ \bibinfo {pages} {792 } (\bibinfo {year} {2010})},\ \bibinfo
  {note} {proceedings of Diamond 2009, The 20th European Conference on Diamond,
  Diamond-Like Materials, Carbon Nanotubes and Nitrides, Part 2}\BibitemShut
  {NoStop}%
\bibitem [{\citenamefont {Umezawa}\ \emph {et~al.}(2012)\citenamefont
  {Umezawa}, \citenamefont {Nagase}, \citenamefont {Kato},\ and\ \citenamefont
  {ichi Shikata}}]{Umezawa}%
  \BibitemOpen
  \bibfield  {author} {\bibinfo {author} {\bibfnamefont {H.}~\bibnamefont
  {Umezawa}}, \bibinfo {author} {\bibfnamefont {M.}~\bibnamefont {Nagase}},
  \bibinfo {author} {\bibfnamefont {Y.}~\bibnamefont {Kato}}, \ and\ \bibinfo
  {author} {\bibfnamefont {S.}~\bibnamefont {ichi Shikata}},\ }\href {\doibase
  http://dx.doi.org/10.1016/j.diamond.2012.01.011} {\bibfield  {journal}
  {\bibinfo  {journal} {Diamond and Related Materials}\ }\textbf {\bibinfo
  {volume} {24}},\ \bibinfo {pages} {201 } (\bibinfo {year}
  {2012})}\BibitemShut {NoStop}%
\bibitem [{\citenamefont {Nawawi}\ \emph {et~al.}(2013)\citenamefont {Nawawi},
  \citenamefont {Tseng}, \citenamefont {Rusli}, \citenamefont {Amaratunga},
  \citenamefont {Umezawa},\ and\ \citenamefont {Shikata}}]{Nawawi}%
  \BibitemOpen
  \bibfield  {author} {\bibinfo {author} {\bibfnamefont {A.}~\bibnamefont
  {Nawawi}}, \bibinfo {author} {\bibfnamefont {K.}~\bibnamefont {Tseng}},
  \bibinfo {author} {\bibnamefont {Rusli}}, \bibinfo {author} {\bibfnamefont
  {G.}~\bibnamefont {Amaratunga}}, \bibinfo {author} {\bibfnamefont
  {H.}~\bibnamefont {Umezawa}}, \ and\ \bibinfo {author} {\bibfnamefont
  {S.}~\bibnamefont {Shikata}},\ }\href {\doibase
  http://dx.doi.org/10.1016/j.diamond.2013.03.002} {\bibfield  {journal}
  {\bibinfo  {journal} {Diamond and Related Materials}\ }\textbf {\bibinfo
  {volume} {35}},\ \bibinfo {pages} {1 } (\bibinfo {year} {2013})}\BibitemShut
  {NoStop}%
\bibitem [{\citenamefont {Ueda}\ \emph {et~al.}(2013)\citenamefont {Ueda},
  \citenamefont {Kawamoto}, \citenamefont {Soumiya},\ and\ \citenamefont
  {Asano}}]{Ueda1}%
  \BibitemOpen
  \bibfield  {author} {\bibinfo {author} {\bibfnamefont {K.}~\bibnamefont
  {Ueda}}, \bibinfo {author} {\bibfnamefont {K.}~\bibnamefont {Kawamoto}},
  \bibinfo {author} {\bibfnamefont {T.}~\bibnamefont {Soumiya}}, \ and\
  \bibinfo {author} {\bibfnamefont {H.}~\bibnamefont {Asano}},\ }\href
  {\doibase http://dx.doi.org/10.1016/j.diamond.2013.06.007} {\bibfield
  {journal} {\bibinfo  {journal} {Diamond and Related Materials}\ }\textbf
  {\bibinfo {volume} {38}},\ \bibinfo {pages} {41 } (\bibinfo {year}
  {2013})}\BibitemShut {NoStop}%
\bibitem [{\citenamefont {Ueda}\ \emph {et~al.}(2014)\citenamefont {Ueda},
  \citenamefont {Kawamoto},\ and\ \citenamefont {Asano}}]{Ueda2}%
  \BibitemOpen
  \bibfield  {author} {\bibinfo {author} {\bibfnamefont {K.}~\bibnamefont
  {Ueda}}, \bibinfo {author} {\bibfnamefont {K.}~\bibnamefont {Kawamoto}}, \
  and\ \bibinfo {author} {\bibfnamefont {H.}~\bibnamefont {Asano}},\ }\href
  {\doibase 10.7567/JJAP.53.04EP05} {\bibfield  {journal} {\bibinfo  {journal}
  {Japanese Journal of Applied Physics}\ }\textbf {\bibinfo {volume} {53}},\
  \bibinfo {pages} {04EP05} (\bibinfo {year} {2014})}\BibitemShut {NoStop}%
\bibitem [{\citenamefont {Ristein}\ \emph {et~al.}(2000)\citenamefont
  {Ristein}, \citenamefont {Maier}, \citenamefont {Riedel}, \citenamefont
  {Cui},\ and\ \citenamefont {Ley}}]{RisMaier}%
  \BibitemOpen
  \bibfield  {author} {\bibinfo {author} {\bibfnamefont {J.}~\bibnamefont
  {Ristein}}, \bibinfo {author} {\bibfnamefont {F.}~\bibnamefont {Maier}},
  \bibinfo {author} {\bibfnamefont {M.}~\bibnamefont {Riedel}}, \bibinfo
  {author} {\bibfnamefont {J.}~\bibnamefont {Cui}}, \ and\ \bibinfo {author}
  {\bibfnamefont {L.}~\bibnamefont {Ley}},\ }\href {\doibase
  10.1002/1521-396X(200009)181:1<65::AID-PSSA65>3.0.CO;2-Z} {\bibfield
  {journal} {\bibinfo  {journal} {physica status solidi (a)}\ }\textbf
  {\bibinfo {volume} {181}},\ \bibinfo {pages} {65} (\bibinfo {year}
  {2000})}\BibitemShut {NoStop}%
\bibitem [{\citenamefont {Tung}(2000)}]{Tung2000}%
  \BibitemOpen
  \bibfield  {author} {\bibinfo {author} {\bibfnamefont {R.~T.}\ \bibnamefont
  {Tung}},\ }\href {\doibase 10.1103/PhysRevLett.84.6078} {\bibfield  {journal}
  {\bibinfo  {journal} {Phys. Rev. Lett.}\ }\textbf {\bibinfo {volume} {84}},\
  \bibinfo {pages} {6078} (\bibinfo {year} {2000})}\BibitemShut {NoStop}%
\bibitem [{\citenamefont {Tung}(1993)}]{Tung1993}%
  \BibitemOpen
  \bibfield  {author} {\bibinfo {author} {\bibfnamefont {R.~T.}\ \bibnamefont
  {Tung}},\ }\href {\doibase http://dx.doi.org/10.1116/1.586967} {\bibfield
  {journal} {\bibinfo  {journal} {Journal of Vacuum Science and Technology B}\
  }\textbf {\bibinfo {volume} {11}},\ \bibinfo {pages} {1546} (\bibinfo {year}
  {1993})}\BibitemShut {NoStop}%
\end{thebibliography}%
    
\end{document}